\documentclass{aa}

\usepackage{graphicx}
\usepackage{siunitx}
\usepackage{natbib}
\usepackage{color}
\usepackage{xcolor} 
\usepackage{pdflscape}
\usepackage{afterpage}
\usepackage{flafter}
\usepackage{fancyhdr}
\usepackage{pdflscape}
\fancypagestyle{lscape}{%
\fancyhf{} 
\fancyfoot[LE]{}
\fancyfoot[LO] {}
 
}
\usepackage{booktabs}

\usepackage{txfonts}
\usepackage{xcolor}
\definecolor{hotpink}{rgb}{1.0, 0.41, 0.71}
\definecolor{dodgerblue}{rgb}{0.12, 0.56, 1.0}
\usepackage[colorlinks=true, citecolor=dodgerblue, linkcolor=hotpink, urlcolor=hotpink]{hyperref}

\begin{document} 

\title{Identifying galaxies, quasars, and stars with machine learning: A new catalogue of classifications for 111 million SDSS sources without spectra}

\titlerunning{Identifying galaxies, quasars, and stars with machine learning}
\authorrunning{A. O. Clarke et al.}


   \author{A. O. Clarke\inst{1},
          A. M. M. Scaife\inst{1}, 
          R. Greenhalgh\inst{1}
          \and
          V. Griguta\inst{1}
          }

   \institute{Jodrell Bank Centre For Astrophysics\\
              \email{alex.clarke-3@manchester.ac.uk} \\
             }

    \date{Received 19 September, 2019; accepted 28 April, 2020}

\abstract{
    We used 3.1 million spectroscopically labelled sources from the Sloan Digital Sky Survey (SDSS) to train an optimised random forest classifier using photometry from the SDSS and the Widefield Infrared Survey Explorer (WISE). We applied this machine learning model to 111 million previously unlabelled sources from the SDSS photometric catalogue which did not have existing spectroscopic observations. Our new catalogue contains 50.4 million galaxies, 2.1 million quasars, and 58.8 million stars. We provide individual classification probabilities for each source, with 6.7 million galaxies (13\%), 0.33 million quasars (15\%), and 41.3 million stars (70\%) having classification probabilities greater than 0.99; and 35.1 million galaxies (70\%), 0.72 million quasars (34\%), and 54.7 million stars (93\%) having classification probabilities greater than 0.9. Precision, Recall, and F$_1$ score were determined as a function of selected features and magnitude error. We investigate the effect of class imbalance on our machine learning model and discuss the implications of transfer learning for populations of sources at fainter magnitudes than the training set. We used a non-linear dimension reduction technique, Uniform Manifold Approximation and Projection (UMAP), in unsupervised, semi-supervised, and fully-supervised schemes to visualise the separation of galaxies, quasars, and stars in a two-dimensional space. When applying this algorithm to the 111 million sources without spectra, it is in strong agreement with the class labels applied by our random forest model. 
    }


   \maketitle

\section{Introduction}

The classification scheme of galaxies, quasars, and stars is one of the most fundamental in astronomy. The early cataloguing of stars and their distribution in the sky has led to the understanding that they make up our own galaxy \citep{herschel1789} and, following the distinction that Andromeda was a separate galaxy to our own \citep{Opik1922, hubble1929}, numerous galaxies began to be surveyed as more powerful telescopes were built. The designation of quasars arose after radio emission was detected from unresolved star-like sources with high redshifts \citep[e.g. 3C48 and 3C273;][]{3C48-1961, 3C48-1963, 3C48-1963B, 3C273-1963, 3C48-3C273-1964}. This emission was later demonstrated to have been produced by accretion disks surrounding super-massive black holes at the centre of some galaxies \citep{AGN-1963, AGN-1984, AGN-radio-1984}. For quasars, the emission from this central region on scales less than a light year, known as an active galactic nucleus \citep[AGN;][]{AGNreview1995}, is expected to dominate over the light from the host galaxy. Large samples of quasars \citep[e.g.][]{SDSS-dr14-quasars-2018} are now routinely selected through the identification of characteristic high-ionisation emission lines in their optical spectrum \citep[e.g. C \textrm{IV}, Mg \textrm{II};][]{quasarspec-1991, quasarspec-SDSS-2001}, as well as via spectroscopic follow-up of optical sources where a radio counterpart is also present \citep{lofar-quasars-2019}. Whilst quasars are typically located at redshifts high enough to be unresolved with optical telescopes, some nearby resolved galaxies \citep[Seyfert galaxies;][]{seyfert1977, seyfert2012} are also labelled as quasars, having bright and compact cores with associated emission lines from AGN, although they are comparatively less bright than high-redshift quasars.

Providing classification labels for astronomical catalogues containing large numbers of sources has a wide range of benefits, both for studies of individual systems and for statistical population analyses. In particular, a significant range of science goals are dependent on large samples of quasars, which are still the minority class. This consideration has been important in motivating the construction of new facilities, such as the Square Kilometre Array \citep[SKA;][]{SKA2015} and the Large Synoptic Survey Telescope \citep[LSST;][]{LSST2009, LSST2012, LSST2019}. Science objectives reliant on quasar samples include Lyman-$\alpha$ forest surveys \citep{quasar-lyaforest-1998, quasar-lyaforest-2007}, cosmic magnetism studies \citep[][]{quasar-cosmicmagnetism-2005}, general cosmology \citep{quasar-powerspectrum-2014, quasar-pol-alignment-2005}, and the evolution of galaxies \citep{quasar-evolution1983, quasar-origins1988, galaxyquasar-evolution2000}, amongst others. 

Millions of sources have already been catalogued from telescopes such as the Sloan Digital Sky Survey \citep[SDSS;][]{SDSSDR152019}, the Wide-field Infrared Survey Explorer \citep[WISE;][]{WISE2010} and the LOw Frequency ARray \citep[LOFAR;][]{LOFAR2013, LOTSS2019}, amongst others. The next generation of telescopes are predicted to significantly increase the size of source catalogues. The LSST is expected to catalogue approximately 20 billion galaxies and a similar number of stars \citep{LSST2019}. Source count predictions for the SKA indicate a source density of around ten galaxies per square arcminute for phase one of the array, and up to 75 galaxies per square arcminute for phase two (assuming a detection threshold of $S_{1~GHz} \geq 100 \mathrm{nJy}$). This results in totals of 1 and 8 billion sources, respectively, for a survey area of $3\pi$ steradians \citep{SKA2015}. Consequently, it is becoming unfeasible for astronomers to manually verify and label individual sources and, whilst efforts such as Galaxy Zoo \citep{galaxyzoo2008, galaxyzoo2011} bring in many more people to help sift through data, this effort alone is not expected to be able to keep up with the source counts anticipated for the next generation of telescopes. For such large datasets, machine learning algorithms are becoming an increasingly valuable tool for analysis and data exploration. The development of such algorithms in computer science fields has accelerated rapidly in the last decade, focusing on processing large datasets in high performance computing workflows and cloud computing systems \citep{MLreview-2014, bigdata2016}.

Although distinguishing astronomical source type is normally straightforward where detailed data are available, such as spectroscopy and multi-wavelength observations, the complexity of obtaining detailed observations for millions of individual sources is time consuming and generally impractical for the largest samples of sources given the survey speeds of the current generation of telescopes. In contrast, classifying sources using only photometry in multiple wavebands, and labelling them based on their colours (differences between pairs of photometry measurements), is comparatively fast. With three orders of magnitude less wavelength coverage than spectroscopy, photometry cannot capture the same detail as spectra, however it can capture the overall shape of the spectrum that distinguishes different types of sources. In the optical, stars show a black-body spectrum, galaxies show the superposition of many black-body spectra, and quasars show a comparatively flatter spectrum due to AGN emission. The most widely used colours for source classification are $u-g$ and $g-r$ in SDSS data, and $w1-w2$ and $w2-w3$ in WISE data \citep{WISE-colours-2014, quasarcolours2015}. Consequently, photometry data have been demonstrated to be useful as machine learning features in source-type classification by a number of studies \citep[e.g.][]{rf-quasars-2015, rf-quasars-2019,rf-BCUs-2019, rf-quasars-KIDS-2019, rf-qsg-2019}. Furthermore, testing whether a source is resolved or unresolved can help distinguish the extended profiles of galaxies from stars and quasars \citep{SDSSDR152019, Baldry2010, xan2018} and serve as a useful machine learning feature.

In this work, we apply supervised, unsupervised and semi-supervised machine learning algorithms to classify galaxies, quasars, and stars using SDSS and WISE data. In Section \ref{section:data}, we introduce the SDSS and WISE data and describe our feature set. In Section \ref{section:ML}, we introduce the machine learning algorithms, describe how the models were optimised using a spectroscopically selected training dataset, and give an in-depth evaluation of the model performance as a function of source magnitude and classification. In Section \ref{section:newsources}, we apply our optimised model to an SDSS photometric catalogue of 111 million previously unclassified sources. In Section \ref{section:discussion}, we discuss the results, and in Section \ref{section:conclusions} we draw our conclusions. All of the code used in this paper to gather, process, and analyse the data is available on our Github repository\footnote{\url{https://www.github.com/informationcake/SDSS-ML/}}.

\section{Data}\label{section:data}

\begin{figure}
\includegraphics[width=\hsize]{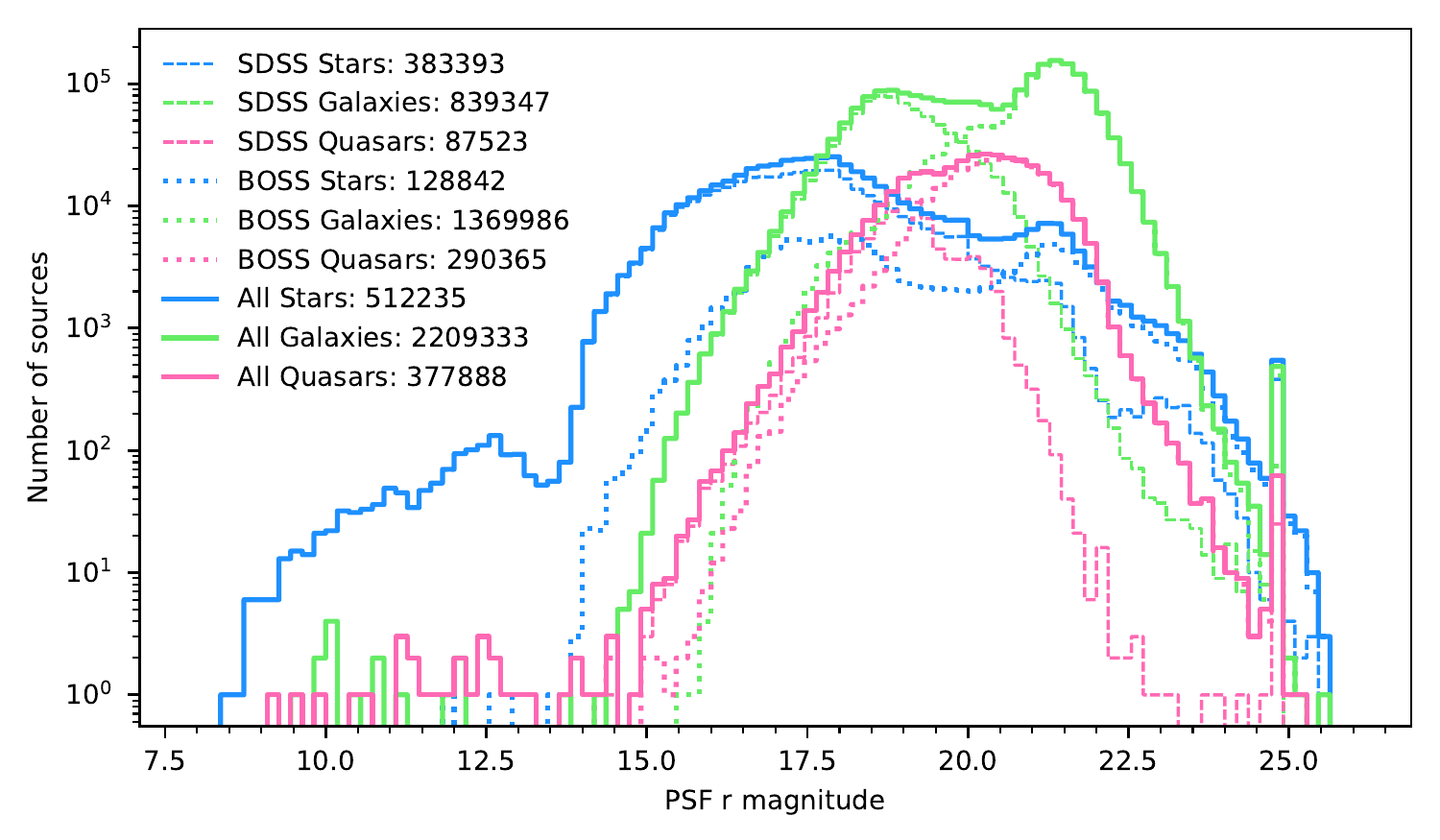}
\caption{Histogram of 3.1 million spectroscopically observed galaxies, quasars, and stars from both the SDSS and BOSS instruments (after removing unclean data). There is a double peaked distribution for galaxies and quasars due to each instrument having different target magnitude selection criteria.}
\label{figure:traintest-hist-rmag}
\end{figure}

Two datasets were constructed from the SDSS Data Release 15 \citep[DR15][]{SDSSDR152019} catalogue. These comprise (i) a labelled dataset of spectroscopically observed sources and (ii) an unlabelled dataset of photometrically observed sources with no associated spectroscopic observation. In both cases we selected sources that have WISE counterparts.

For SDSS, photometric measurements are provided in five optical bands: $u \, (\lambda = 0.355 \,\mu\mathrm{m})$, $g \, (\lambda = 0.477\,\mu\mathrm{m}$), $r \, (\lambda = 0.623\,\mu\mathrm{m})$, $i \, (\lambda = 0.762\,\mu\mathrm{m})$ and $z \, (\lambda = 0.913\,\mu\mathrm{m})$, with  associated errors. WISE provides photometric measurements in four infrared bands: $ \mathit{w1} \, (\lambda = 3.4\,\mu\mathrm{m})$, $\mathit{w2} \, (\lambda = 4.6\,\mu\mathrm{m})$, $\mathit{w3} \, (\lambda = 12\,\mu\mathrm{m})$ and $ \mathit{w4} \, (\lambda = 22\,\mu\mathrm{m})$ with their associated errors. SDSS photometric measurements, as described in \citep[][]{SDSS_catdescription2002}, are optimised for different types of source. For unresolved point sources (e.g. stars and quasars) that are well-described by the point spread function (PSF), the best measure of the total flux is determined by fitting a PSF model to the source, referred to as \textit{psfMag}. However, for resolved sources such as galaxies this is not the case, and a better measure of total flux comes from a model fitted to a source's radial profile. There are several of these model magnitudes associated with each catalog source. The \textit{devMag} and \textit{expMag} magnitudes are associated with de Vaucouleurs and exponential model fits, respectively. These magnitudes are calculated from independent models in each of the five bands. In addition, there is the \textit{modelMag}, which uses the better of the two fits in the \textit{r}-band as a matched aperture to calculate the flux in all bands. For extended sources, this option provides the best measurement of colours due to the flux being measured over equivalent apertures across all bands. A composite model magnitude \textit{cmodelMag} is also defined, taking the best fit from either the de Vaucouleurs or the exponential model in each band and obtaining a linear combination of the two that best fits the image. In practice, \textit{cmodelMag} is the optimum total flux indicator, agreeing well with \textit{psfMag} and \textit{modelMag}. However, it does not result in as high a signal-to-noise measurement for colours compared to \textit{modelMag}.

For each source we determine if it is resolved or point-like by calculating the difference between the \textit{cmodel} magnitude and the PSF magnitude in the \textit{r}-band: 

\begin{equation}
resolved_r = | psf_r - cmod_r | .
\label{equation:resolvedr}
\end{equation}

\subsection{Spectroscopic data}
\label{sec:spectro}

The SDSS data release 15 includes 3\,238\,003 unique sources, spectroscopically observed and cross-matched with detections in the WISE catalogue. Each source in the catalogue is labelled as "STAR", "GALAXY" or "QSO" (quasar) depending on the outcome of fitting various pre-defined models to the spectrum of each source \citep{SDSS_class_pipeline2012}. We obtained the DR15 data by submitting Structured Query Language (SQL) jobs to the CasJobs component of SkyServer API\footnote{\url{http://skyserver.sdss.org/casjobs/}}.
We then applied various conditions upon retrieving the data to ensure the dataset is clean, with minimal contamination from sources which could have incorrect classifications or problematic photometry.

Firstly, 21\,086 sources have more than one WISE match within five arc-seconds. We removed the duplicate entries, only keeping the entry with the closest WISE match. Next, we removed 130 sources that have -9999 for the \textit{cmodel} magnitudes where the fit failed. Sources with spectra that are known to have problems are indicated in the DR15 dataset by the \textit{zwarning} flag. We only selected sources where this flag has a value of either 0 or 16. If the flag is zero, this indicates the spectra has no problems. If the flag is 16, this indicates the "MANY\_OUTLIERS" warning which is only present for data taken with the SDSS spectrograph and not with the BOSS spectrograph, and usually indicates a high signal-to-noise spectrum or broad emission lines in a galaxy. Consequently, it rarely signifies a true error. These \textit{zwarning} conditions removed 116\,026 sources, 92\,349 of which are due to the "SMALL\_DELTA\_CHI2" warning ($\mathit{zwarning}=4$), which indicates that the chi-squared value of the best fit spectrum is too close to that of the second best. Finally, we removed 1304 sources where the value of any of the WISE band magnitudes were set to 9999, where extracting the magnitude failed. 

In total the cleaning process described above removed 138\,546 sources: 90\,969 galaxies, 38\,890 quasars, and 8687 stars. This left a total of 3\,099\,457 sources, with 2\,209\,333 galaxies, 377\,888 quasars, and 512\,236 stars. The class ratio is approximately 5.8:1.0:1.4. Within the resulting dataset, 1\,789\,194 sources were observed using the BOSS spectrograph (58\%) and 1\,310\,263 were observed using the SDSS spectrograph (42\%). Due to the different survey goals for each of these spectrographs, there is a difference in the magnitude distribution for the sources measured in each case. This is illustrated in Figure \ref{figure:traintest-hist-rmag}, where a double peaked distribution for both galaxies and quasars is evident from the dotted and dashed lines of the BOSS and SDSS spectrographs respectively.
We note that due to WISE having a slightly lower resolution than SDSS, 13\,904 sources in our spectroscopic dataset have a duplicate WISE match due to multiple SDSS sources being in close proximity. For these SDSS sources in close proximity, their mean separation is 2.36 arcseconds, with a standard deviation of 1.38 arcseconds.

\subsection{Photometric data}
\label{sec:photo}

The complete SDSS photometric catalogue contains 1\,231\,051\,050 entries, including repeat observations. To create the catalogue of previously unclassified photometrically observed sources, we used sources from the SDSS \textit{PhotoPrimary} table with WISE matches that have no associated spectroscopic observation and clean photometry (where the flag \textit{clean}=1). We limited the dataset to have SDSS (and WISE) magnitudes between 0 and 35, which prevented contamination from spurious nonphysical values. This resulted in a catalogue of 111\,447\,786 sources. 52\,318 SDSS sources were found to be associated with more than one WISE source. We removed duplicate WISE associations by selecting the WISE source closest in angular separation to the SDSS position. Following these selection steps, the remaining dataset contained 111\,395\,468 unclassified photometrically observed sources. We note that 14\,768\,549 of these sources have a duplicate WISE match due to multiple SDSS sources being in close proximity. For these SDSS sources, their mean separation is 2.40 arcseconds, with a standard deviation of 1.38 arcseconds.

\subsection{Feature Set}
We used SDSS and WISE photometry, and the $resolved_r$ parameter, as features in our machine learning models. We did not explicitly calculate colours, and instead gave all five SDSS bands, and all four WISE bands as individual features. We used the dataset of spectroscopically observed sources to train, validate, and test a model to predict the class labels which are spectroscopically confirmed (we do not use the spectra itself as a feature or in the analysis). We used this model to predict the class labels on the second dataset of unlabelled photometrically observed sources that do not have spectra in SDSS.

\section{The machine learning model} \label{section:ML}

In this work we use a random forest \citep[thoroughly reviewed in][]{RF2014} as a supervised learning algorithm to classify galaxies, quasars, and stars. A random forest is an ensemble of independent decision trees where each individual tree is trained on a random subset of both features and data samples. The predicted classification for a new data sample comes from a majority consensus classification across the full set of approximately uncorrelated machine learning models from all the trees in the forest. These principles make a random forest robust to over-fitting, and minimise the variance and bias in its predictions. In addition, random forests are a commonly used algorithm for supervised learning problems due to a number of other strengths: they can deal with numerical and categorical features over different scales, are effective at multi-class problems, and naturally return classification probabilities and feature importance rankings.

We divide our spectroscopically-classified dataset into a training dataset and a testing dataset. The fraction of the complete training set used for each category is discussed in Section~\ref{sec:split}. The training set is used to train the random forest classifier and fit the machine learning model. It must be large enough for the random forest to extract a sufficiency of information on the expected types of galaxies, quasars, and stars and their relation to the features we are using. The test dataset is used to derive a variety of performance metrics which assess how the machine learning model would perform on unseen data. These metrics will be used to assess the confidence of the classifications when applying the model to unlabelled sources without spectra. When creating these two subsets of data, we ensure that each contains the same ratio (5.8:1.0:1.4) of the three different classes, and that they have the same distribution in feature space, in order not to bias training, validation or testing.

Instead of splitting out a fixed validation dataset we implement a cross-validation scheme during training to tune the hyper-parameters, described in more detail in Section~\ref{sec:hyper}.
Validation of this type is used to ensure that the fitted machine learning model does not over- or under-fit the training data.

We use precision, recall, and F$_1$ score as metrics to assess the performance of the model:
\begin{equation}
Precision = \frac{\mathrm{TP}}{\mathrm{TP} + \mathrm{FP}},
\label{equation:precision}
\end{equation}
indicating how good the classifier is at identifying true positives (TP), which are the correctly identified sources. A low precision for an individual class would indicate a low fraction of positive identifications.

\begin{equation}
Recall = \frac{\mathrm{TP}}{\mathrm{TP} + \mathrm{FN}}
\label{equation:recall}
\end{equation}
indicates how good the classifier is at minimising false negatives. A low recall for an individual class would indicate it is often misclassified as another class.

\begin{equation}
F_1 =  \frac{2\mathrm{TP}}{2\mathrm{TP} + \mathrm{FP} + \mathrm{FN}}
\label{equation:f1}
\end{equation}
is the harmonic mean of precision and recall and is used as an overall performance metric. In this multi-class scheme these metrics are calculated per class to show the relative performance of each class.

\subsection{Training data volume}
\label{sec:split}

\begin{figure}
\includegraphics[width=\hsize]{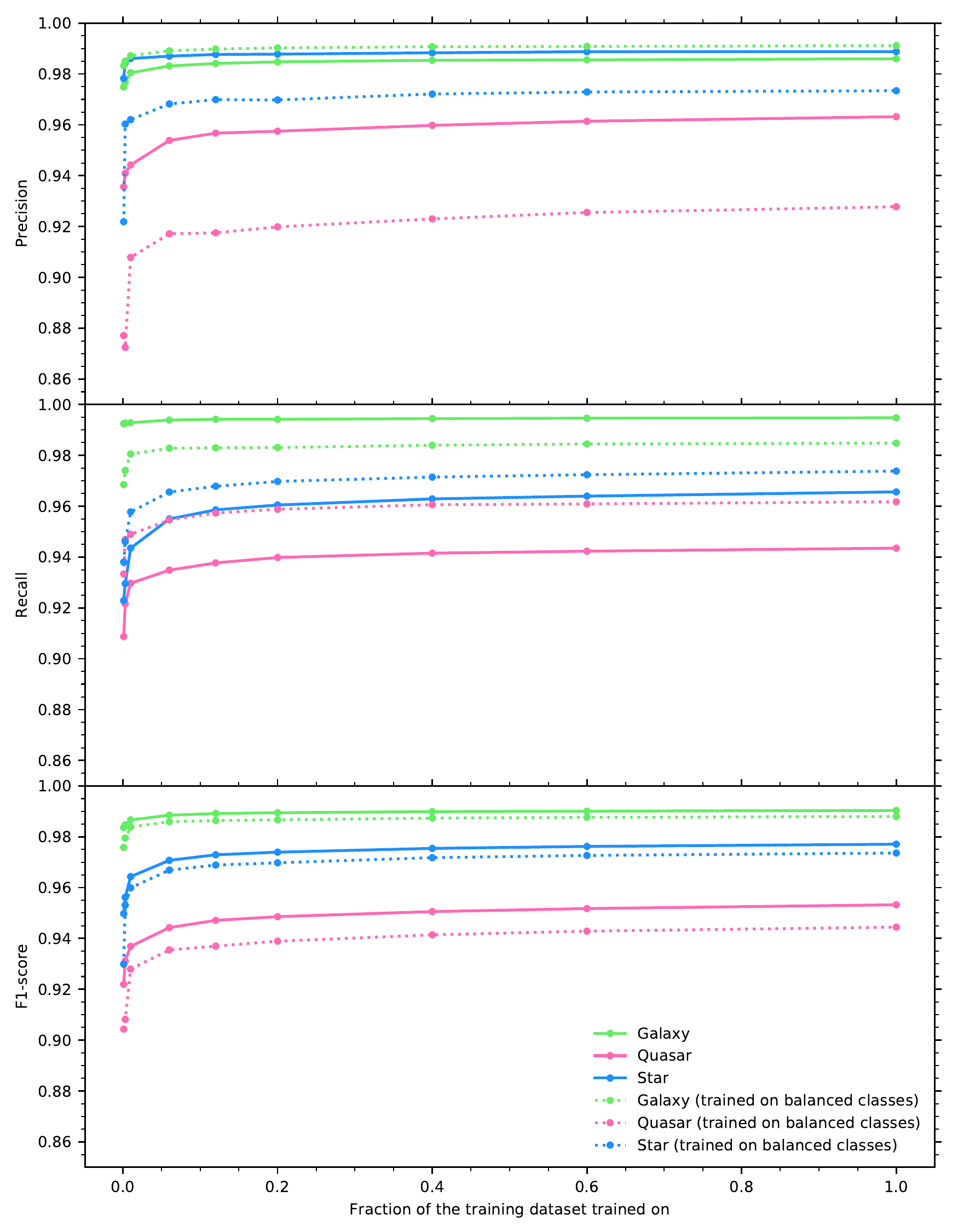}
\caption{Precision (\textit{top}), recall (\textit{middle}), and F$_1$ score (\textit{bottom}) per class as a function of the fraction of the training dataset (1.55 million sources) used to train the random forest. Balancing the classes was done by taking 20\% of the galaxies in the training set. All models were evaluated on the test dataset of 1.55 million spectroscopically confirmed sources, without balancing the classes. Class weights inversely proportional to the class frequency were used in the training in all cases.}
\label{figure:train-vs-f1score}
\end{figure}

Given the large number of sources available to us in the spectroscopic catalogue, we initially split the full dataset into two halves, using one half as a preliminary training set, and the other as a preliminary test set. By taking increasing fractions of the full training set to train a random forest model, we can use these models to predict the class labels of the sources in a consistent test dataset and assess how the volume of data in the training set affects the model performance.

During training we set a weight per class which is inversely proportional to the frequency of that class in the dataset. This is used to prevent the imbalance between the different classes, primarily caused by an excess of galaxies, influencing the fit. 
Figure~\ref{figure:train-vs-f1score} shows how the F$_1$ score per class varies as a function of the fraction of the full training set used to fit the model. The performance increase becomes linear for all classes when using more than 60\% of the of the full training dataset, equivalent to 30\% of all sources with spectra. When using fewer than 60\% of the samples in the full training set the classes show a non-linear increase in performance that is more significant for the minority classes (quasars and stars). This effect is due to the training data subset not capturing a broad enough range of examples for all classes and indicates that a minimum of 60\% of the training dataset, equivalent to 30\% of the full spectroscopic catalogue, should be used to fit the machine learning model in order to avoid biased results. 

Given the class imbalance in our dataset that arises from there being approximately 4 times more galaxies than stars, and 6 times more than quasars, we repeat the procedure described above including only 20\% of the galaxies in the training set. Balancing the three classes in the training dataset whilst maintaining the same unbalanced test dataset gives the result shown by a dashed line in Figure~\ref{figure:train-vs-f1score}. The resulting F$_1$ scores for the model are lower for all classes, decreasing by the same fraction for all fractions of the training dataset. There is a small increase in precision for galaxies, but a larger decrease in precision for quasars and stars. There is a decrease in recall for galaxies, but an increase in recall for quasars and stars. By limiting the number of galaxies available to train on, the model is hindered on all classes, with a particularly large drop in the precision of quasars (a large increase in false positives in this class from galaxies misclassified as quasars). Overall our approach of including weights in the fitting of the random forest is robust to the galaxy class imbalance without requiring us to sub-sample this class. Reducing the number of galaxies in the training set results in a poorer model for all classes.

\subsection{Hyper-parameter optimisation}
\label{sec:hyper}

Random forest algorithms are inherently designed to prevent over-fitting through their selection of random subsets of data and features during training. However, tuning hyper-parameters in the random forest model can lead to over- or under-fitting if set incorrectly, and so we use a validation dataset when optimising the model. To do this we use cross-validation scheme \citep{cross-validation1968} with a five-fold split.
This means that our training dataset is randomly split up into five parts, and each part is used validate the model trained on the other four parts combined. Thus, the model is trained and validated on five different subsets of the training dataset, where each one is referred to as a fold. 80\% of our full training dataset (40\% of the full spectroscopic catalogue) is used to fit the machine learning model in each fold, and 20\% of our training set (10\% of the full spectroscopic catalogue) is used for validating each fold.  Given the results described in Section~\ref{sec:split}, this reduction in the full training dataset will not have an adverse effect on the performance of our model applied to each fold.

We optimize over three main hyper-parameters that influence the fitting of the random forest model: the number of trees in the forest (\textit{n\_estimators)}, the maximum number of features per decision tree (\textit{max\_features}), and the minimum number of samples required for a decision tree to split (\textit{min\_samples\_leaf}). All other hyper-parameters are left at their default values whilst tuning these parameters one at a time. 

We found that using 200 estimators in the random forest gave the best performance whilst minimising run-time. Using 1000 trees did not provide a notable increase in performance over 200, but required five times longer to train the model. This is consistent with the expected computational complexity for random forests, which scales linearly with the number of estimators \citep{RF2014}. Increasing the \textit{min\_samples\_leaf} parameter in the range from one to 500 always results in a large drop in the F$_1$ score, and so we keep it set to one. Adjusting the \textit{max\_features} parameter in the range from two to six results in a very small increase in F$_1$ score. However, given that we only have ten features, substantially increasing this parameter will lead to over-fitting. We therefore use the the default value which is the square root of the number of features, which is rounded down to three in our case.

After optimising the hyper-parameters individually, we verify the optimisation by re-fitting the random forest using all three optimised hyper-parameters together, whilst also implementing five-fold cross-validation. This demonstrated that the model had not over- or under-fitted the training data, and would generalise to an unseen dataset. 

Finally, to maximise the volume of data available for training, we re-fit the machine learning model without implementing a cross-validation scheme. This uses the complete training dataset (50\% of the full spectroscopic catalogue) and the random forest with optimised hyper-parameters. The resulting machine learning model is our production model.

\begin{table*}
\caption{Performance metrics derived from applying the random forest model to the test dataset of 1.55 million spectroscopically confirmed sources. SDSS PSF magnitudes are in bands \textit{u, g, r, i, z}, WISE magnitudes are in bands \textit{w1, w2, w3, w4}.}\label{table:performance}
\centering
\begin{tabular}{lccccccccc} 
\toprule
Features & \multicolumn{3}{c}{Precision}  & \multicolumn{3}{c}{Recall} & \multicolumn{3}{c}{F$_1$ score} \\
\cmidrule(lr){2-4} \cmidrule(lr){5-7} \cmidrule(lr){8-10}
 & Galaxy & Quasar & Star & Galaxy & Quasar & Star & Galaxy & Quasar & Star \\
\midrule
  SDSS + WISE + resolved$_{\rm r}$ & 0.987 & 0.961 & 0.991 & 0.995 & 0.944 & 0.965 & 0.991 & 0.952 & 0.978 \\
  SDSS + WISE  & 0.981 & 0.957 & 0.989 & 0.993 & 0.929 & 0.958 & 0.987 & 0.943 & 0.973 \\
  SDSS  & 0.956 & 0.905 & 0.958 & 0.980 & 0.865 & 0.888 & 0.968 & 0.885 & 0.922 \\
  SDSS + resolved$_{\rm r}$  & 0.981 & 0.948 & 0.979 & 0.993 & 0.924 & 0.945 & 0.987 & 0.936 & 0.962 \\
  WISE & 0.880 & 0.886 & 0.711 & 0.954 & 0.862 & 0.466 & 0.915 & 0.874 & 0.563  \\
\midrule
\bottomrule
\end{tabular}
\end{table*}

\subsection{Feature optimisation}
\label{sec:features}

A number of other measurements are available for each source in the SDSS catalogue, which might alternatively be used as machine learning features. In order to establish that there was no advantage in using these alternative features, we evaluated the model performance for a variety of feature iterations using the test dataset. For each of these feature iterations, we also repeated the procedures in Section~\ref{sec:hyper} to establish that the changes to the feature set were not significant enough to affect the optimal hyper-parameter values. The precision, recall, and F$_1$ scores per class are shown in Table~\ref{table:performance} for each different set of features. 

Including the $resolved_r$ parameter (Equation \ref{equation:resolvedr}) as a feature gives an improved classification performance for all classes, mostly for the recall of quasars (Table~\ref{table:performance}). This is illustrated in Figure~\ref{figure:resolved-features} where we see a large improvement in the classification of unresolved stars and quasars, and resolved galaxies. We use the \textit{r}-band magnitude for this measure as it has the highest signal to noise. We see the performance decrease when using either \textit{u}- or \textit{z}-bands for this purpose due to the poorer signal to noise in these bands, agreeing with similar work by \citep{xan2018}. Using the three high signal to noise bands, \textit{g, r, i}, to calculate the resolved nature of a source as three separate features does not increase performance as this does not provide the classifier with any additional information. Any increase in signal to noise from doing this is insignificant when also using the nine other photometric features. Consequently we use only the \textit{r}-band value in order to minimise the number of features.

Excluding WISE magnitudes from the features gives worse results for all classes. This affects quasars the most with the F$_1$ score dropping from 0.952 to 0.936, which is due to an increase in false negatives (quasars classified as galaxies or stars) with recall being lower by 0.02, and an increase in false positives (stars and galaxies classified as quasars), with precision lower by 0.014. Only using WISE magnitudes as features significantly lowers the F$_1$ scores for all classes, particularly so for stars.

Whilst a correction for total Galactic extinction is provided per band for SDSS sources \citep{extinction1998}, with a median and standard deviation of $ 0.13 \pm 0.30$, $0.10 \pm 0.24$, $0.07 \pm 0.16$, $0.05 \pm 0.12$, $0.04 \pm 0.09$ for the \textit{u}, \textit{g}, \textit{r}, \textit{i}, \textit{z} bands, respectively, applying this correction does not alter the performance of the model for any class. However, this correction is only appropriate for extra-galactic sources and when we apply our model to the unlabelled SDSS photometric catalogue we will not know if sources are galactic or extra-galactic. Therefore we do not include SDSS extinction-corrected magnitudes in our feature set.

Using \textit{cmodel} magnitudes in place of PSF magnitudes for the SDSS bands gives a poorer performance across all target classes, with the most significant effect evident for stars. Using a combination of cmodel and PSF magnitudes provides no improvement over using PSF magnitudes alone.

\begin{figure}
\includegraphics[width=\hsize]{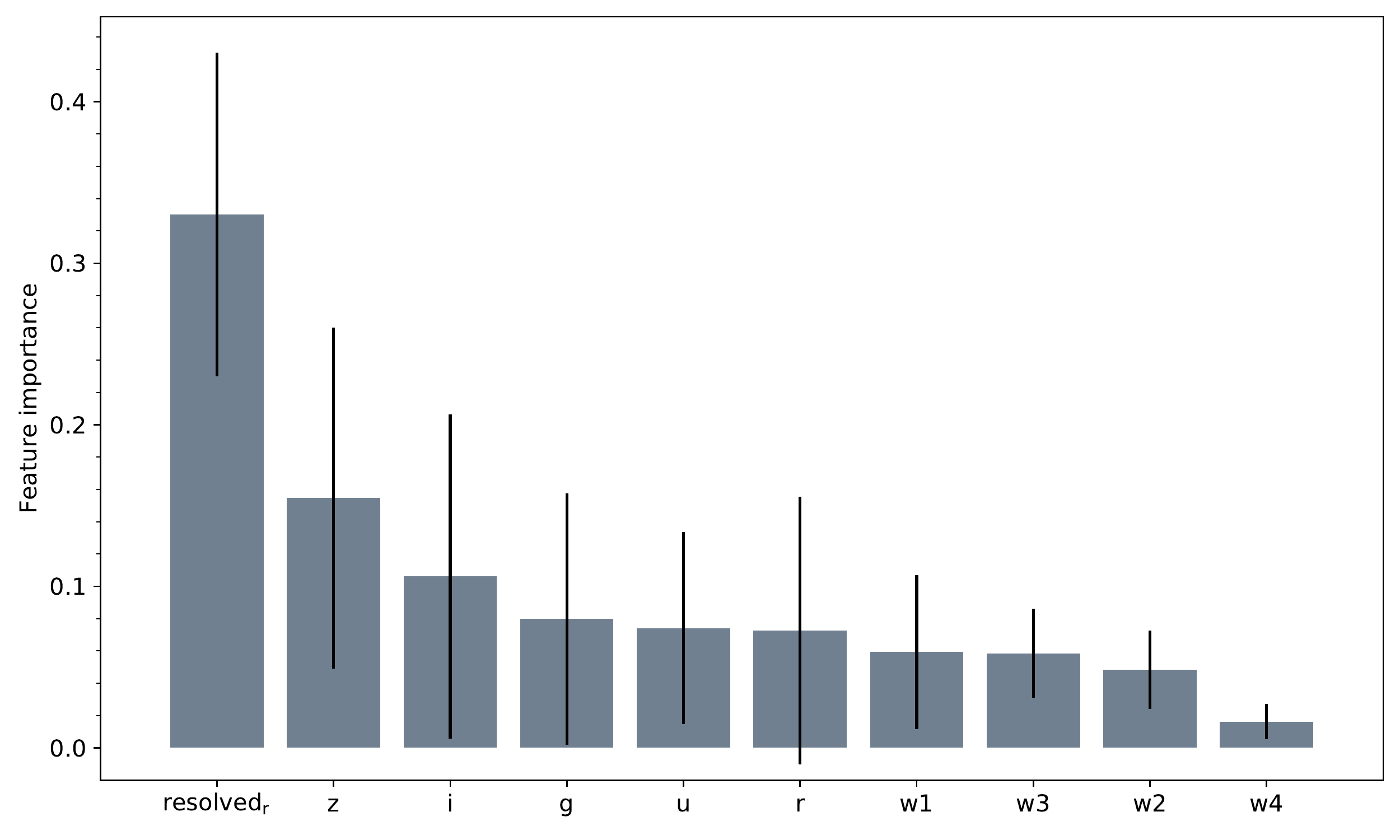}
\caption{Relative feature importances and their one standard deviation returned by the random forest classifier.}
\label{figure:feature-ranking}
\end{figure}

Overall, the optimal combination of features includes the SDSS PSF magnitudes, the WISE magnitudes and the $resolved_r$ parameter. The relative weight assigned to each feature from the resulting model is shown in Figure~\ref{figure:feature-ranking}.

\subsection{Performance evaluation}
\label{sec:performance}

\begin{figure*}
\includegraphics[width=\hsize]{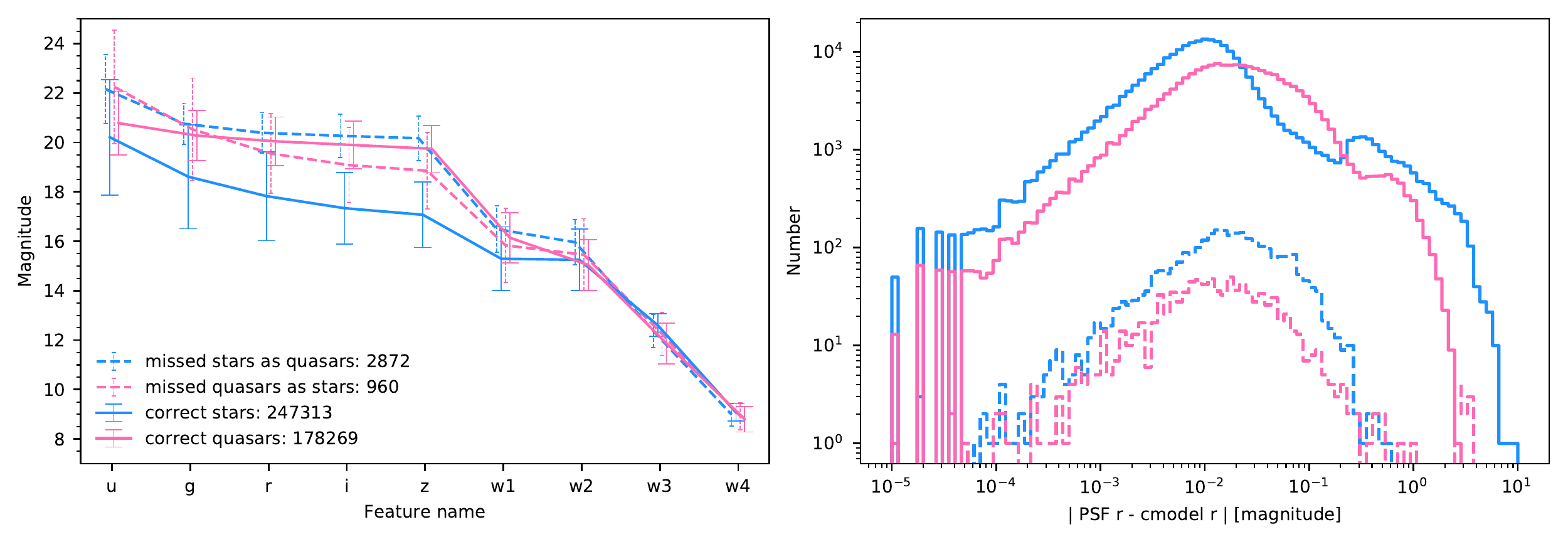}
\includegraphics[width=\hsize]{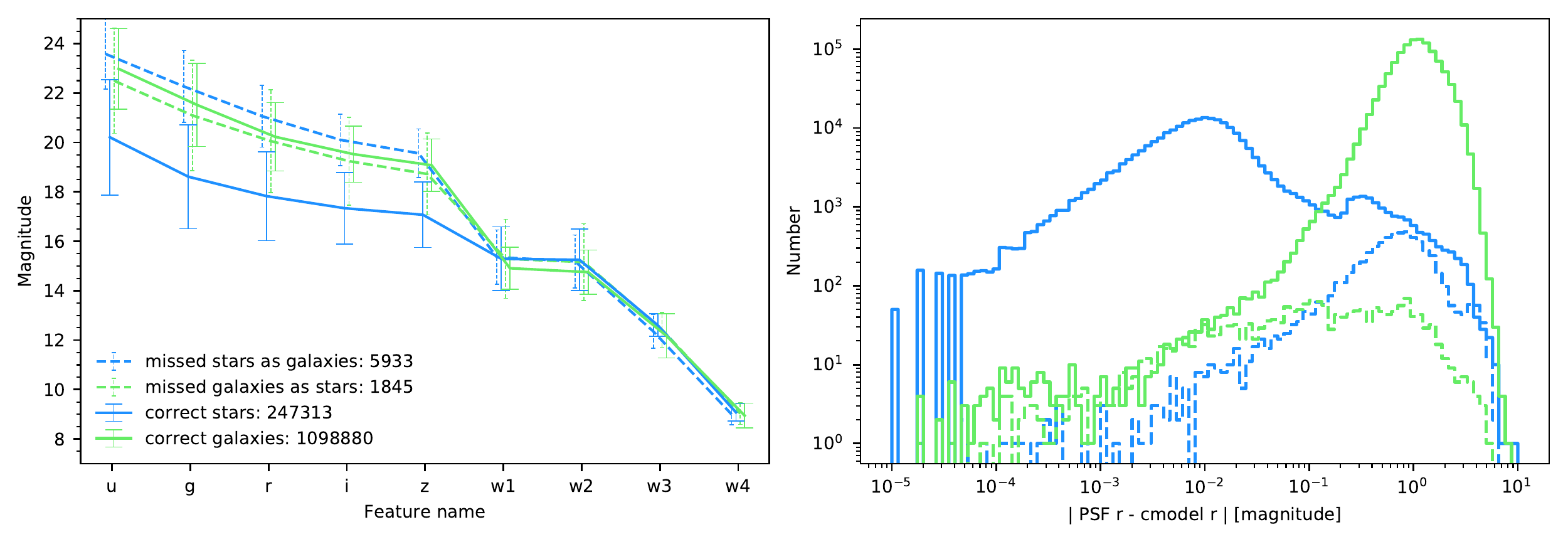}
\includegraphics[width=\hsize]{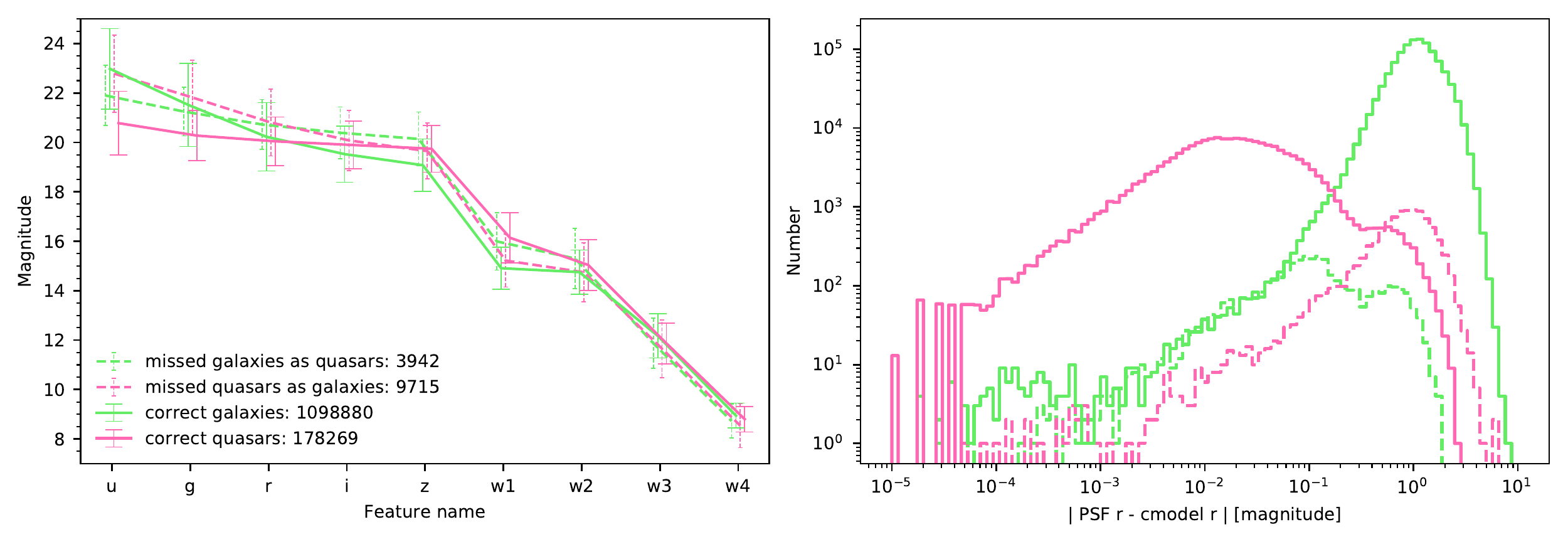}
\caption{\textit{Left}: Average magnitude and one standard deviation error bar for each feature (waveband) for correct and missed sources, per class. Error bars and lines are offset in the x-axis per feature for clarity. The resolved feature was also used in the model. The F$_1$ score is 0.990 for galaxies, 0.953 for quasars and 0.977 for stars. \textit{Right}: Histogram of the $resolved_r$ parameter per class, for correct and missed sources.}
\label{figure:features}
\end{figure*}

\begin{figure}
\includegraphics[width=\hsize]{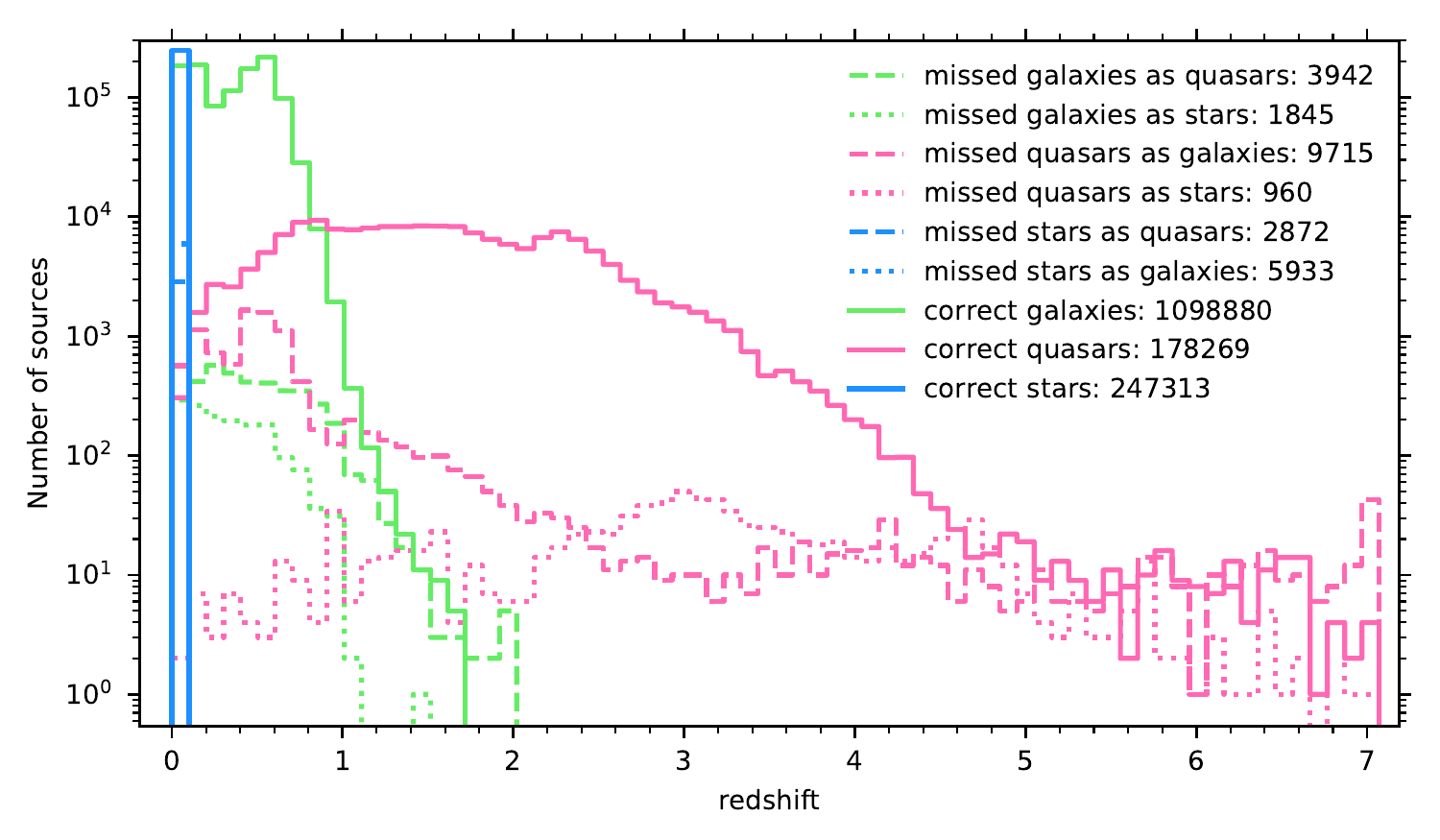}
\caption{Histogram of spectroscopic redshifts for correct and missed sources per class from the random forest model applied to the test dataset of 1.55 million spectroscopically confirmed sources.}
\label{figure:z-hist}
\end{figure}

The random forest model derived in Section \ref{sec:hyper} was applied to our test dataset of 1.55 million spectroscopically confirmed sources (50\% of our full spectroscopic catalogue) using the optimal combination of features described in Section~\ref{sec:features}. Figure~\ref{figure:features} shows the how the features vary for correctly identified (correct) and missed sources per class.
There are 1\,098\,880 correct galaxies, 178\,269 correct quasars, and 247\,313 correct stars. There are 3942 galaxies that were missed as quasars, 9715 quasars missed as galaxies, 5933 stars missed as galaxies, 1845 galaxies missed as stars, 2872 stars missed as quasars, and 960 quasars missed as stars. Figure~\ref{figure:features} shows the distribution of correctly identified and missed sources as a function of feature space. The average magnitude and one standard deviation range is shown for the SDSS and WISE bands on the left side in Figure~\ref{figure:features}, and a histogram of the $resolved_r$ parameter is shown on the right side. This is shown pairwise for all classes (top: quasar/star, middle: galaxy/star, bottom: galaxy/quasar) indicating which feature ranges contain particular sources that have been misclassified as others. Furthermore, a histogram of each of the SDSS and WISE magnitude features is shown in Figure~\ref{figure:histmatrix-mag}, per class, for correct and missed sources. In summary, it can be seen that correctly identified galaxies and stars show a similarly shaped spectrum across the SDSS bands, with galaxies being brighter in the WISE bands. Correctly identified quasars show a flatter spectrum across the SDSS bands compared to correctly identified galaxies and stars. Correctly identified quasars also have a much larger $w1-w2$ difference than galaxies and stars. Correctly identified quasars and stars are mostly unresolved, with similar distributions, whilst correctly identified galaxies are mostly resolved.

\begin{figure*}
\centering
\includegraphics[width=0.16\hsize]{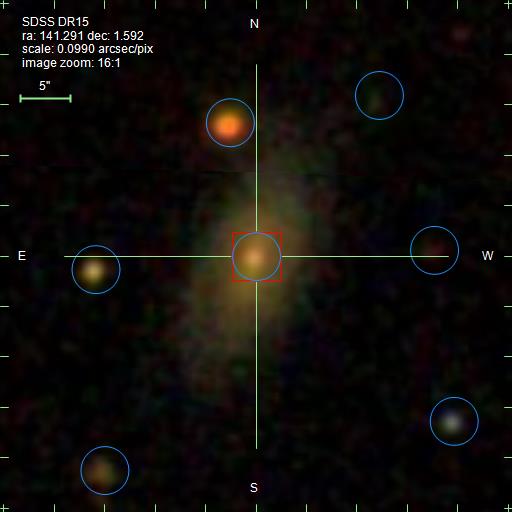}
\includegraphics[width=0.16\hsize]{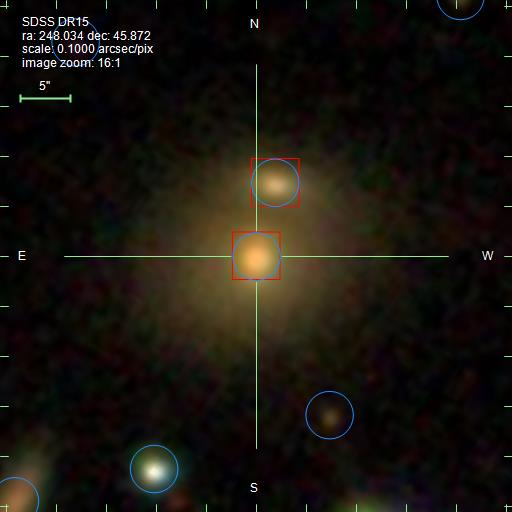}
\includegraphics[width=0.16\hsize]{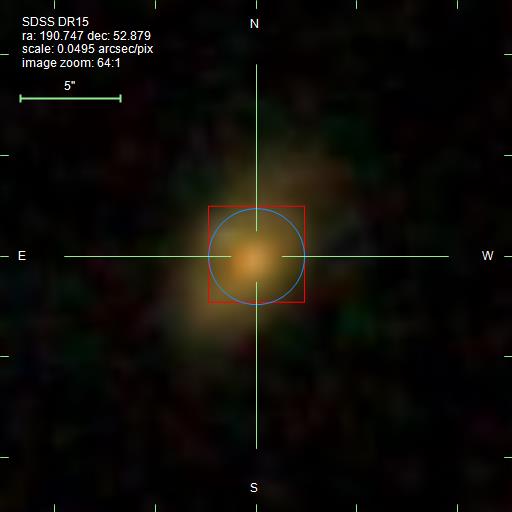}
\includegraphics[width=0.16\hsize]{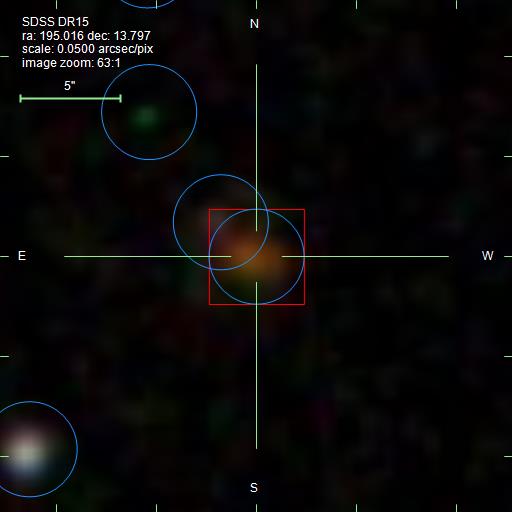}
\includegraphics[width=0.16\hsize]{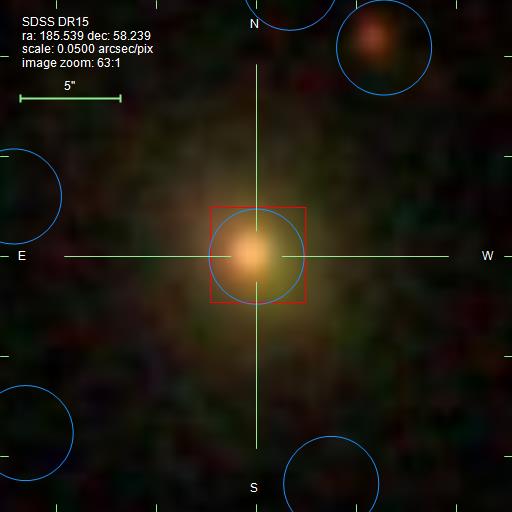}
\includegraphics[width=0.16\hsize]{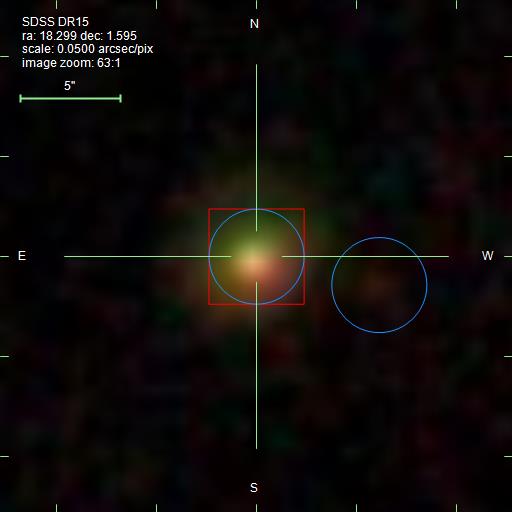}
\includegraphics[width=0.16\hsize]{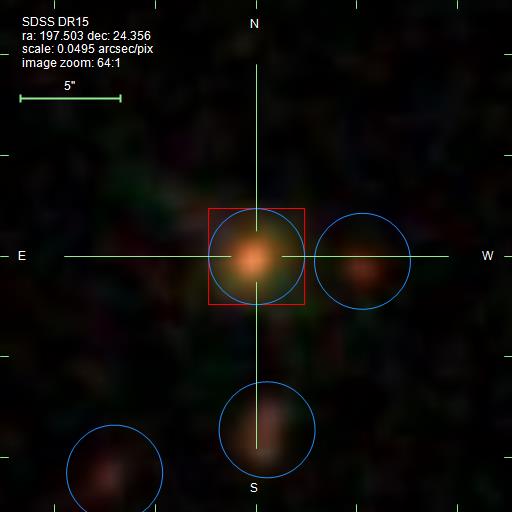}
\includegraphics[width=0.16\hsize]{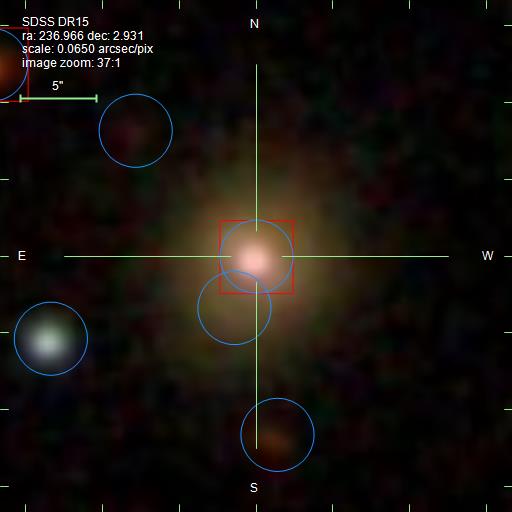}
\includegraphics[width=0.16\hsize]{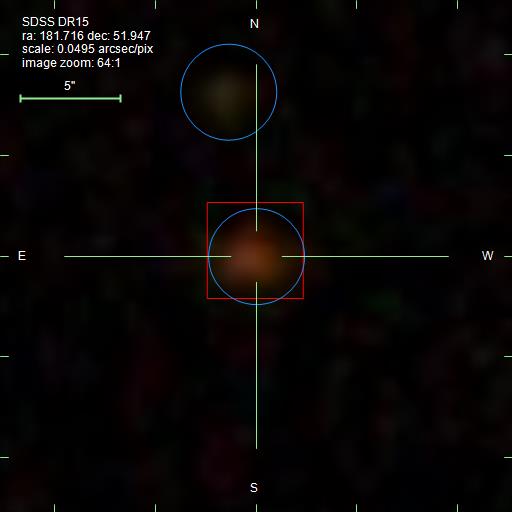}
\includegraphics[width=0.16\hsize]{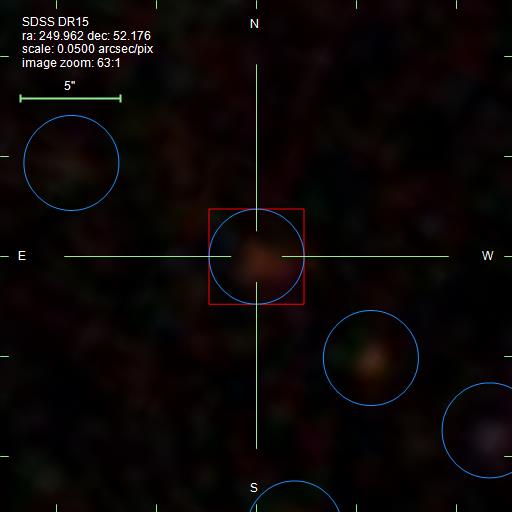}
\includegraphics[width=0.16\hsize]{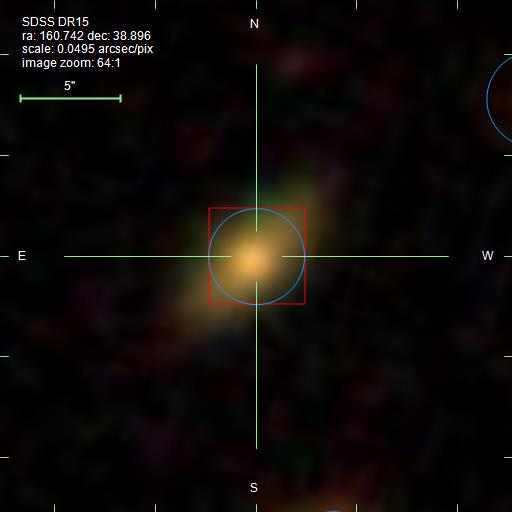}
\includegraphics[width=0.16\hsize]{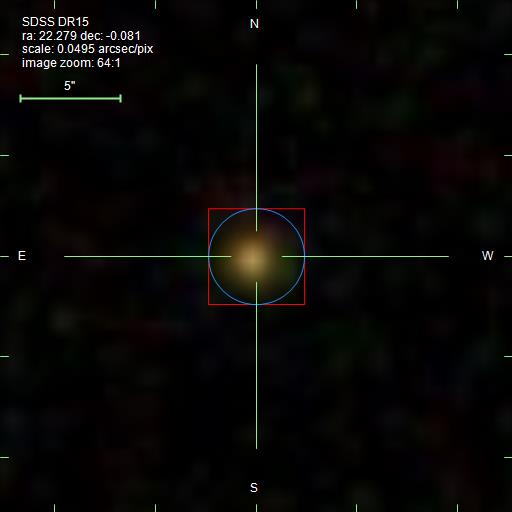}
\caption{Examples of quasars missed as galaxies. The spectroscopically observed target is at the centre in a red box, and photometrically observed sources are circled in blue. From top left to bottom right in order of how resolved they are, their SpecObjIDs, $resolved_r$ parameters and redshifts are:
5331182333197262848, 2.544 (z=0.137); 706011020694415360, 1.655 (z=0.058); 996574564462913536, 1.603 (z=0.170); 6097903957849382912, 1.347 (z=0.527); 1481775858556364800, 1.163 (z=0.100); 8848470304915169280, 1.047 (z=0.238); 6729619569105674240, 0.97151 (z=0.142); 668808218473949184, 0.968 (z=0.098); 7523363890043789312, 0.899 (z=0.592); 9072765177980493824, 0.841 (z=0.802); 1612303285509711872, 0.782 (z=0.107) and 9897866743979679744, 0.631 (z=0.337).}

\label{figure:QasGexamples}
\end{figure*}

Quasars missed as galaxies (9715 sources) appear similar to galaxies across both the SDSS and WISE bands, and are more often resolved (bottom row of Figures~\ref{figure:features} and \ref{figure:histmatrix-mag}). Most of these sources have redshifts less than 1 (see Figure~\ref{figure:z-hist}). They can appear as resolved galaxies where the light from the galaxy dominates over the AGN, giving a galaxy-like spectrum across the SDSS and WISE bands whilst there is still an AGN present (e.g. the top row of Figure~\ref{figure:QasGexamples}). Alternatively, they can be more unresolved galaxies that are much redder (e.g. the bottom row of Figure~\ref{figure:QasGexamples}), although their $resolved_r$ parameter is still much larger on average than point source quasars and stars (shown by the dashed pink line in the bottom right panel of Figure \ref{figure:features}). Their distribution per magnitude feature in SDSS and WISE is shown in Figure~\ref{figure:histmatrix-mag}. In particular, their distribution has a shape that follows that of the correctly identified galaxies (shown by the dashed pink line in the bottom row of Figure \ref{figure:histmatrix-mag}). Overall, quasars missed as galaxies tend to be resolved objects (at low redshifts) which are redder in colour.

Galaxies missed as quasars (3942 sources) appear similar to quasars in the WISE bands, and have a spectrum across the SDSS bands of a similar shape to quasars but are fainter than quasars on average. Compared to correctly identified galaxies, they tend to be unresolved (shown by the dashed green line in the bottom right panel of Figure \ref{figure:features}), and have a similar distribution in redshift from $z=0$ to $z=1$ (Figure \ref{figure:z-hist}). Whilst number density of galaxies drops significantly above $z=1$, higher redshift galaxies are more often misclassified as quasars than stars. Whilst we do not use redshift as a feature, we expect the photometry to be correlated with redshift, and therefore for this effect to be seen in our classification scheme.  In the SDSS bands galaxies missed as quasars appear bluer on average than correctly identified galaxies, with a flatter spectrum. Figure~\ref{figure:GasQexamples} shows a selection of examples of galaxies missed as quasars. Their distribution per magnitude feature in SDSS and WISE is shown in the bottom row of Figure~\ref{figure:histmatrix-mag}.

Galaxies missed as stars (1845 sources) look similar to correctly identified galaxies in the SDSS bands, but are fainter in the WISE bands (middle row of Figures~\ref{figure:features} and \ref{figure:histmatrix-mag}). Figure~\ref{figure:GasSexamples} shows a selection of examples of these sources. The top-left image in Figure~\ref{figure:GasSexamples} is a case where the SDSS classification pipeline is incorrect, and our model has correctly identified it as a star. There may be a small number other cases like this, although without manually inspecting all of them (which we have not done), it is not possible to find these cases automatically with this dataset. Their distribution in redshift and the $resolved_r$ parameter is similar to correctly identified galaxies (middle row of Figures~\ref{figure:features} and \ref{figure:z-hist}).

Stars missed as galaxies (5933 sources) have a similarly shaped spectrum to stars and galaxies within the SDSS and WISE bands, but are much fainter in the SDSS bands, and mostly resolved (middle row of Figures \ref{figure:features} and \ref{figure:histmatrix-mag}). The top row of Figure~\ref{figure:SasGexamples} shows examples of these resolved sources with $resolved_r > 1$. The first image shows a galaxy with a foreground star at its centre. The second image shows a galaxy with a foreground star just off-centre, which is close and bright enough to dominate the spectrograph, hence why the SDSS pipeline has incorrectly labelled it as a star whilst our algorithm has correctly labelled it as a galaxy. The fifth image shows a foreground star on top of a galaxy where the SDSS pipeline has correctly labelled both the star and the background galaxy (which was in our training set) from their spectra. However our algorithm has missed this star as a galaxy, likely due to contamination from the background galaxy, which is 0.35 magnitudes brighter in the PSF \textit{r}-band, and the large value of the $resolved_r$ parameter. The bottom row of Figure~\ref{figure:SasGexamples} shows the rarer examples of more unresolved sources ($resolved_r < 0.3$), where the reason for the misclassification is predominantly due to these stars having a spectrum across the SDSS and WISE bands that is very similar to galaxies.

Quasars missed as stars (960 sources) tend to be fainter in the $w2$-band than correctly identified quasars, which results in a smaller $w1-w2$ difference, and have a spectrum across the SDSS bands similar to stars (top row of Figures \ref{figure:features} and \ref{figure:histmatrix-mag}). They are much fainter in the \textit{u}-band than correct quasars making them appear redder on average. They have a similar distribution in the $resolve_r$ parameter to correct quasars, correct stars and missed stars (top right panel of Figure \ref{figure:features}). They tend to be at higher redshifts since lower redshift quasars are more likely to be resolved and misclassified as galaxies (shown in Figure \ref{figure:z-hist}). The main reason for their misclassification is due to the shape of their spectrum within the SDSS and WISE bands, and being very faint in the \textit{u}-band. Some examples of these are shown in Figure~\ref{figure:QasSexamples}.

Stars missed as quasars (2872 sources) tend to be much fainter across all bands than correctly identified stars and have a larger $w1-w2$ difference (top row of Figures \ref{figure:features} and \ref{figure:histmatrix-mag}). They are significantly fainter in the \textit{z}-band than correct stars, giving them a flatter spectrum across the SDSS bands. They have a similar distribution in the $resolved_r$ parameter to correct stars, correct quasars and missed quasars. The main reason for their misclassification is due to the shape of their spectrum across the SDSS and WISE bands, and being very faint sources where there are much fewer stars in the training set. Some examples of these are shown in Figure~\ref{figure:QasSexamples}.

\subsubsection{Precision, recall, and F$_1$ score}

The average precision, recall, and F$_1$ scores per class are shown in Table \ref{table:performance}. These metrics represent the overall performance of the model on the test dataset of 1.55 million spectroscopically confirmed sources (50\% of our full spectroscopic catalogue). Furthermore, we calculate precision, recall, and F$_1$ score for sources binned along various variables to assess the model performance throughout the 10-dimensional feature space.

\begin{figure*}[h!]
\includegraphics[width=\hsize]{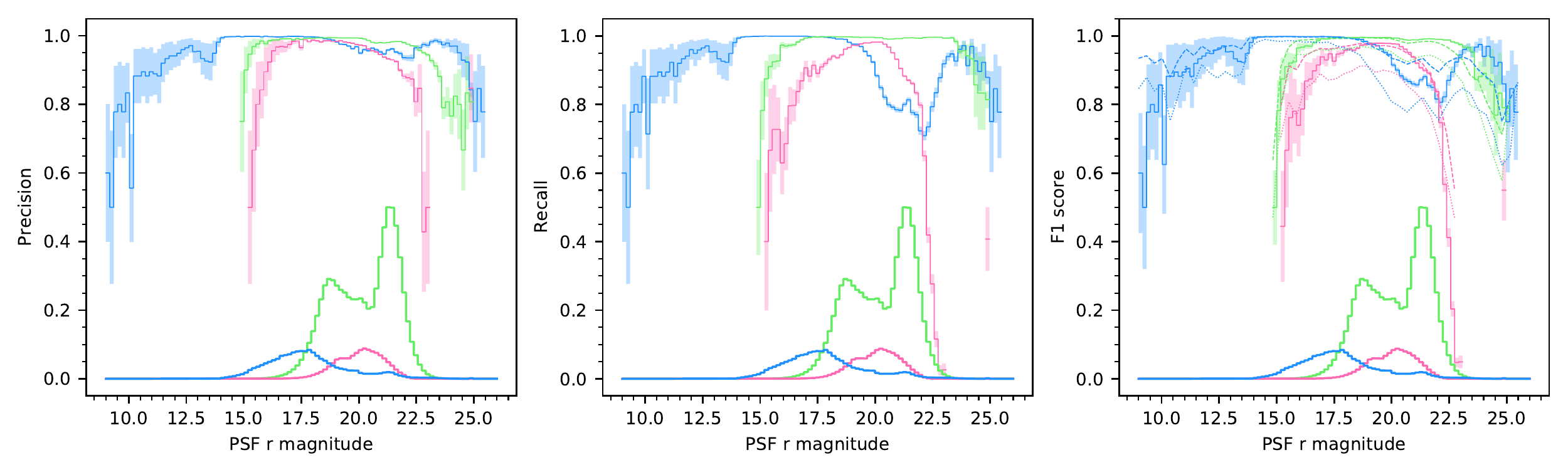}
\includegraphics[width=\hsize]{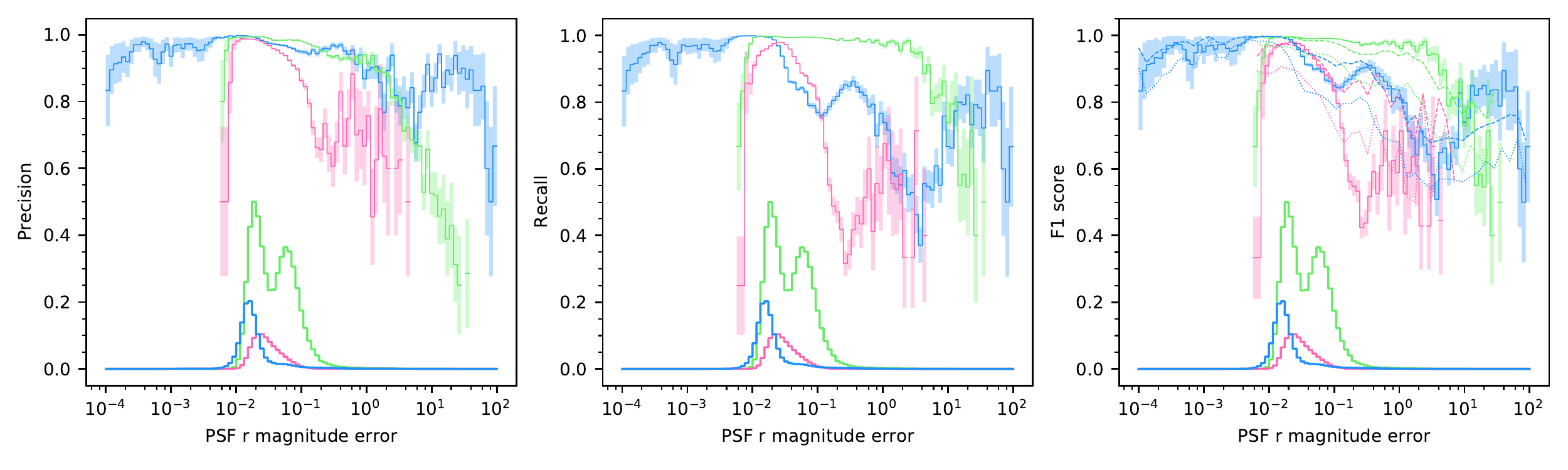}
\includegraphics[width=\hsize]{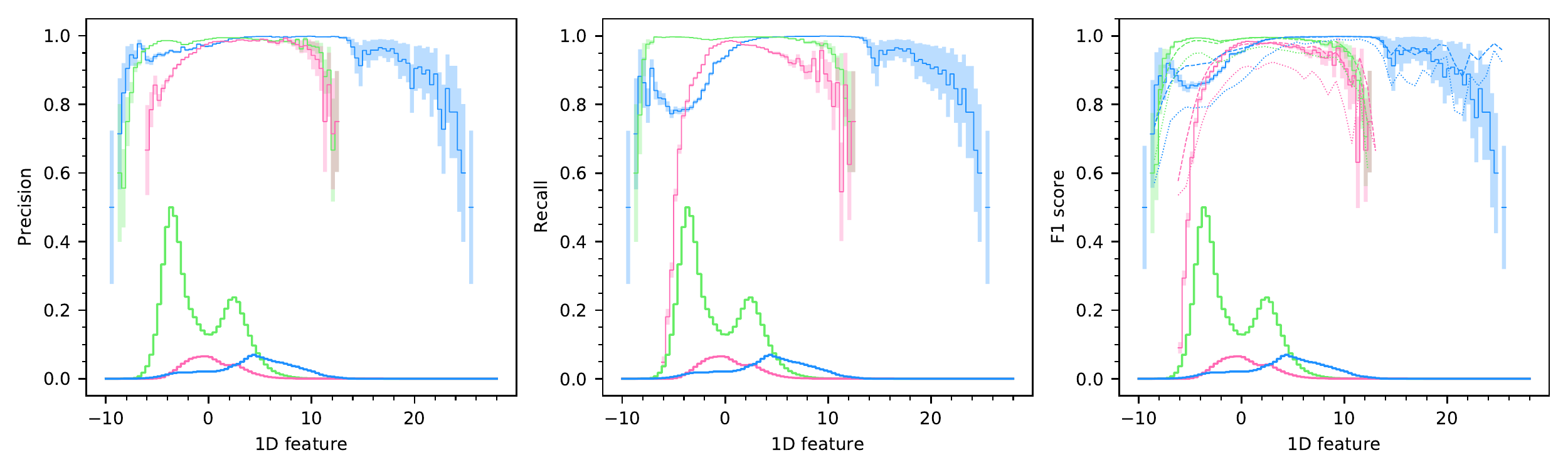}
\includegraphics[width=\hsize]{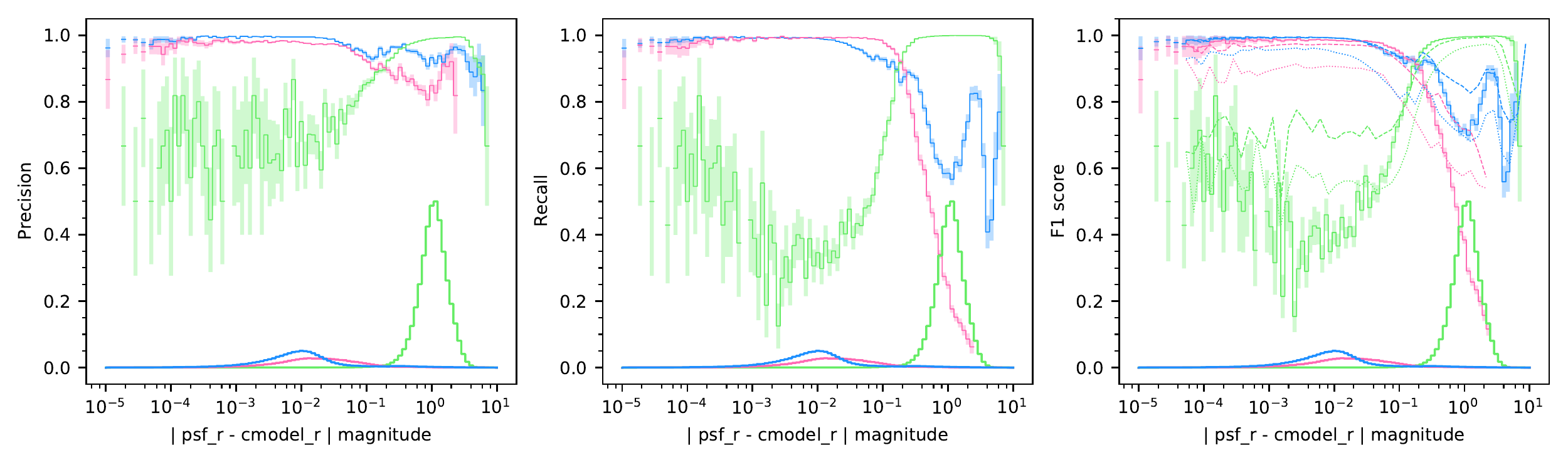}
\caption{Precision, recall, and F${_1}$ score as a function of: PSF \textit{r} magnitude (\textit{first row}), PSF \textit{r} magnitude error (\textit{second row}), features transformed into one dimension (\textit{third row}), and the $resolved_r$ parameter (\textit{fourth row}). These metrics are calculated from applying the random forest model to the test dataset of 1.55 million spectroscopically labelled sources. The shaded area shows a one standard deviation confidence interval calculated per bin from the Wilson score interval. A histogram per class is shown normalised relative to galaxies which peaks at a half. 
In the last column the dashed lines shows the mean classification probability per bin for the predicted classes, and the dotted lines shows one standard deviation below this mean. Galaxies are in green, quasars in pink and stars in blue.}
\label{figure:metric-curves}
\end{figure*}

Figure~\ref{figure:metric-curves} shows precision, recall, and F$_1$ score calculated as a function of the binned PSF \textit{r}-band magnitude, PSF \textit{r}-band error, one dimensional feature, and $resolved_r$ parameter. The shaded area shows a one standard deviation confidence interval calculated per bin from the Wilson score interval \citep{wilson1927}. The Wilson interval score behaves well for probability distributions with many values close to zero or one (unlike a Normal distribution) or with small sample sizes. A histogram per class is also shown to guide where the source counts drop, normalised relative to the galaxy class which peaks at a half. Furthermore, Figure~\ref{figure:histmatrix-mag-metrics} shows histograms of precision, recall, and F$_1$ score over each of the individual SDSS and WISE magnitude bands.

Stars are classified correctly with very high precision as a function of the PSF \textit{r}-band magnitude. The precision has values of $0.99-1$ between PSF \textit{r}-band magnitudes of $14 - 18$, dropping to 0.95 at a magnitude of 21, before rising again and then tailing off as the source density of stars decreases. The recall for stars is higher than 0.95 in the magnitude range $14-19$, but drops to 0.73 at magnitude 22 before rising again. This drop in recall is due to an increase in false negatives. These stars missed as galaxies (4071 sources) are mostly stars super-posed along the line of sight towards background galaxies, increasing their resolved parameter. The precision, recall, and F$_1$ score as a function of each magnitude feature in SDSS and WISE is shown in Figure~\ref{figure:histmatrix-mag-metrics}. The other source of false negatives, the stars missed as quasars (2328 sources), are mostly faint unresolved sources with a distribution per magnitude feature in SDSS and WISE shown in Figure~\ref{figure:histmatrix-mag-metrics}. Overall, the F$_1$ score for stars is greater than 0.8 for the majority of sources, excluding the very high and low magnitude limits where the number of sources is low. The F$_1$ score drops at magnitudes $20-22.5$ due to an increase in false negatives. As a function of the $resolved_r$ parameter (bottom row in Figure~\ref{figure:metric-curves}) the precision stays above 0.9, dropping when $resolved_r$ is greater than 4. The recall drops significantly for resolved stars.

Galaxies are classified correctly with high precision as a function of the PSF \textit{r}-band magnitude. This is higher than 0.95 from magnitude 15.5 to 23, and reduces to 0.8 from magnitude 23 to 25 as the source density falls. The recall for galaxies is also very high, being above 0.98 from magnitude 16 to 23.5, only dropping either side of this as the source density significantly decreases. Overall the F$_1$ score is very high over the entire magnitude range, only dropping when the source density significantly falls. As a function of the $resolved_r$ parameter, the precision and recall for galaxies drops below 0.95 when $resolved_r$ is less than 0.2 or greater than 6, as the source density significantly drops.

Quasars are classified correctly with high precision as a function of the PSF \textit{r}-band magnitude, and is greater than 0.95 between magnitude $15.7$ and $20.7$. Precision falls either side of this as the source density drops. The recall for quasars is above $0.95$ between magnitude 18.6 and $20.7$. Outside of this range the recall drops due to an increased number of false negatives. These are mostly due to quasars missed as galaxies (9715 sources) while only a small number (960 sources) are due to quasars missed as stars. The bottom row in Figure~\ref{figure:metric-curves} shows how the precision and recall drop below 0.95 when the $resolvedr_r$ parameter is greater than 0.8 and 1.1, respectively. Whilst the precision stays high over the majority of the source population, the recall drops significantly for resolved sources due to resolved galaxies being misclassified as quasars. 

Whilst fainter sources generally have larger PSF \textit{r} magnitude errors we do not see evidence that this affects the performance of the classifier (second row of Figure~\ref{figure:metric-curves}). For example, the recall for stars falls as the PSF magnitude error increases, however the precision remains high. This drop in recall for stars (due to an increase in misclassified stars) is seen at fainter magnitudes from 20 to 22, where the classifier confuses them with galaxies and quasars. In general for all three classes, the F$_1$ scores only drop as a function of PSF \textit{r} error when the source density falls significantly.

\subsubsection{Classification probabilities}

\begin{figure*}
\includegraphics[width=\hsize]{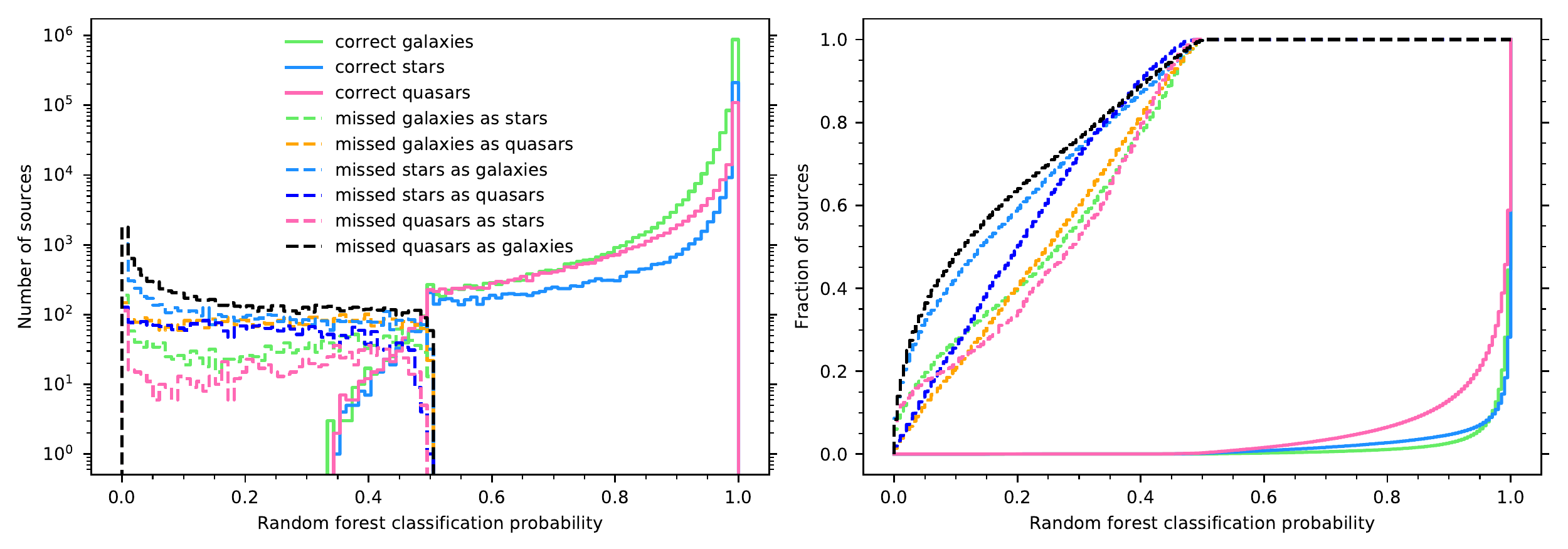}
\caption{Histogram (\textit{left}) and cumulative normalised histogram (\textit{right}) of the random forest classification probabilities using a bin size of 0.005. 71\%, 52\% and 81\% of the correctly classified galaxies, quasars, and stars have classification probabilities greater than 0.99. Correct classifications must have a probability greater than one third and incorrect classifications must have a probability less than one half. In general correctly classified quasars have lower classification probabilities than correctly classified stars or galaxies.}
\label{figure:prob-hist}
\end{figure*}

The random forest algorithm naturally returns the probabilities per class of a classification, which sum to one over the number of possible classes. The class with the largest probability is the predicted class assigned by the algorithm. The left of Figure \ref{figure:prob-hist} shows a histogram per classification possibility over the probabilities for the assigned class. For example, the dashed pink line shows quasars that the random forest predicted were stars, and is a histogram of the probability that the random forest assigned to it being a quasar. Incorrect classes must have probabilities less than one half, whilst correct classes must have probabilities greater than one third. The right side of Figure \ref{figure:prob-hist} shows the same but as a cumulative normalised histogram.

The 0.99-1 bin in Figure \ref{figure:prob-hist} contains 71\%, 52\% and 81\% of the correctly classified galaxies, quasars, and stars, respectively. Furthermore, 96\% of galaxies, 84\% of quasars and 94\% of stars have classification probabilities above 0.9. Quasars are the weakest performing class in this regard, where in general the correctly classified quasars have lower classification probabilities than correctly classified stars or galaxies. This indicates that this class is the most difficult to classify.

The last column in Figure \ref{figure:metric-curves} has the classification probabilities (dashed lines) and one standard deviation (dotted lines) plotted along with the F$_1$ scores. Overall they follow the trend in the F$_1$ scores for all classes. They only start to decrease in areas with very low numbers of sources (for example at the edges of the plots).

\begin{figure*}
\includegraphics[width=\hsize]{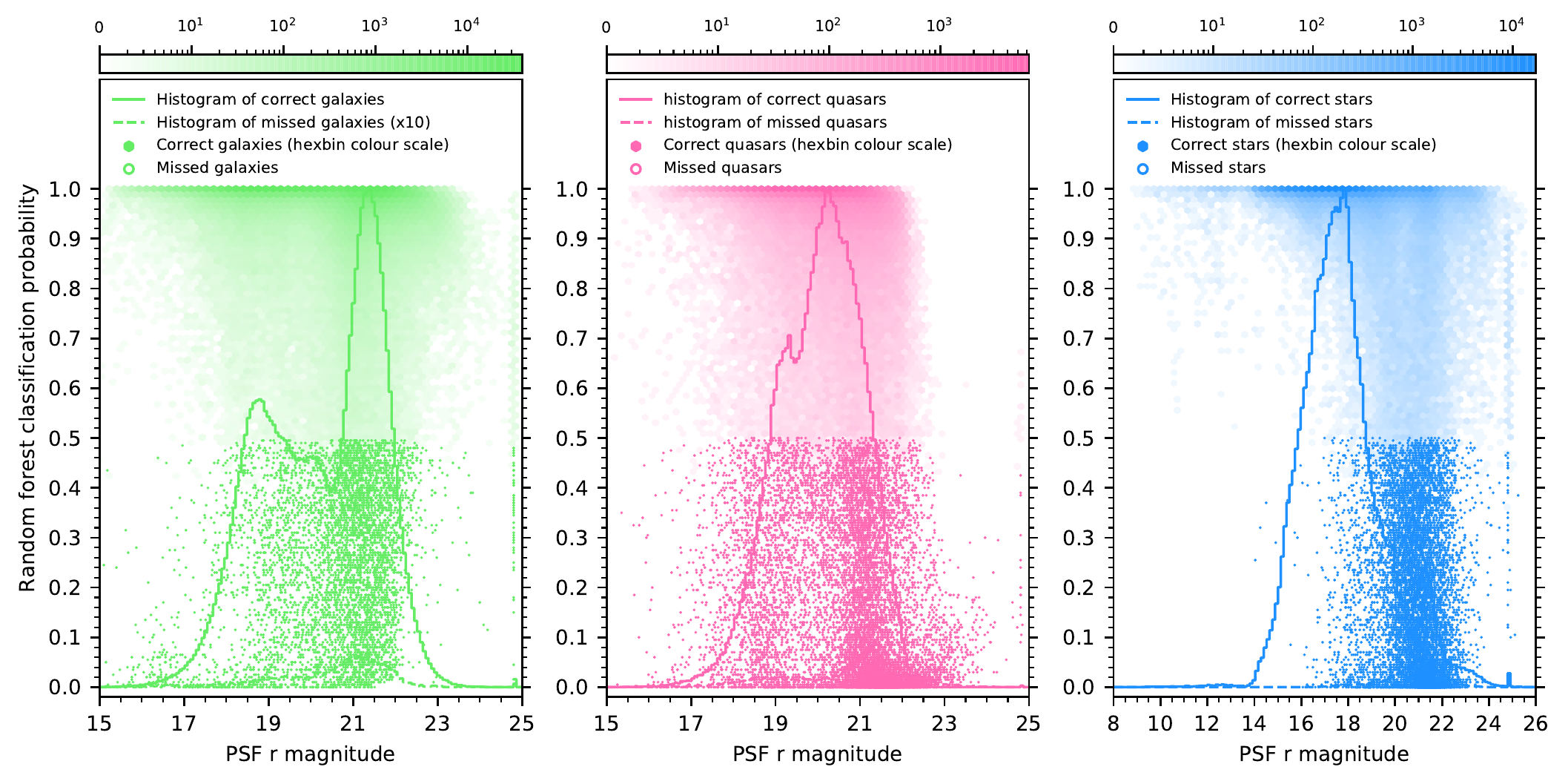}
\caption{Random forest classification probabilities plotted as a function of PSF \textit{r} magnitude. A normalised histogram of the PSF \textit{r} magnitude is also overlaid, though note that the missed galaxies histogram is multiplied by a factor of ten to be visible. 71\%, 52\% and 81\% of the correctly classified galaxies, quasars, and stars have classification probabilities greater than 0.99 (as detailed in Figure \ref{figure:prob-hist}).}
\label{figure:prob-hist-hexbin}
\end{figure*}

Figure \ref{figure:prob-hist-hexbin} shows a 2-D histogram of the PSF \textit{r} magnitude and classification probability. Misclassified sources are also plotted individually and a normalised 1-D histogram of the PSF \textit{r} magnitude is overlaid. Overall the random forest probabilities show a similar distribution over the whole PSF \textit{r} magnitude range. Misclassified sources are mostly found with \textit{r} magnitudes from 20 to 22.

\subsubsection{Limiting the training set}
\label{section:training-mag-limit}

One of the primary intended uses for machine learning algorithms is to classify sources detected in new astronomical surveys based on models trained from existing data, a form of transfer learning \citep[e.g.][]{pratt1991, tang2019}. As telescopes become more powerful, new surveys typically become progressively deeper in sensitivity and recover fainter populations of sources. One possible limitation for this form of transfer learning is the introduction of biases when deploying a model on a population of sources that are fainter than the population used for training. Here we consider how the performance of our machine learning model changes as a function of source brightness.

\begin{figure*}
\includegraphics[width=\hsize]{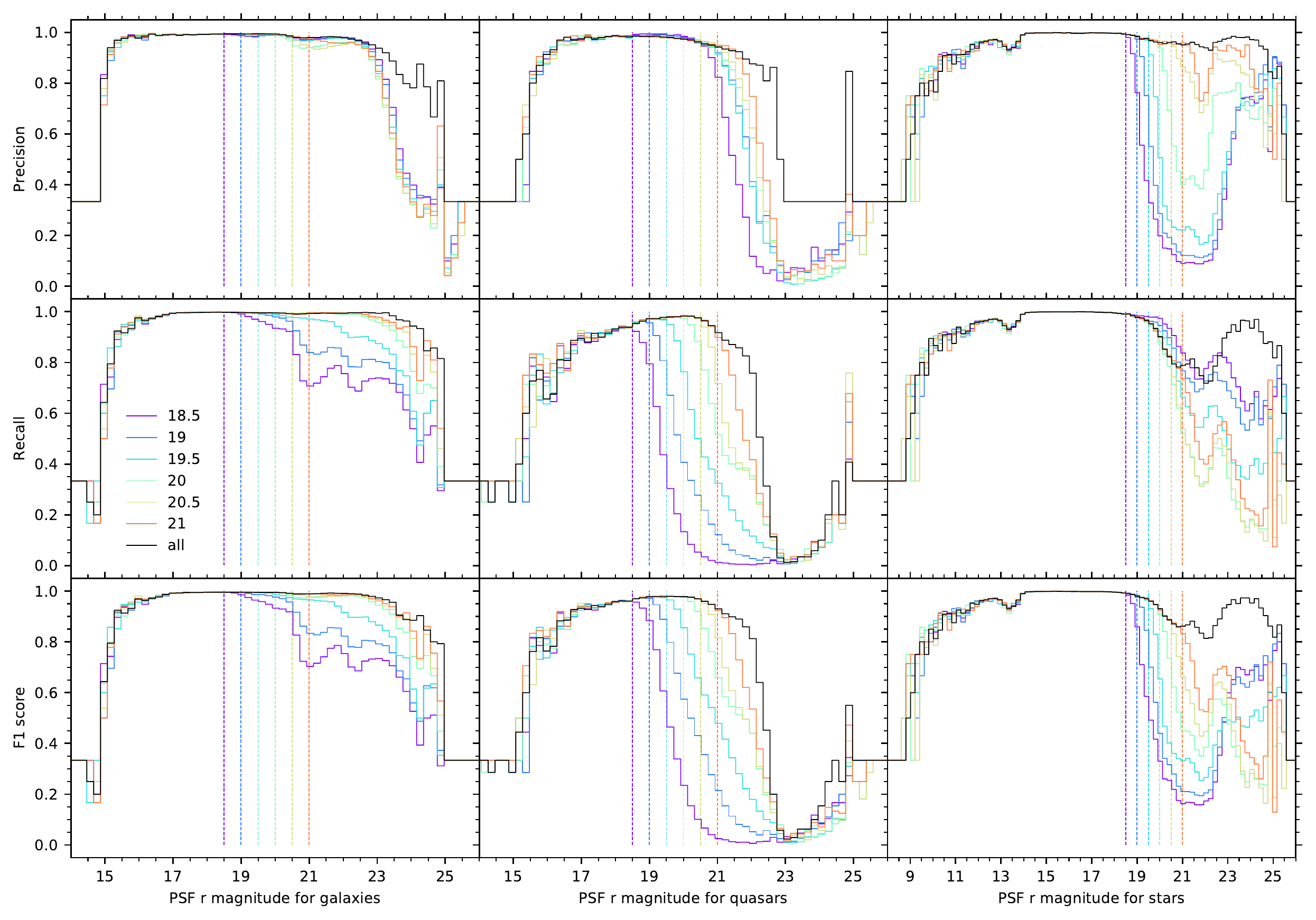}
\caption{Precision, recall, and F$_1$ score plotted as a function of PSF \textit{r} magnitude for galaxies (\textit{left}), quasars (\textit{middle}), and stars (\textit{right}). Each of the coloured lines depicts the upper limit of the \textit{r} magnitude for the random forest training. Below each of these limits 50\% of the dataset was used for training.}
\label{figure:metrics-train-maglim}
\end{figure*}

\begin{figure*}
\includegraphics[width=\hsize]{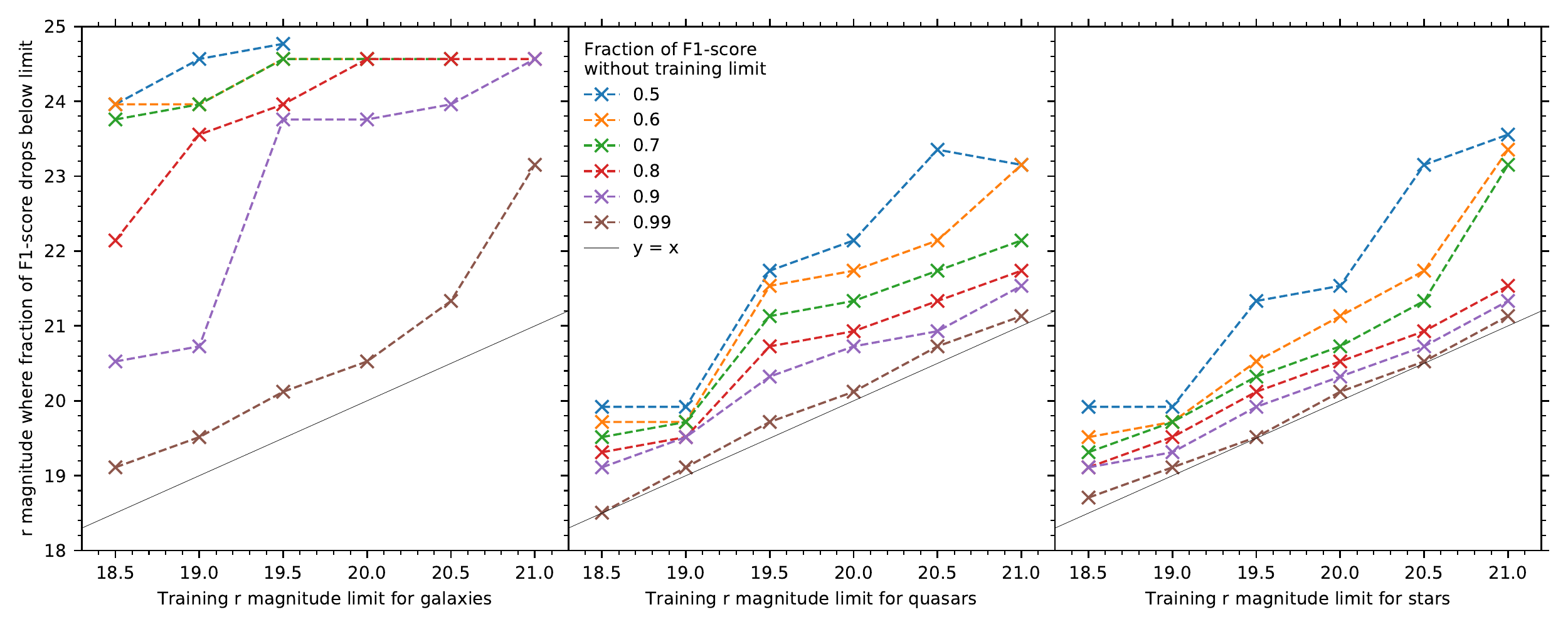}
\caption{The \textit{r} magnitude training upper limit plotted against the \textit{r} magnitude at which the F$_1$ score drops to a fraction of the F$_1$ score (shown in the legend) without a \textit{r} magnitude training limit. Below this limit 50\% of the data were used for training. 
A linear relationship is maintained for stars and quasars, while galaxies show a non-linear change between \textit{r} magnitude of 19 and 20 for F$_1$ score fractions less than 0.9 (also seen in Figure \ref{figure:metrics-train-maglim}).
}
\label{figure:metrics-train-maglim-fraction}
\end{figure*}

Figure~\ref{figure:metrics-train-maglim} shows the precision, recall, and F$_1$ score when upper limits were set on the \textit{r} magnitude in the training set. For a given limit, 50\% of all sources brighter than the limit were used to train the model, and all remaining sources were used to test the model. Figure~\ref{figure:metrics-train-maglim-fraction} shows the relationship between the \textit{r} magnitude training limit and the \textit{r} magnitude at which the F$_1$ score drops below a certain fraction of the original value. These figures show that the model maintains its performance up to the training magnitude limit as expected, however, beyond this limit the model weakens at various rates for the three classes of source. For galaxies and quasars, the recall is significantly affected, due to an increase in false negatives. For stars, the precision is significantly affected, due to an increase in false positives.

For galaxies, the precision is not strongly affected by imposing any magnitude limit in the training until above magnitude 23. The the recall (and therefore F$_1$ score) is immediately affected above the training limit, but improves significantly as the training magnitude limit increases beyond 19.5. Overall, the F$_1$ score (dominated by the poor recall) shows a jump in performance above magnitude 19.5 (shown in Figure \ref{figure:metrics-train-maglim-fraction}).

For quasars, the precision is not strongly affected by any magnitude limit in the training until above magnitude 21. However, the recall is immediately affected above the training limits imposed. Overall, the F$_1$ score (dominated by the poor recall) improves linearly as the training magnitude limit increases (shown in Figure \ref{figure:metrics-train-maglim-fraction}).

For stars, the recall is not strongly affected by any magnitude limit in the training until above magnitude 21. However, the precision is immediately affected above the training limits imposed. Overall, the F$_1$ score (dominated by the poor precision) improves linearly as the training magnitude limit increases (shown in Figure \ref{figure:metrics-train-maglim-fraction}).

When imposing the \textit{r} magnitude training limit at its lowest value of 18.5, the overall fraction of each class used to fit the model is 6.5\%, 3.3\% and 34.6\% for galaxies, quasars, and stars, respectively. Comparing Figure~\ref{figure:train-vs-f1score} and Figure~\ref{figure:metrics-train-maglim} indicates that on average the F$_1$ scores should still be much higher despite this reduction in the number of training samples. Furthermore, the F$_1$ score as a function of the PSF \textit{r} magnitude when training on 1\% of sources shows the same distribution as that shown in the first row of Figure~\ref{figure:metric-curves}, which used 50\% of the data for training. In other words, training on 1\% of each class gives a distribution in \textit{r} magnitude of F$_1$ scores that is similar to training on 50\% of each class. Overall this demonstrates that when limiting the training set to sources below magnitude 19, the training set does not contain enough examples of each class that are representative out to fainter magnitudes. This degraded performance is alleviated significantly once the training set goes up to and beyond magnitude 19.5, but overall the models cannot be applied to sources at much fainter magnitudes than were included in the training set without resulting in a significantly poorer performance. Furthermore, this decrease in performance will not be quantifiable for a distribution of sources at magnitudes fainter than those in the training dataset. Consequently, we consider that the metrics used to quantify the performance of the random forest model are only relevant within the magnitude range explored by our training dataset, as shown in Figure~\ref{figure:metric-curves}.

\subsubsection{Confidence of classification for individual sources} \label{section:10d}

\begin{figure*}
\includegraphics[width=\hsize]{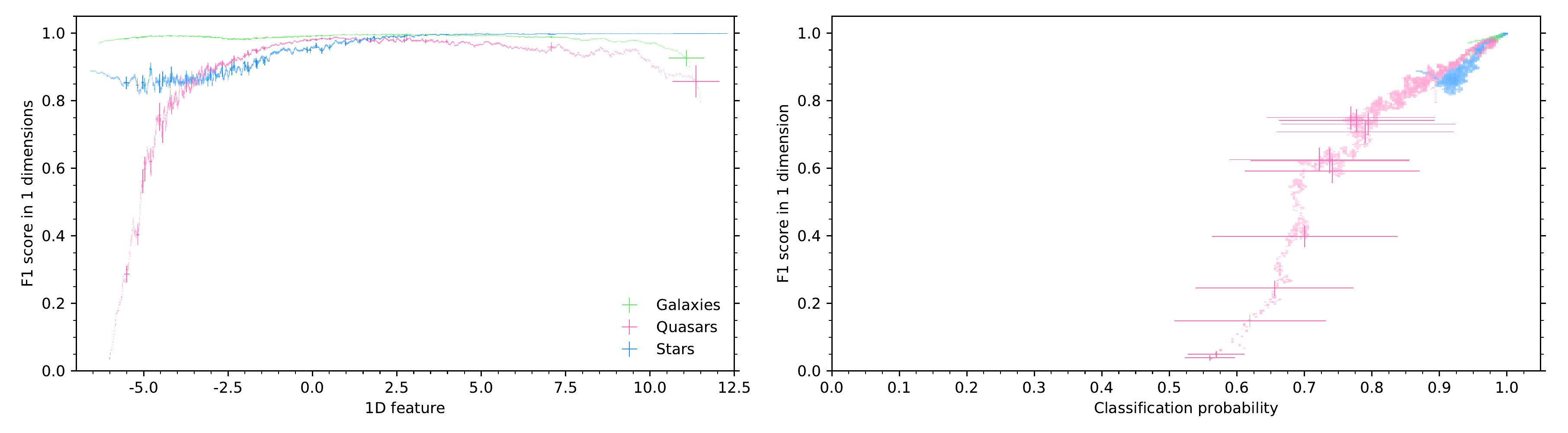}
\includegraphics[width=\hsize]{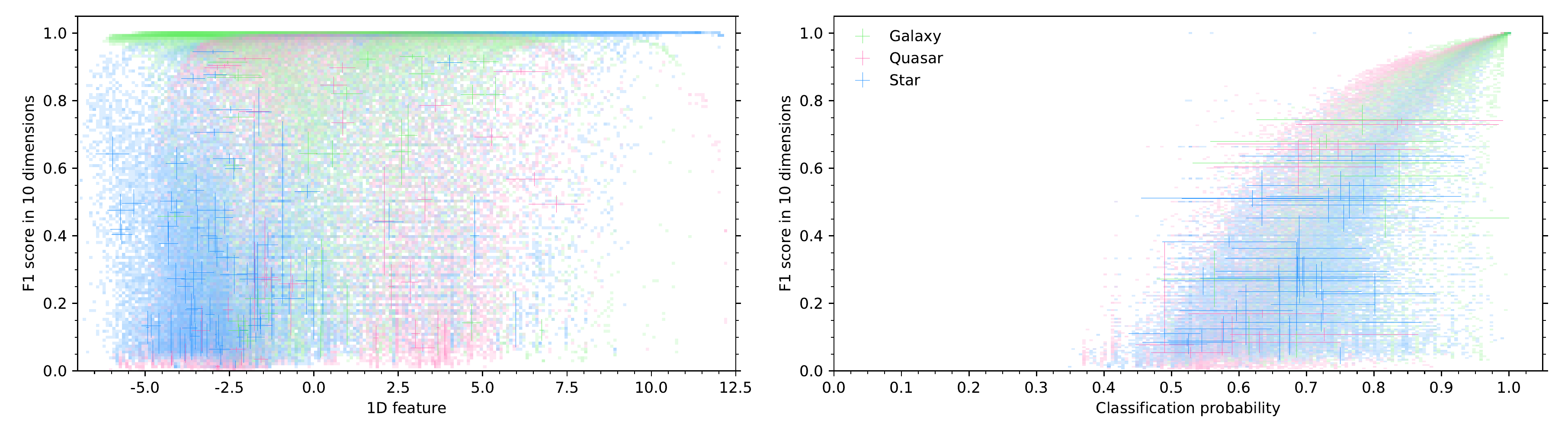}
\caption{100\,000 sources from our spectroscopic test dataset, the F$_1$ score is calculated from 10\,000 nearest neighbours in 1-D (\textit{top}) and 10-D (\textit{bottom}) feature spaces, plotted against the mean 1-D feature (\textit{left}) and mean classification probability (\textit{right}) for those neighbours. Sources are binned per pixel and colours combined proportional to how many of each class are in that pixel bin. The brightness corresponds to the total source count in that pixel on a logarithmic scale. A small sample of individual data points with their error bars are also overlaid.}
\label{figure:nn-1and10D}
\end{figure*}

In the previous section, we make approximations in order to specify how the model performs as a function of individual features, reducing the full 10-D space used by the performance metrics. However these feature-specific metrics are still only correct on average within a specific range of feature space, and cannot be used to quantify performance for individual sources. In order to narrow further the range of feature space that a performance metric represents, we implement a nearest neighbour approach for a subset of sources in one and ten dimensions, and we investigate the relation between these metrics and the classification probabilities returned by the random forest for individual sources.

For each source, the ten features are transformed into one single feature by using PCA (as seen in the third row of Figure~\ref{figure:metric-curves}). In this single feature, 75\% of the variance is retained from the set of 10 features. Given the computationally intensive nature of finding large numbers of nearest neighbours for many sources, we find nearest neighbours for only 100\,000 sources from the test dataset. For each of these sources we find the 10\,000 nearest neighbours in the 1-D feature space, calculating the F$_1$ score from these nearest neighbours. The top left panel in Figure~\ref{figure:nn-1and10D} shows the F$_1$ score as a function of the 1-D feature for these 100\,000 sources. Whilst in Figure~\ref{figure:metric-curves} the bin size is constant, this nearest neighbours approach uses a variable bin size, defined by the requirement to have 10\,000 nearest neighbours within the bin. This is represented by the horizontal error bars in the top left panel of Figure~\ref{figure:nn-1and10D}, calculated as one standard deviation from the 10\,000 nearest neighbours. These error bars are small where source density is high, and only begin to become visible where source density drops at high and low values of the 1-D feature. The error in the F$_1$ score is calculated using the Wilson score interval, and is only significant when the source density drops at the edges of the plot. In the top right panel of Figure~\ref{figure:nn-1and10D} we show the classification probabilities for the same sources. As shown in Figure~\ref{figure:prob-hist}, quasars generally have lower classification probabilities, whilst galaxies and stars have higher values. The probabilities for all classes are positively correlated with the F$_1$ score calculated from the 10\,000 nearest neighbours.
Whilst this nearest neighbour method does not fully explore the 1-D feature space (only using 6.5\% of the test dataset), the F$_1$ scores and probabilities are in agreement with those shown in Figure~\ref{figure:metric-curves} (the final panel of the third row).

We repeat this process using all of the features, finding the 10\,000 nearest neighbours for each of 100\,000 sources in the original 10-D feature space (bottom row of Figure~\ref{figure:nn-1and10D}). The error bars on the 1-D feature are now much larger, as a sample of 10\,000 nearest neighbours selected in 10-D will have a wider distribution than in 1-D. A broadly similar trend in F$_1$ scores is seen when compared with that for the 1-D case, see the top left panel of Figure~\ref{figure:nn-1and10D} and Figure~\ref{figure:metric-curves}. However, we now see a much larger scatter in the F$_1$ scores for each class as a result of exploring more sparsely populated areas of the 10-D feature space. In particular, there are regions in the 10-D feature space where the F$_1$ scores are significantly below the average, most notably for stars and quasars. This occurs because the source density in the 10-D space can drop significantly, even when the corresponding 1-D feature is densely populated with a high F$_1$ score. In the bottom right panel of Figure~\ref{figure:nn-1and10D} we show the mean classification probability per source calculated from its nearest neighbours in 10-D. There are significantly more galaxies, quasars, and stars with lower probabilities due to the lower source density in some regions of the 10-D feature space. Despite this, the classification probabilities are still positively correlated with the F$_1$ score. In other words, areas of the 10-D feature space which had low F$_1$ scores calculated from their nearest neighbours in 10-D also have low classification probabilities.

Overall, Figure~\ref{figure:nn-1and10D} shows that the random forest probabilities are positively correlated with the calculated F$_1$ scores across the full 10-D feature space. This means that even in sparsely populated regions of the 10-D feature space, the random forest probabilities reflect this uncertainty in the classification for individual sources.

Calculating F$_1$ scores in the 10-D space like this is highly computationally intensive. For each of our 111\,million photometrically observed sources that we classify in Section~\ref{section:newsources}, finding the 10\,000 nearest neighbours from the 1.55\,million sources in our spectroscopic test dataset and calculating F$_1$ scores localised to each newly classified source would require significant computational resources. Whilst this would be beneficial when using classification algorithms in higher dimensions that do not return a classification confidence per source, the classification probabilities returned by the random forest provide a more efficient solution. Furthermore, through testing a subset of 100\,000 sources we have demonstrated that the random forest probabilities are in agreement with the localised F$_1$ scores. Overall, the random forest probabilities are an effective method to estimate the reliability of a classification, representing the confidence of classifications in that particular area of 10-D feature space.

\subsection{Unsupervised clustering with UMAP}

\begin{figure*}
\includegraphics[width=0.5\hsize]{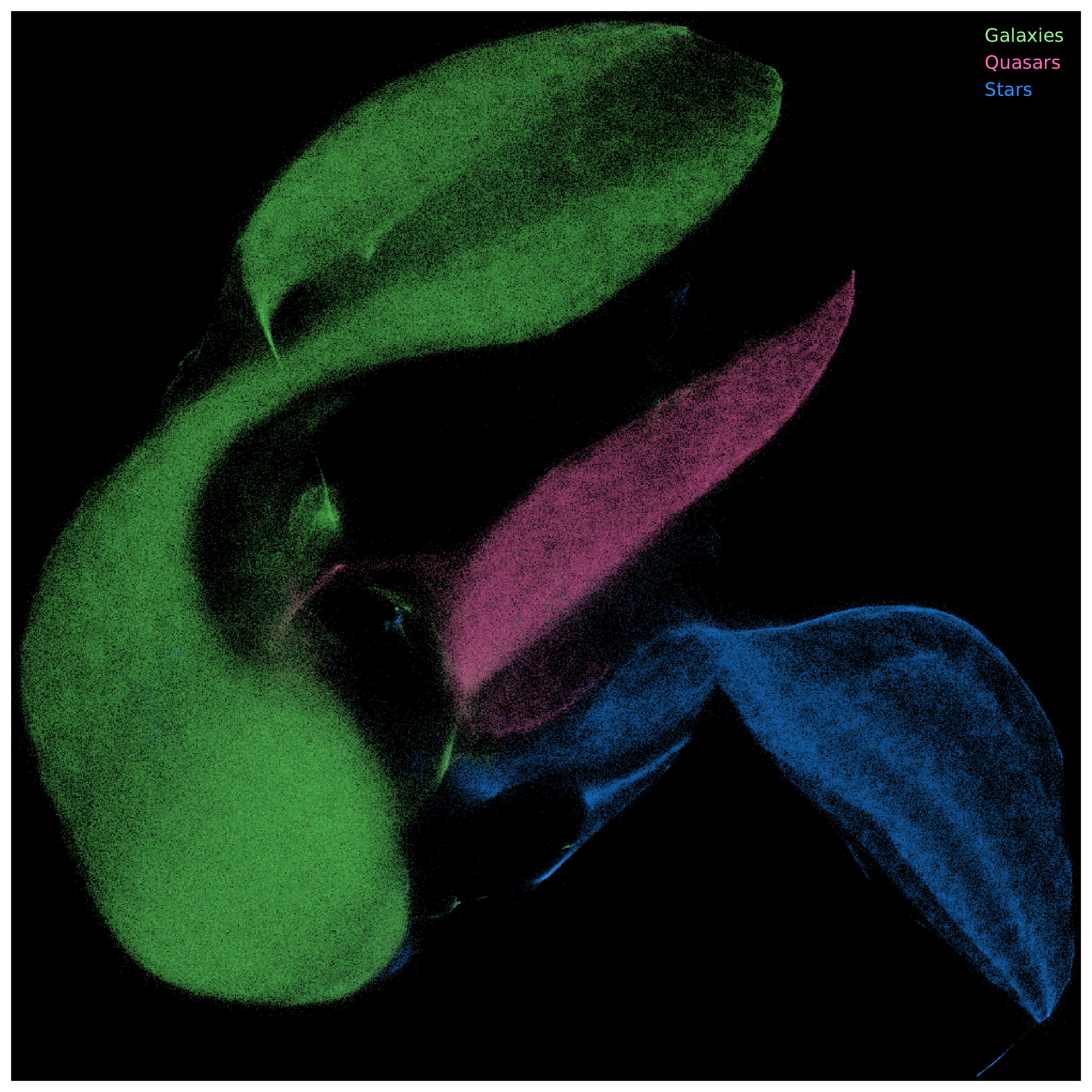}
\includegraphics[width=0.5\hsize]{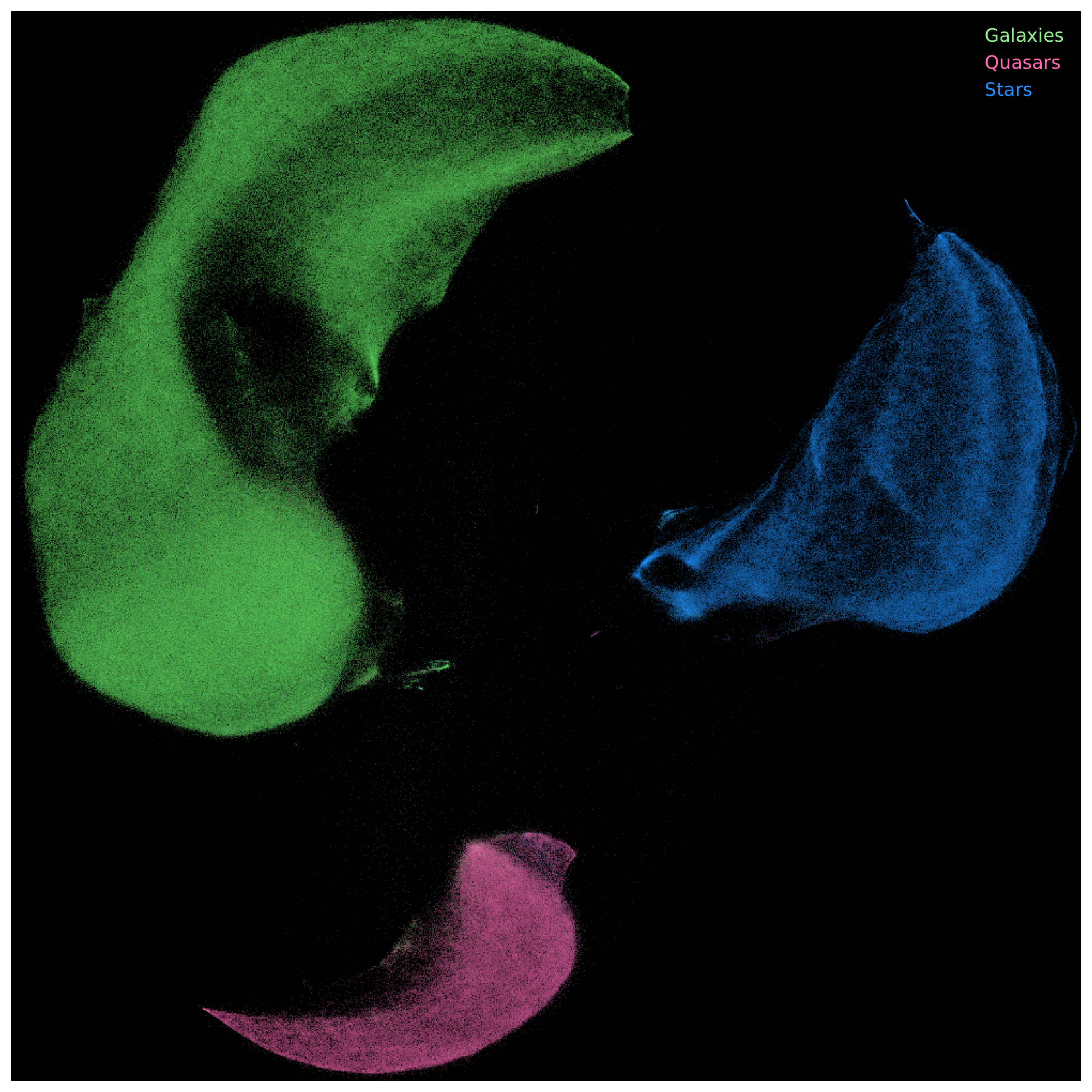}

\includegraphics[width=0.247\hsize]{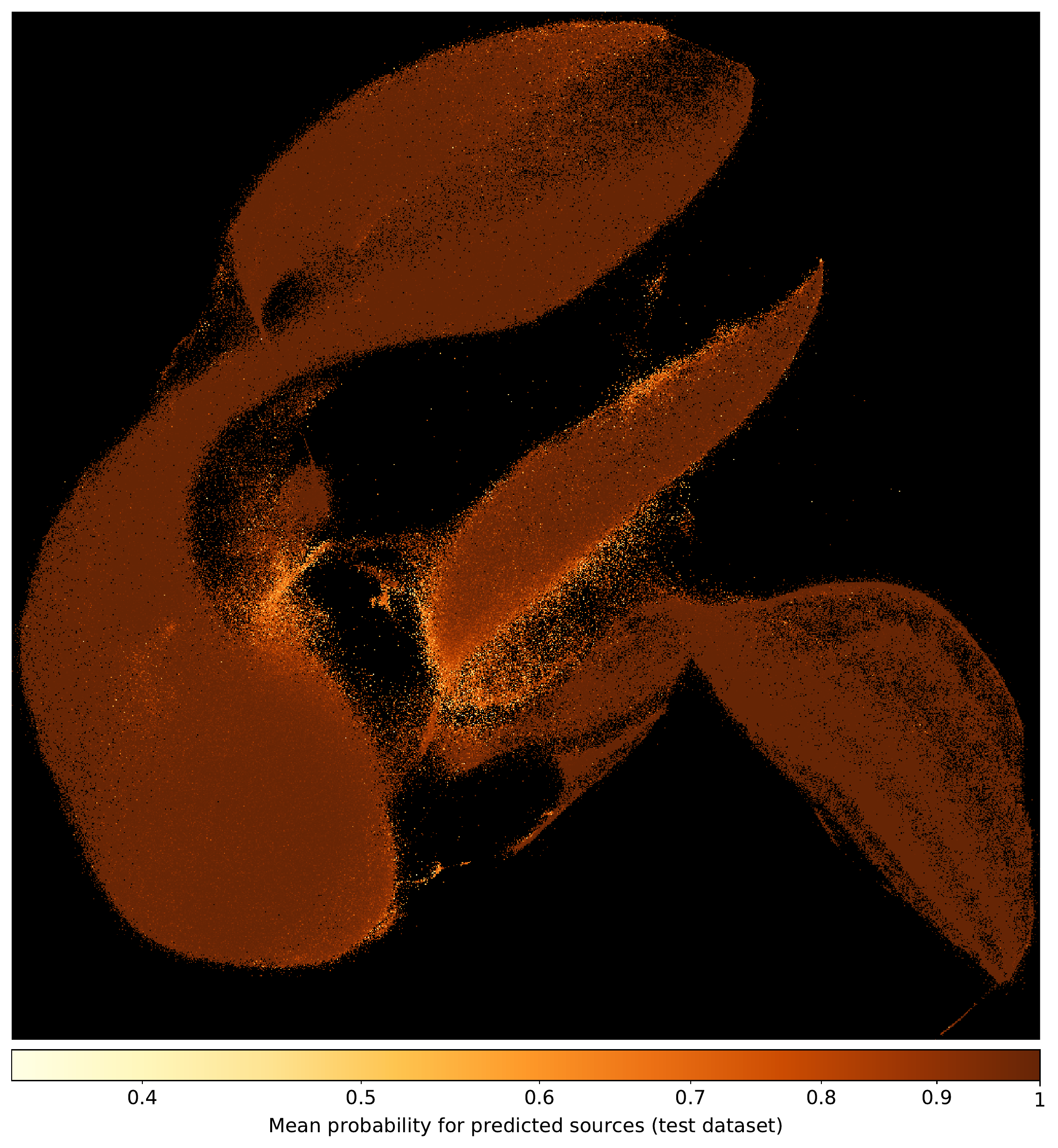}
\includegraphics[width=0.245\hsize]{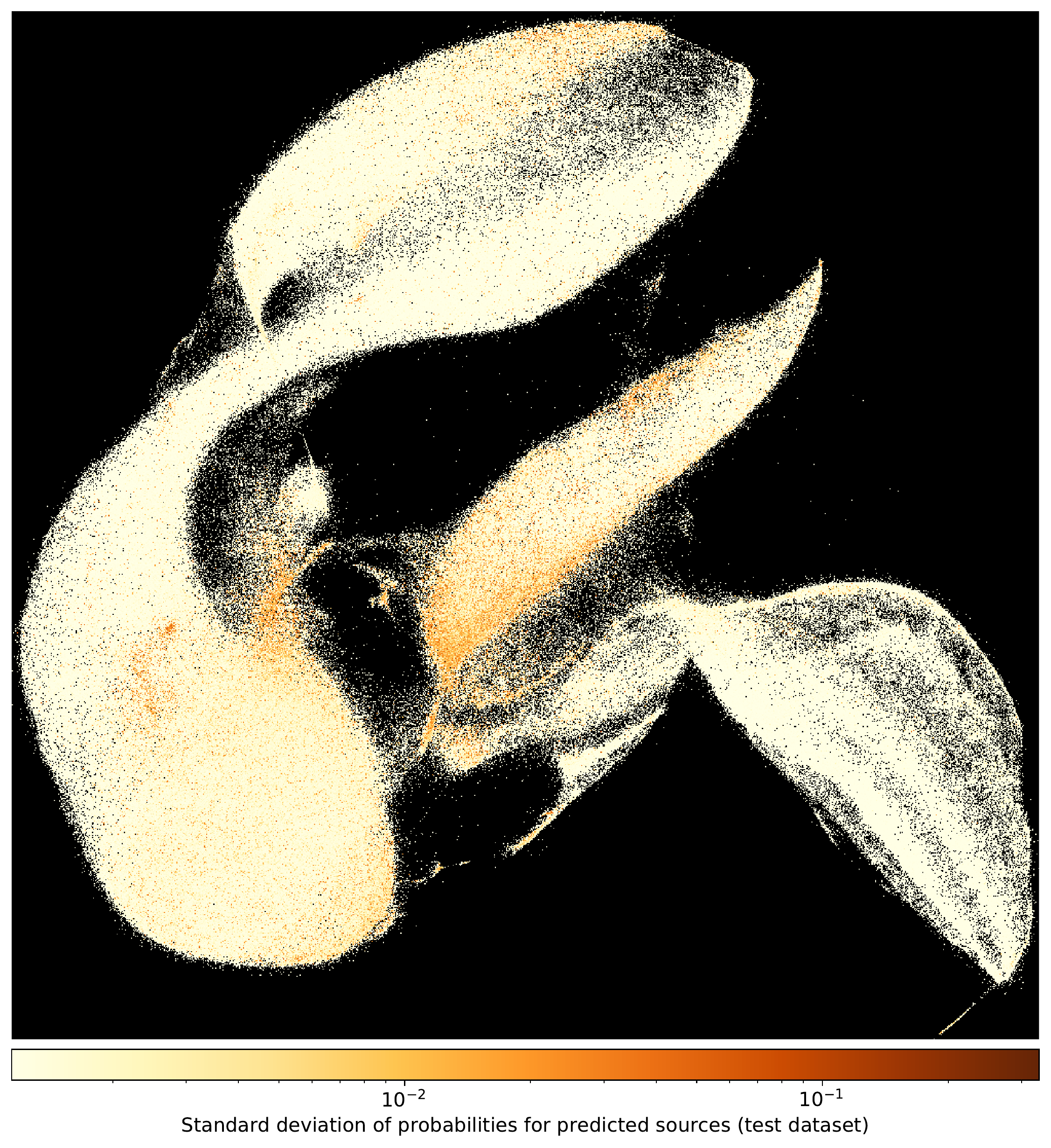}
\includegraphics[width=0.247\hsize]{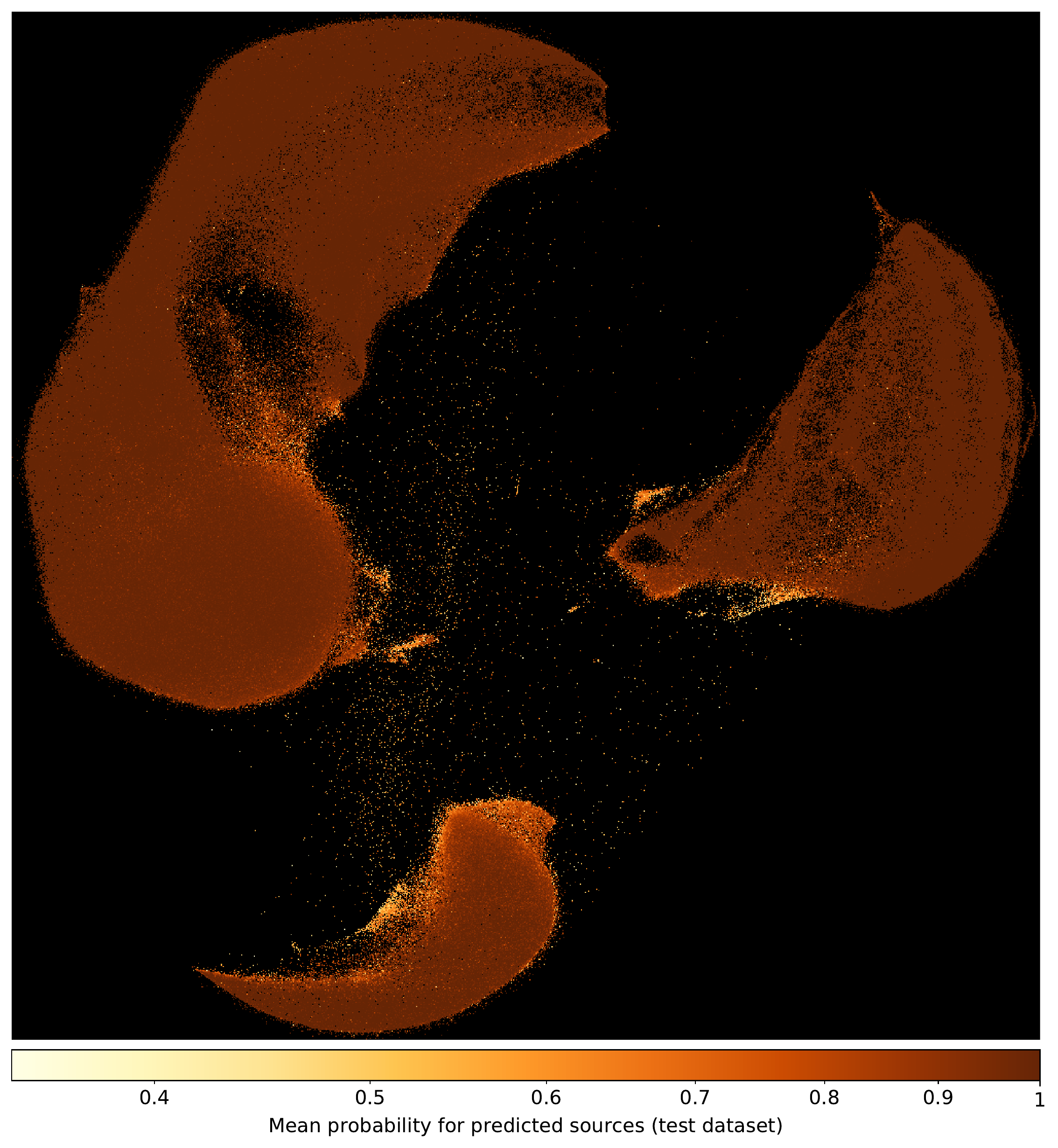}
\includegraphics[width=0.245\hsize]{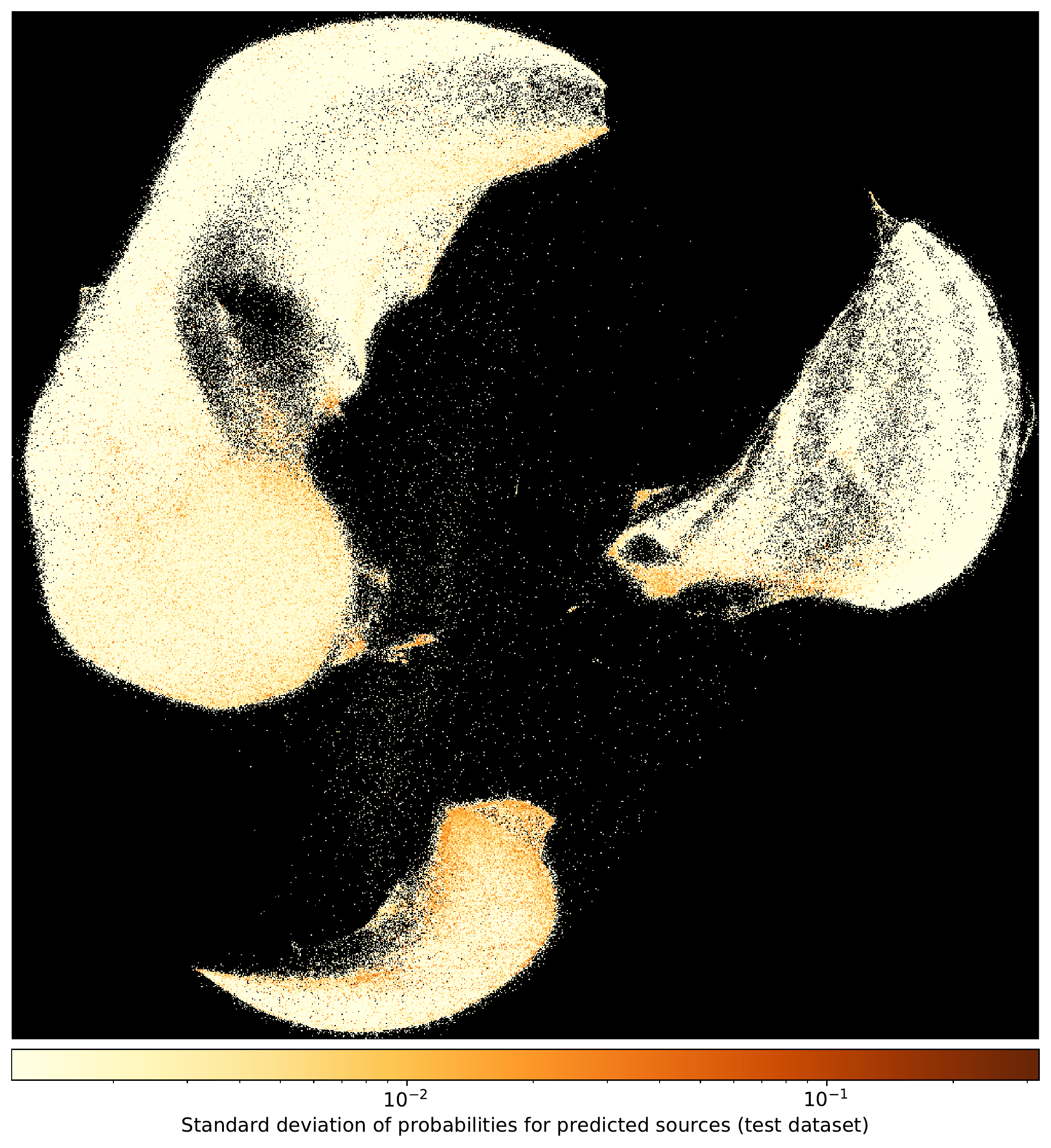}

\caption{Test dataset of 1.55 million spectroscopically observed sources processed with UMAP in unsupervised (top left) and supervised (top right) schemes to reduce from ten features to two. The resulting two dimensions are plotted, with sources binned per pixel and colours combined proportional to how many of each class are in that pixel bin. Galaxies are shown in green, quasars in pink and stars in blue. The brightness corresponds to the total source count in that pixel on a logarithmic scale. Whilst the axis have no physical meaning in this 2-D space, UMAP returns proportionate distances between points and clusters, effectively displaying local and global structures in both supervised and unsupervised schemes. The bottom row shows the mean and standard deviation of the random forest classification probabilities.}
\label{figure:umap-spec}
\end{figure*}

\begin{figure*}
\includegraphics[width=0.33\hsize]{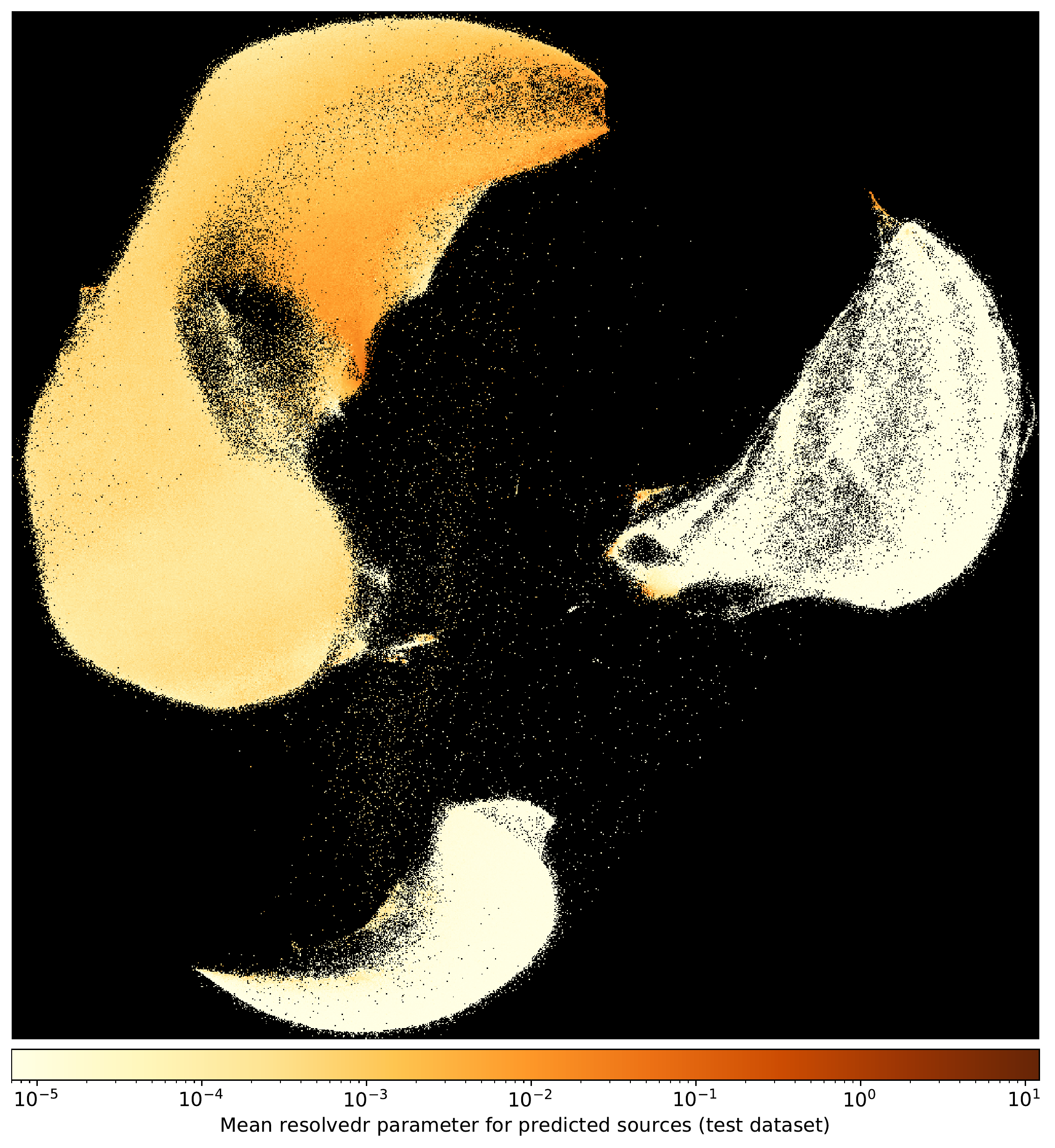}
\includegraphics[width=0.33\hsize]{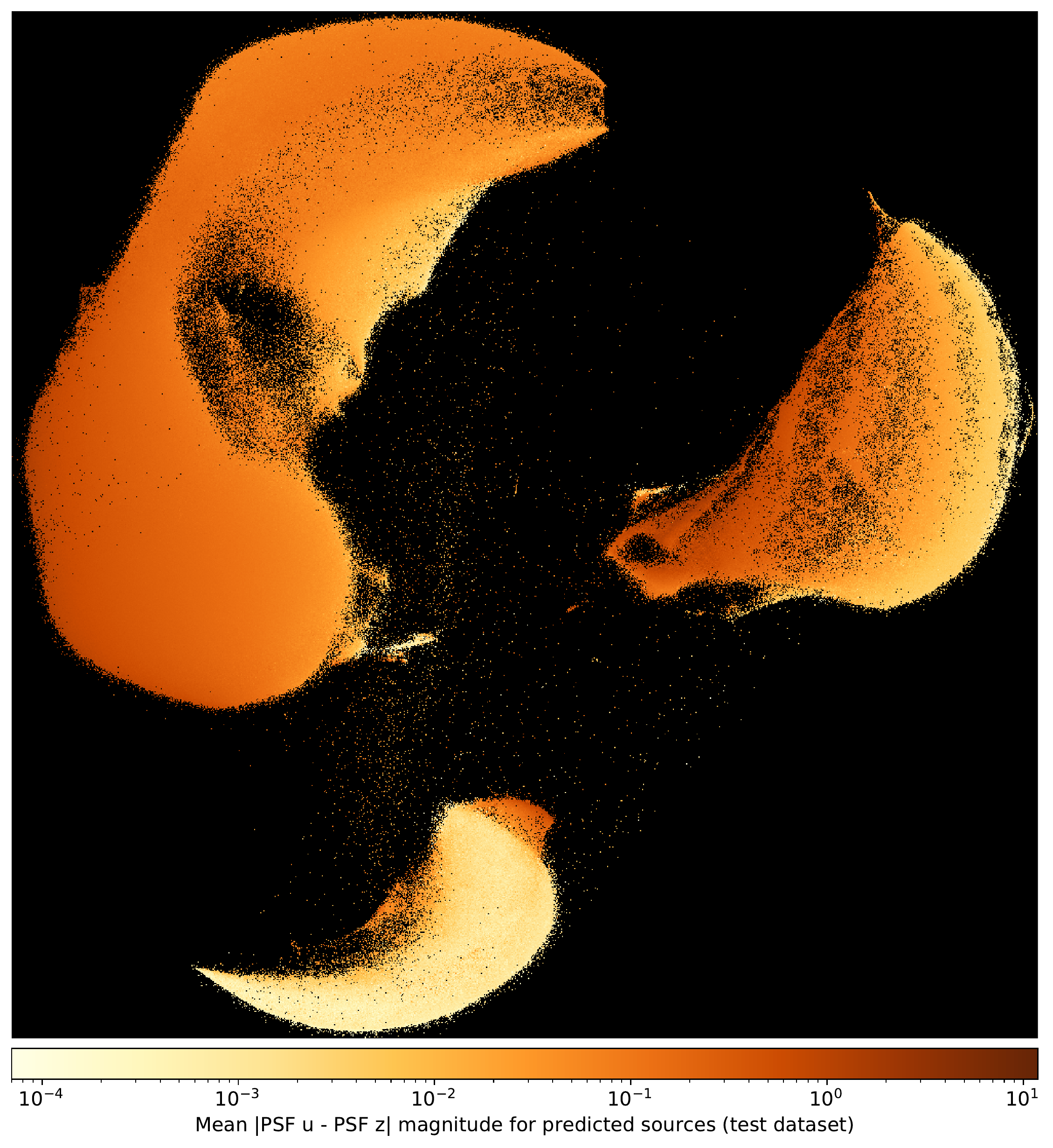}
\includegraphics[width=0.33\hsize]{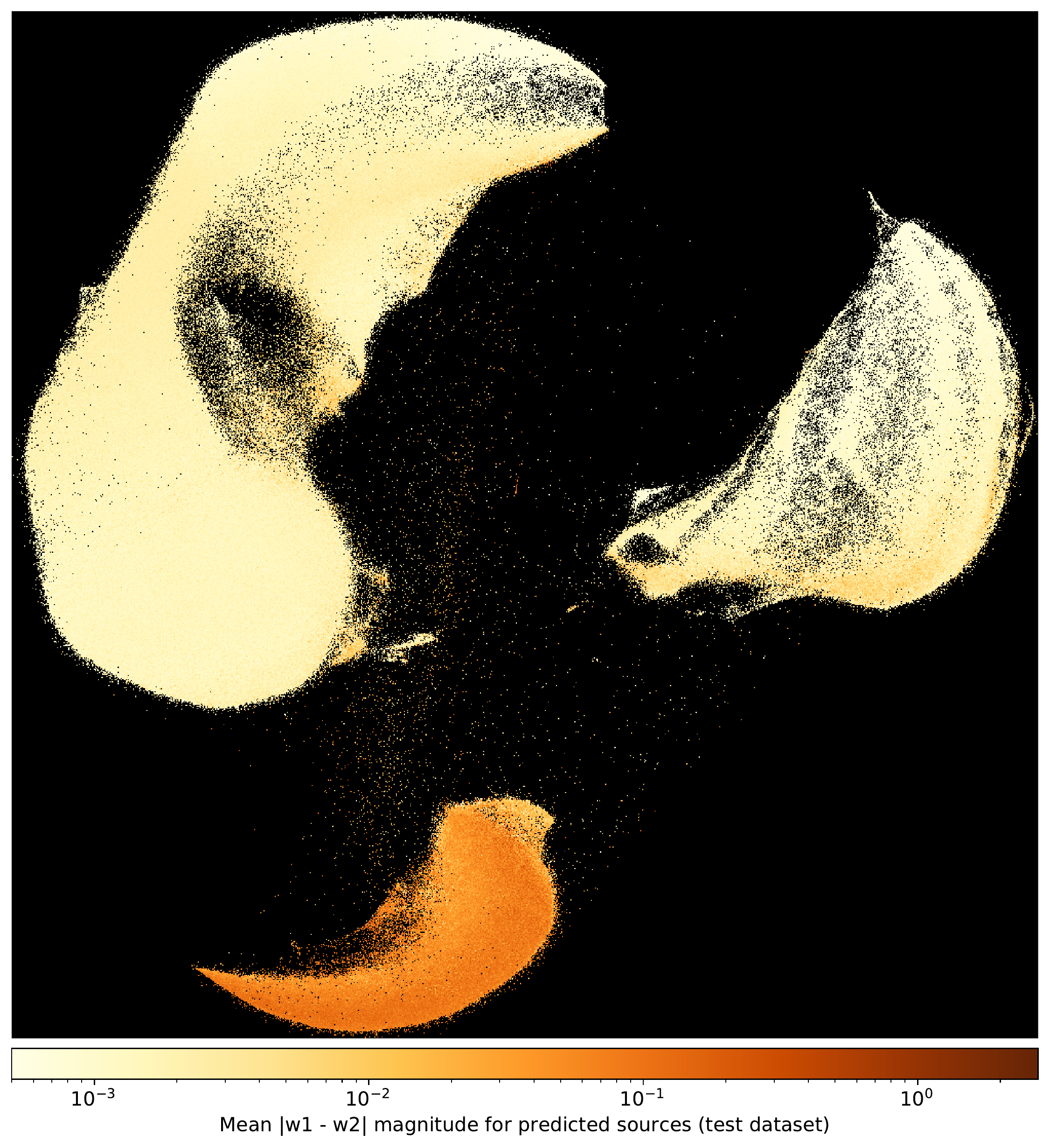}
\includegraphics[width=0.33\hsize]{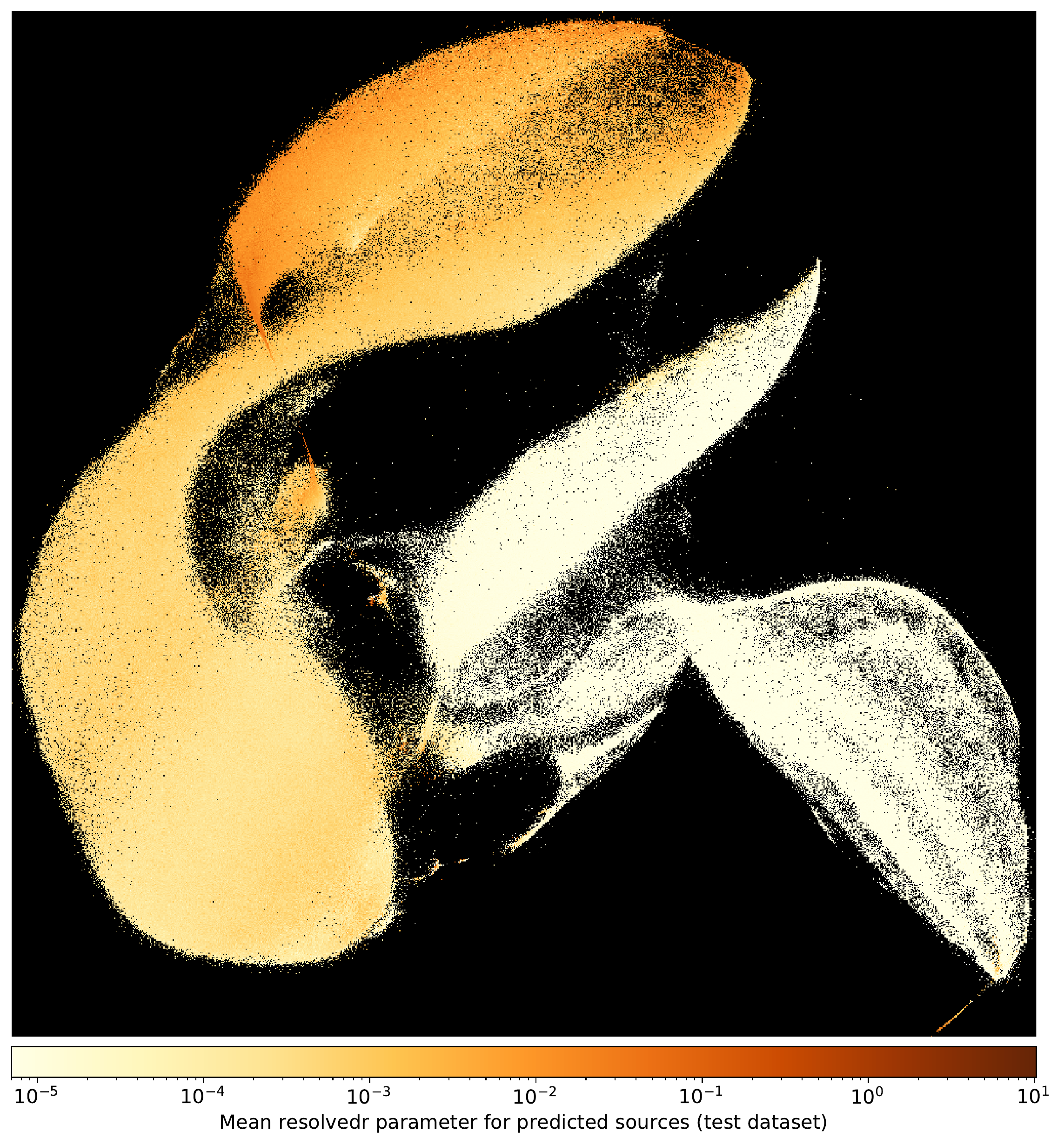}
\includegraphics[width=0.33\hsize]{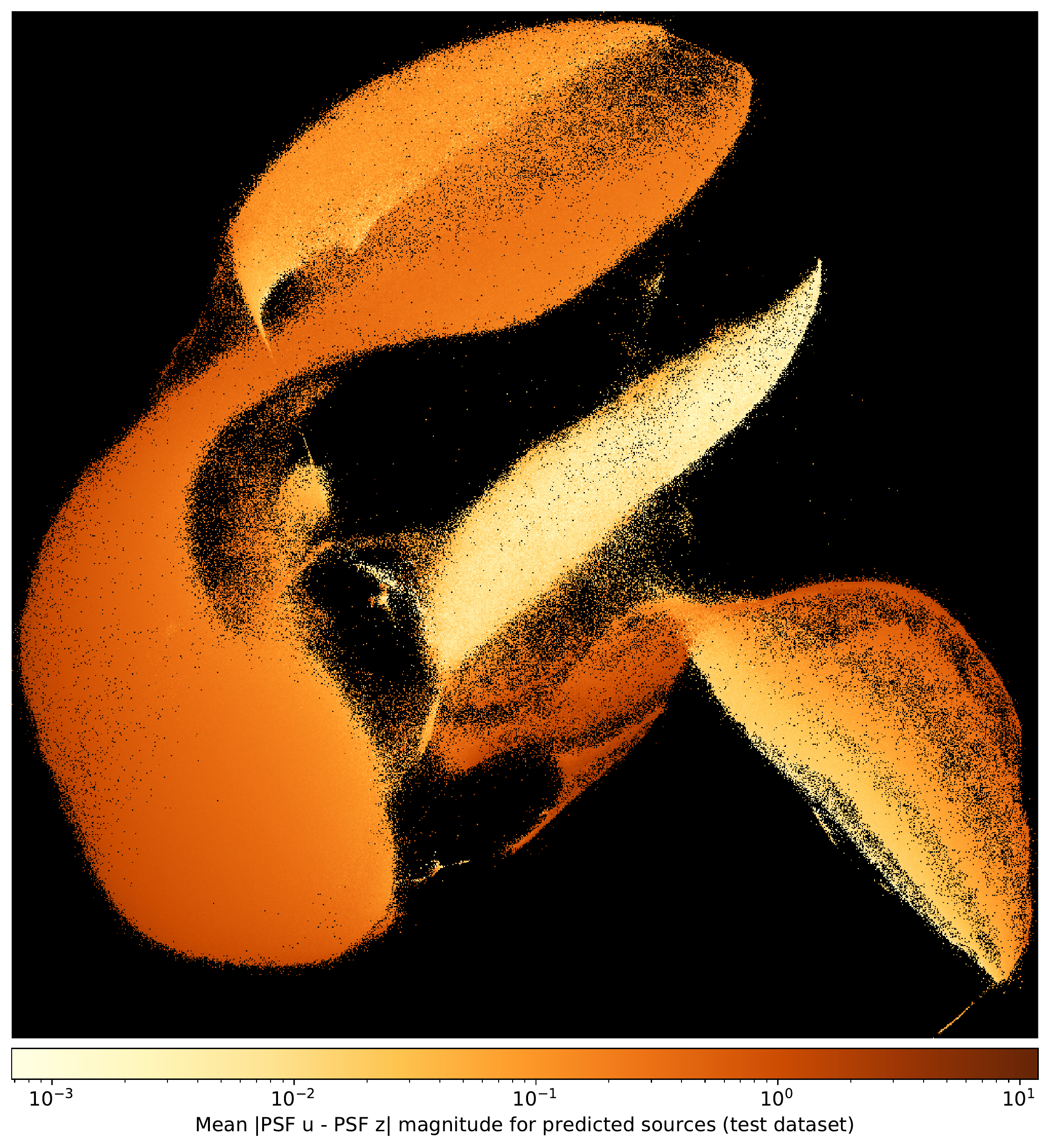}
\includegraphics[width=0.33\hsize]{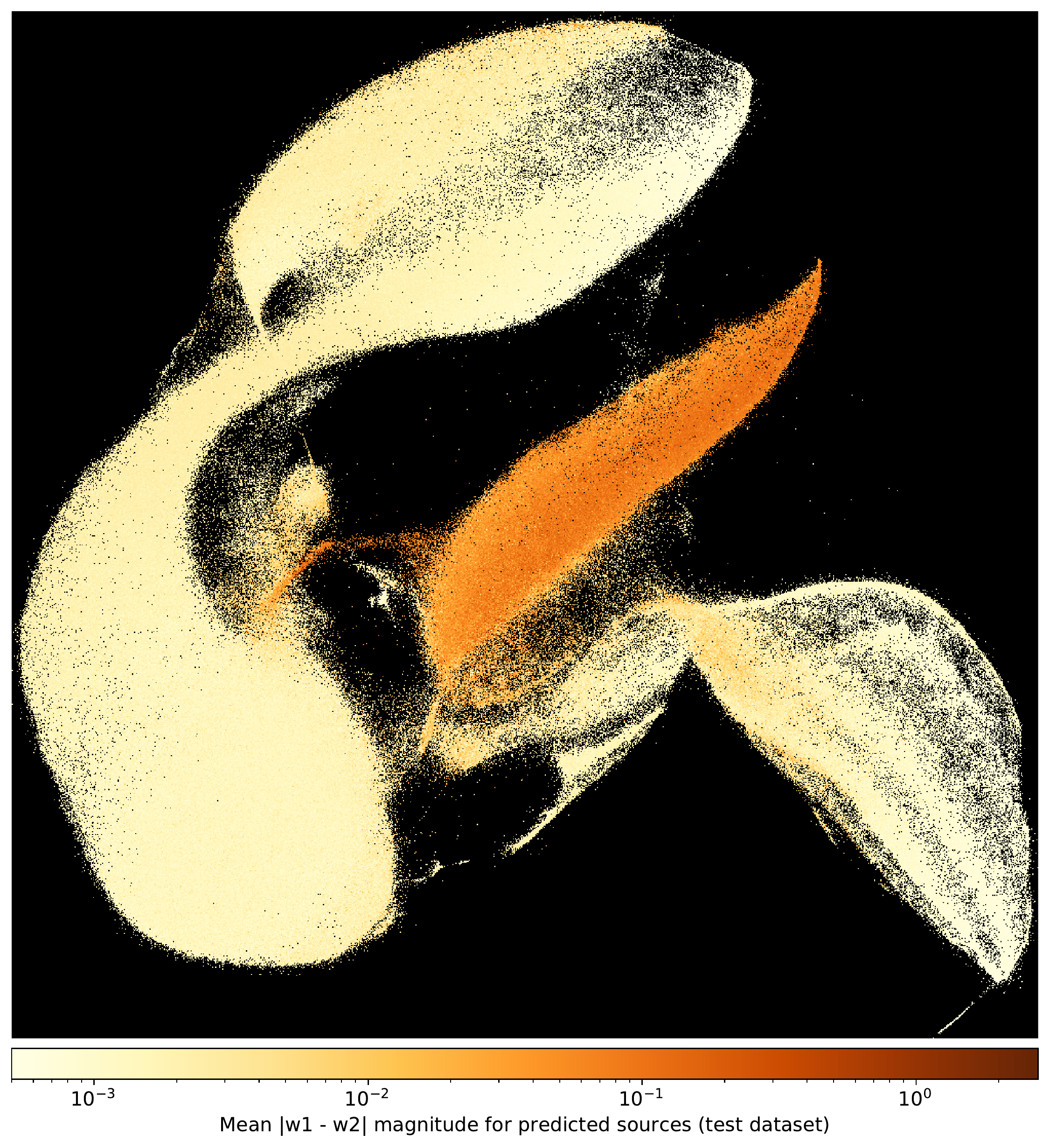}

\caption{Test dataset of 1.55 million spectroscopically observed sources processed with UMAP to reduce from ten features to two in supervised (\textit{top row}) and unsupervised (\textit{bottom row}) schemes. The resulting two dimensions are plotted, with sources binned per pixel and coloured by the $resolvedr$ parameter (\textit{left column}), an SDSS colour $|\mathrm{PSF}\, u - \mathrm{PSF}\, z|$ (\textit{middle column}) and a WISE colour $|w1 - w2|$ (\textit{right column}). Whilst the axis have no physical meaning in this two dimensional space, UMAP returns proportionate distances between points and clusters, effectively displaying local and global structures in both supervised and unsupervised schemes. Galaxies can be seen distributed by how resolved they are, separating nearby extended galaxies from distant point sources. Furthermore, red galaxies can be distinguished from blue ones and green valley galaxies can be seen connecting the two regions. Stars are arranged in an alternate representation of a Hertzsprung Russell diagram, distinguishing their colour and luminosity. Quasars are predominantly unresolved bluer sources, though redder quasars are seen on the outskirts of the cluster. }
\label{figure:umap-spec-colours}
\end{figure*}

In order to interrogate the performance of our machine learning model further, we use a form of unsupervised machine learning. Unsupervised machine learning algorithms have no prior knowledge of target class labels, and attempt to group sources by inferring their similarities in feature space. Here we use the UMAP \citep[Uniform Manifold Approximation and Projection:][]{UMAP} non-linear dimension reduction algorithm to explore potential patterns or clustering within our dataset. UMAP is a non-linear dimension reduction algorithm based on manifold learning techniques and topological data analysis. It is designed to preserve structure information on local scales, but is also effective at preserving structure on global scales, particularly as compared to alternative non-linear algorithms such as t-distributed Stochastic Neighbour Embedding \cite[t-SNE:][]{tsne2008}. It also has significantly superior scaling performance to t-SNE \citep[][Figures 5, 6, and 7]{UMAP}, allowing for much larger samples sizes and number of dimensions. UMAP can also be semi-supervised or fully-supervised, and allows for metric learning where a model can be applied to unseen data. Furthermore, unlike t-SNE it returns meaningful distances between clusters of points. However, the dimensions returned in embedded space have no specific meaning, unlike principal component analysis, for example, where the returned dimensions are the direction of greatest variance in the original data.

In an unsupervised scheme, we applied UMAP to our test dataset (1.55 million spectroscopically observed sources) without supplying the class labels to the algorithm. UMAP transforms the 10-D data into 2-D, and the resulting embedding is shown in the top left panel of Figure~\ref{figure:umap-spec}, where each point has been colour coded by its class label, and where multiple classes are present in a single pixel bin those colours are combined proportionally. The brightness of the pixel represents the number of sources within that bin, and is on a logarithmic scale.

In a supervised scheme, we used our training dataset (1.55 million spectroscopically observed sources), to train a UMAP model embedding the 10-D data into two dimensions. In this way, UMAP sees the labels of these sources whilst fitting the embedding which maps them into the 2-D space. Once trained, we applied the model to the unseen test dataset of 1.55\,million sources. This is shown in the top right panel of Figure \ref{figure:umap-spec}. The same result is achieved in a semi-supervised scheme where only half of the class labels in the training dataset are given to UMAP.

In both supervised and unsupervised scenarios, UMAP has effectively separated the data into distinct classes in this 2-D space. This is achieved much clearer in the supervised scheme. Furthermore, UMAP has distributed the classes in a way that represent more than just their class label. The structures present in Figure \ref{figure:umap-spec} represent their magnitude, the shape of the spectrum across the SDSS and WISE bands, plus how resolved the sources are. Figure~\ref{figure:umap-spec-colours} is coloured by these parameters as a guide. Galaxies that are predominantly star forming (termed the blue cloud in colour-magnitude diagrams) are distinguished as a semi-separate cluster with smaller $| \mathrm{PSF}\,u - \mathrm{PSF}\,z |$ values (bluer colours). These also tend to be more resolved galaxies. From these bluer galaxies, the source density drops as their colours become greener. This is referred to as the green valley \citep{greenvalley2014, greenvalley2019}, which contains galaxies where star-formation has been quenched. Galaxies then become redder in the left of the structure which contains old passive galaxies (termed red sequence galaxies). Stars are arranged in an alternate representation of a Hertzsprung Russell diagram, distinguishing their colour and luminosity. Quasars are predominantly unresolved bluer sources, though redder quasars are seen on the outskirts of this cluster.

When taking a closer view of the clusters in Figure \ref{figure:umap-spec} it is noticeable that some regions are contaminated at a low level by colours from other classes. For example the region dominated by galaxies is contaminated at a low level by stars and quasars (shown by pink and blue colours within the green). This contamination is most noticeably so for the galaxy region, less so for the quasar region, and barely noticeably so in the star region. This qualitatively reflects the precision and recall for each class presented by the random forest. Stars have the highest precision (few false positives in the blue region of Figure \ref{figure:umap-spec}), followed by galaxies then quasars. Galaxies have the best recall (few false negatives in the pink and blue region of Figure \ref{figure:umap-spec}), however stars and quasars have a poorer recall, shown by blue and pink appearing outside of their respective classes.

The bottom row of Figure~\ref{figure:umap-spec} shows the mean and standard deviation of the random forest classification probabilities in the supervised and unsupervised metric-learning schemes for the spectroscopic test dataset. The mean classification probabilities are very high (also seen in Figure~\ref{figure:prob-hist}) throughout this 2-D space, and only decrease in regions where classes overlap in the unsupervised case, and on the outskirts of clusters in the supervised case. The standard deviation of the classification probabilities is very low, varying slightly throughout the galaxy and quasar clusters, but remains low for most of the cluster of stars. Overall, a source on the outskirts of a cluster in these UMAP projections does not always correlate with it having a lower random forest classification probability.

Whilst we have used UMAP as a qualitative assessment of the class labels, it affirms that using UMAP in an unsupervised, semi-supervised, or fully-supervised scheme is an effective tool in understanding the distribution of sources where labels are not available. In Section \ref{section:newsources-umap} we explore using UMAP on the 111 million un-labelled photometrically observed sources.


\begin{figure}
\includegraphics[width=\hsize]{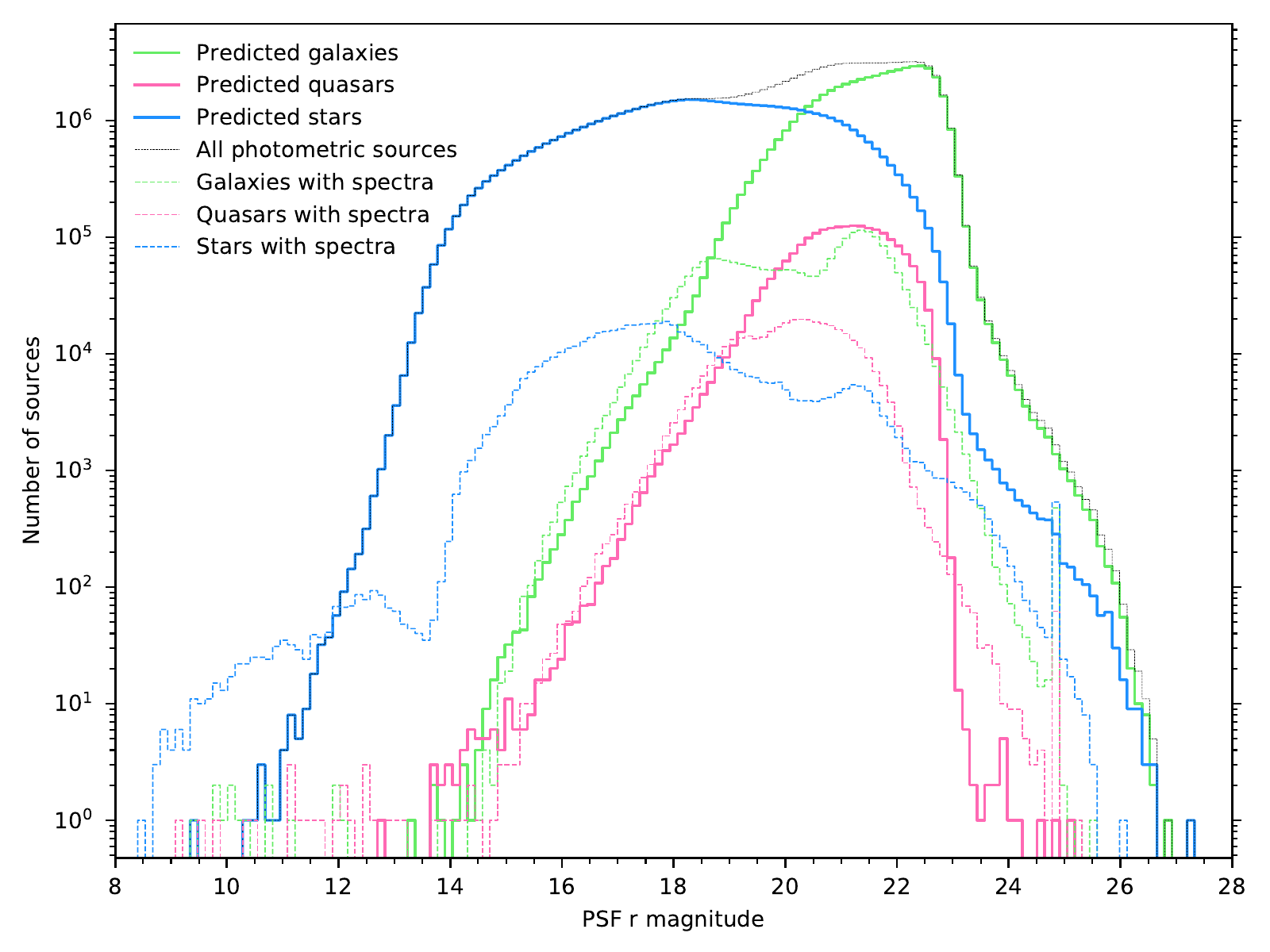}
\caption{Histogram of the PSF \textit{r} magnitude for all 111 million newly classified sources. Spectroscopically observed sources are also plotted.}
\label{figure:new-sources-hist-psfr}
\end{figure}

\begin{figure*}
\includegraphics[width=\hsize]{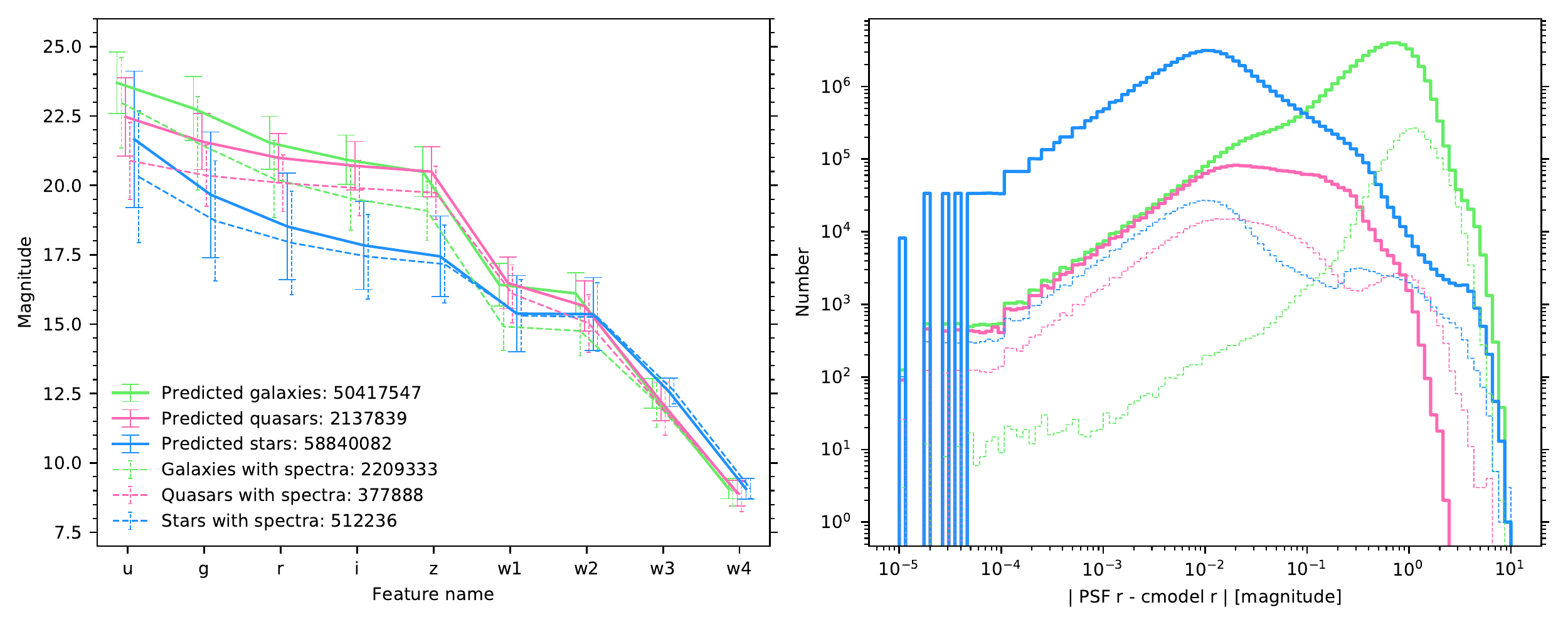}
\caption{\textit{Left:} Average magnitude and one standard deviation error bar for each feature (waveband) for 111 million newly classified sources. Error bars and plot lines are offset in the x-axis for clarity. \textit{Right:} Histogram per class over a measure of how resolved the source is for the newly classified sources (solid line), and for our dataset of 3.1 million sources with spectra (dashed line).}
\label{figure:new-sources-features}
\end{figure*}

\begin{figure*}
\includegraphics[width=\hsize]{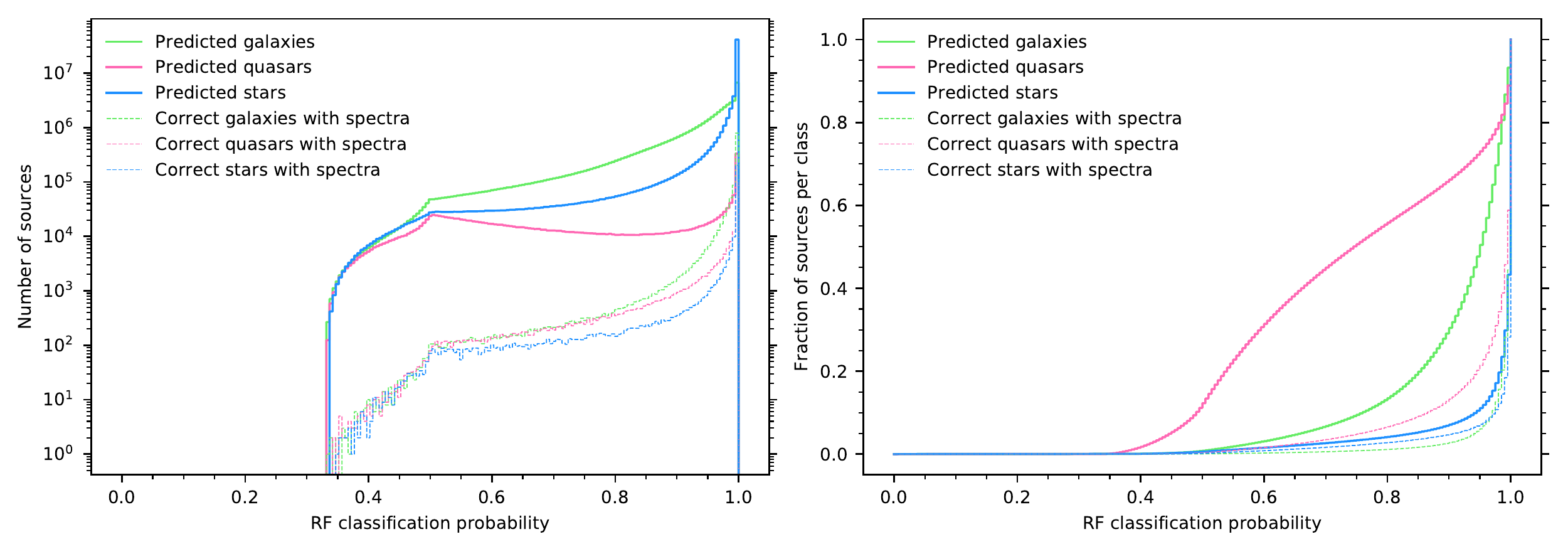}
\caption{Histogram (\textit{left}) and cumulative normalised histogram (\textit{right}) of the random forest classification probabilities using a bin size of 0.005. 13\%, 15\% and 70\% of the 111 million classified galaxies, quasars, and stars have classification probabilities greater than 0.99. 70\%, 34\% and 93\% of galaxies, quasars, and stars have classification possibilities greater than 0.9.}
\label{figure:new-sources-probhist}
\end{figure*}

\begin{figure*}
\includegraphics[width=\hsize]{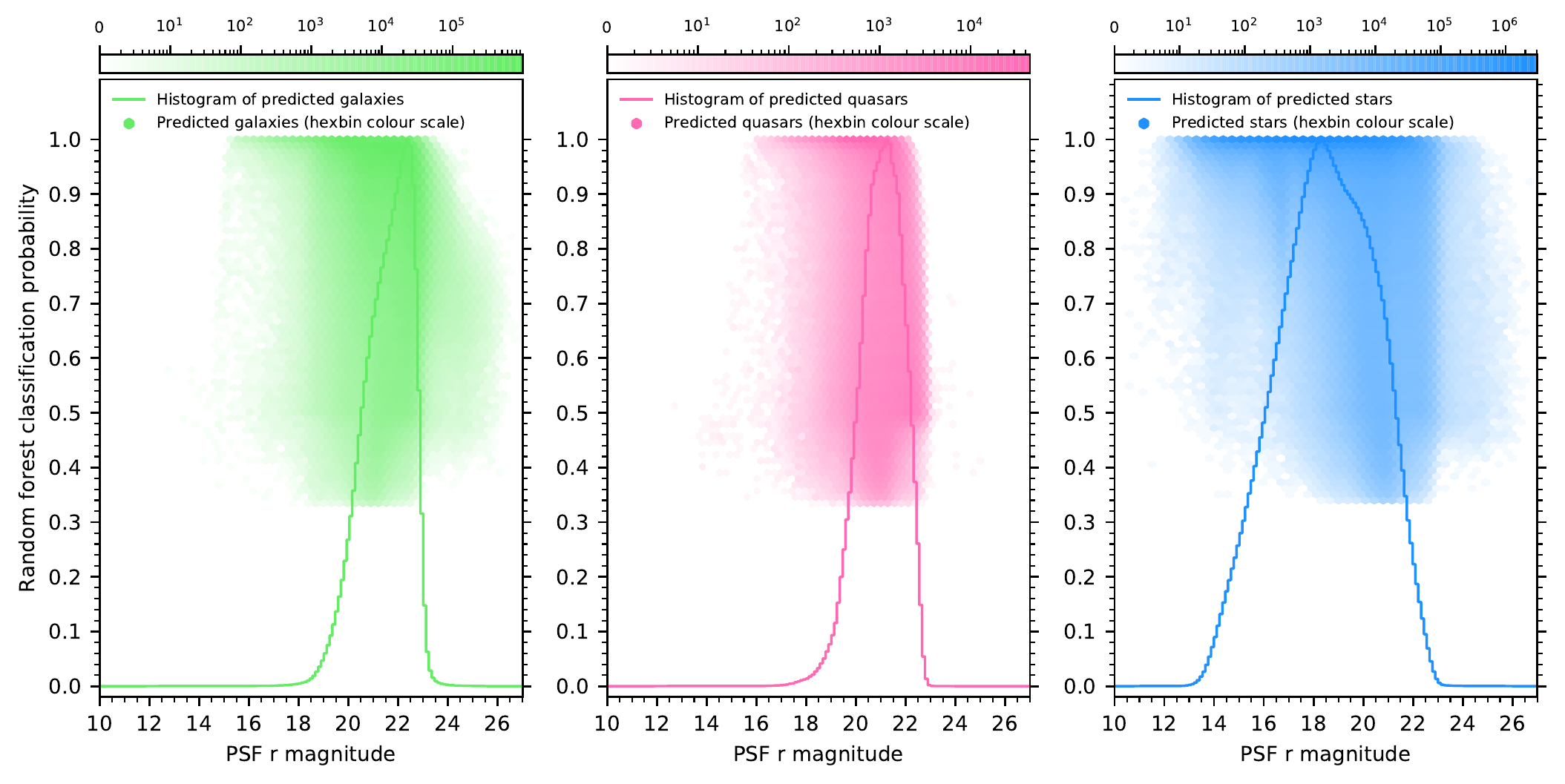}
\caption{Random forest classification probabilities plotted as a function of the PSF \textit{r} magnitude for the 111 million newly classified sources. A histogram of the PSF \textit{r} magnitude is also overlaid normalised for each class. 13\%, 15\% and 70\% of the galaxies, quasars, and stars have classification probabilities greater than 0.99. 70\%, 34\% and 93\% of galaxies, quasars, and stars have classification possibilities greater than 0.9 (as detailed in Figure \ref{figure:new-sources-probhist}).}
\label{figure:new-sources-prob-hexbin}
\end{figure*}

\begin{figure}
\includegraphics[width=\hsize]{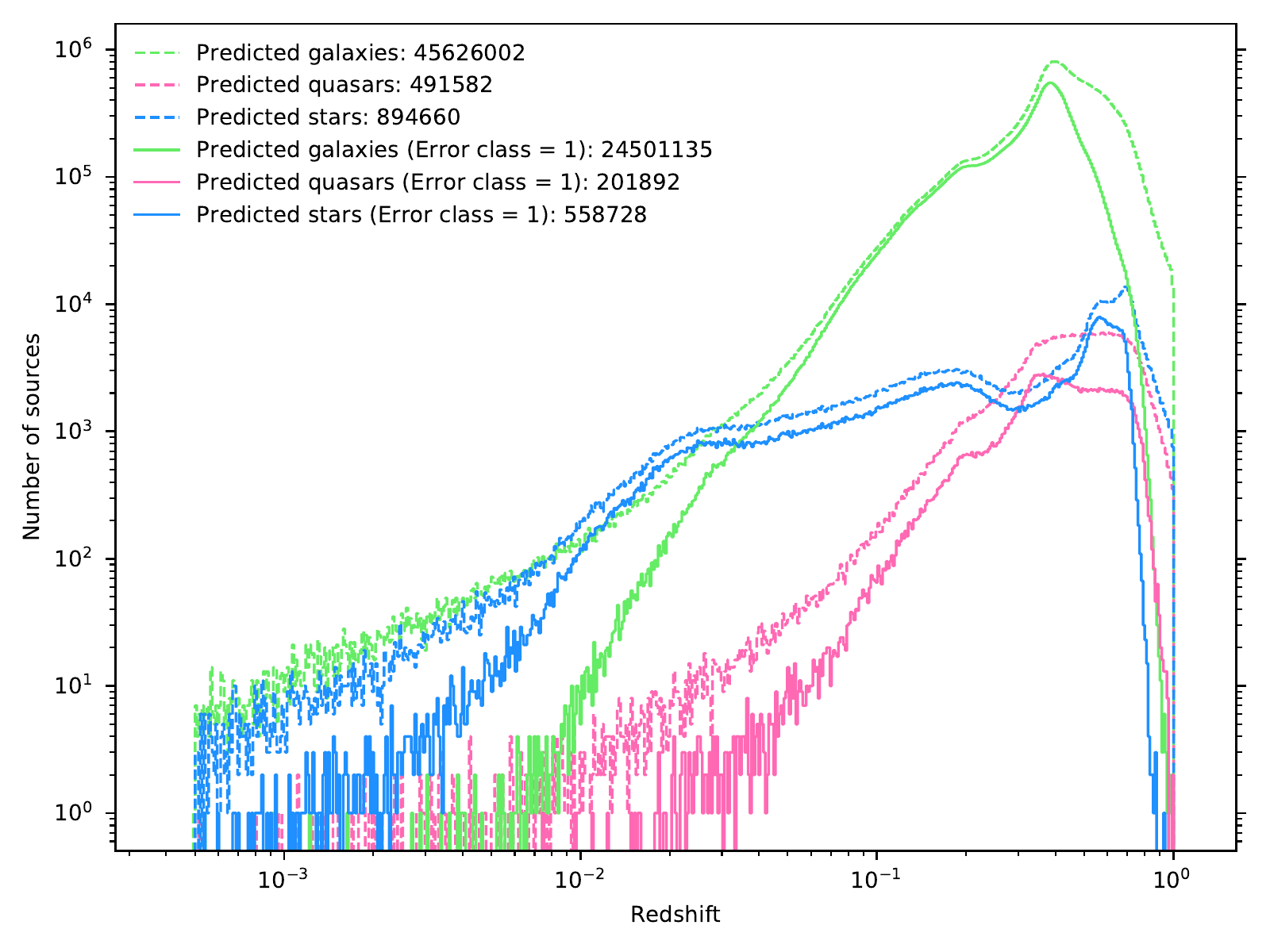}
\caption{Histogram of photometric redshifts for newly classified sources. 47 million (42\%) have available photometric redshifts. We note the large portion of stars with large photometric redshifts, even when limiting to those with the most confident fits (error class = 1). The distribution of classification probabilities for these sources is not shifted towards lower values, and we suggest it is likely due to the photometric redshift pipeline incorrectly identifying galaxies as stars.}
\label{figure:new-sources-redshifts}
\end{figure}

\begin{figure}
\includegraphics[width=\hsize]{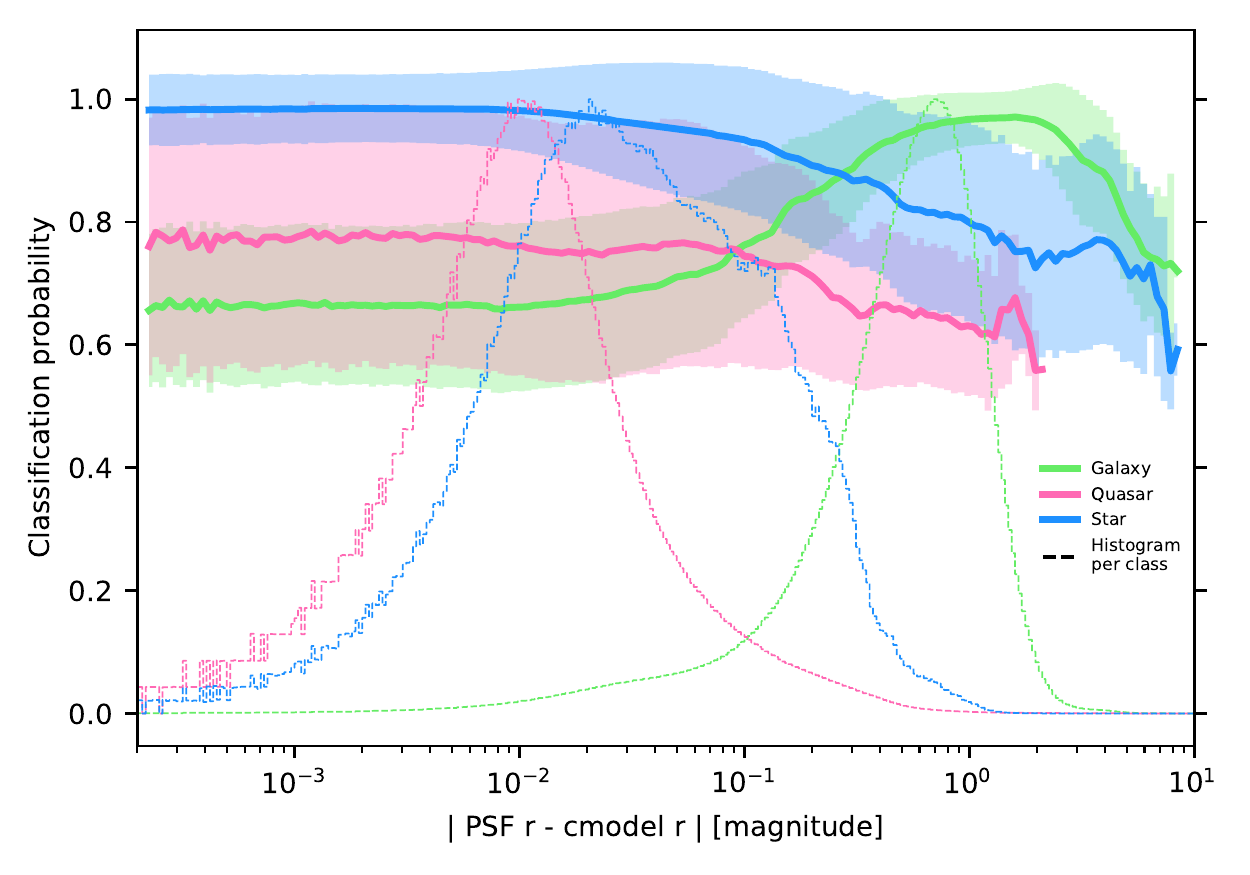}
\caption{Classification probabilities for 111 million newly classified sources as a function of the $resolved_r$ parameter. One standard deviation is shaded, though note that the distribution is not Gaussian (as shown in Figure~\ref{figure:new-sources-probhist}) and as such one standard deviation can exceed a probability of 1. Dashed lines show normalised histograms per class to highlight the source density as a function of the $resolved_r$ parameter.}
\label{figure:new-sources-resolvedr-probs}
\end{figure}

\section{Classifying new sources} \label{section:newsources}

\subsection{Applying the random forest model}

Using the machine learning model described in Section~\ref{section:ML}, we classify the 111\,395\,468 previously unlabelled SDSS photometric sources described in Section~\ref{sec:photo} as either a galaxy, quasar or star. This returns 50\,417\,547 galaxies, 2\,137\,839 quasars, and 58\,840\,082 stars, with their distribution shown in Figure~\ref{figure:new-sources-hist-psfr}.

Figure~\ref{figure:new-sources-features} shows that the average shape of a spectrum across the SDSS and WISE bands for these newly labelled photometrically observed sources is similar to that of the spectroscopically observed sources. Newly labelled galaxies and quasars are fainter across all bands (apart from \textit{w3} and \textit{w4}), whilst newly labelled stars are only significantly fainter in the shorter wavelength SDSS bands. The $resolved_r$ feature effectively distinguishes stars from galaxies for these new sources, shown in the right plot of Figure~\ref{figure:new-sources-features}, although there are significantly more newly labelled galaxies that are unresolved point sources than were in the spectroscopically observed dataset.

Figure~\ref{figure:new-sources-probhist} shows the distribution of classification probabilities for the newly labelled sources. The 0.99-1 probability bin is the most populated bin, containing 6\,683\,526 galaxies (13\%), 330\,666 quasars (15\%), and 41\,279\,349 stars (70\%). There are 35\,075\,918 galaxies (70\%), 722\,159 quasars (34\%), and 54\,673\,689 stars (93\%) with probabilities greater than 0.9, which is a lower fraction than those in our spectroscopically observed test dataset (96\%, 84\%, 94\%). The cumulative histogram in the right of Figure~\ref{figure:new-sources-probhist} shows that the probabilities of newly labelled stars have a similar distribution to those with spectra. However, newly labelled galaxies and quasars have significantly more sources with lower classification probabilities than those with spectra. This indicates that stars are easier to classify than the other two classes, and quasars are the most difficult, having lower classification probabilities on average. 

Figure~\ref{figure:new-sources-prob-hexbin} shows how the classification probabilities vary as a function of the PSF \textit{r} magnitude. For stars fainter than a PSF \textit{r} magnitude of 18, the classification probabilities are very high. However, above magnitudes of 18, there are many more stars with lower classification probabilities. This is likely because stars at fainter magnitudes are more easily confused with the higher number density of galaxies and quasars at fainter magnitudes. Most of the newly labelled quasars with lower classification probabilities have magnitudes above 20. Above a magnitude of 23, new sources have low classification probabilities mainly split between galaxies and stars.

Figure~\ref{figure:new-sources-resolvedr-probs} shows how the classification probabilities vary as a function of the $resolved_r$ parameter for these newly classified sources.
The number density of galaxies drops at smaller $resolved_r$ values, with 8.1 million galaxies having $resolved_r < 0.2$, and 4.2 million with $resolved_r < 0.1$ (also seen in Figure \ref{figure:new-sources-features}). Whilst 70\% of galaxies have probabilities greater than 0.9 and the peak in the normalised histogram for galaxies overlaid on Figure~\ref{figure:new-sources-resolvedr-probs} is at $resolved_r=0.7$, for unresolved galaxies the classification probabilities are much lower, plateauing at 0.66 when $resolved_r<0.01$. This is also reflected in the training and testing datasets where there are significantly less unresovled galaxies to train on, and therefore the their classification probabilities are much lower (Figure \ref{figure:metric-curves}).

There are photometric redshifts available for 47 million sources in our catalogue \citep{beck2016photoz}. Figure \ref{figure:new-sources-redshifts} shows their distribution, with quasars and galaxies having distributions peaked above $z=0.1$, and stars having a broader distribution cover lower redshifts. When looking at the classification probabilities for these stars with photometric redshifts, we see no difference to that of the overall distribution of probabilities. In other words, it is unlikely that these stars are all misclassified by our model. It is likely that the SDSS photometric redshift pipeline has fitted incorrect photometric redshifts to these sources which they have misclassified as stars.

\subsection{Clustering with UMAP} \label{section:newsources-umap}

\begin{figure*}
\includegraphics[width=0.33\hsize]{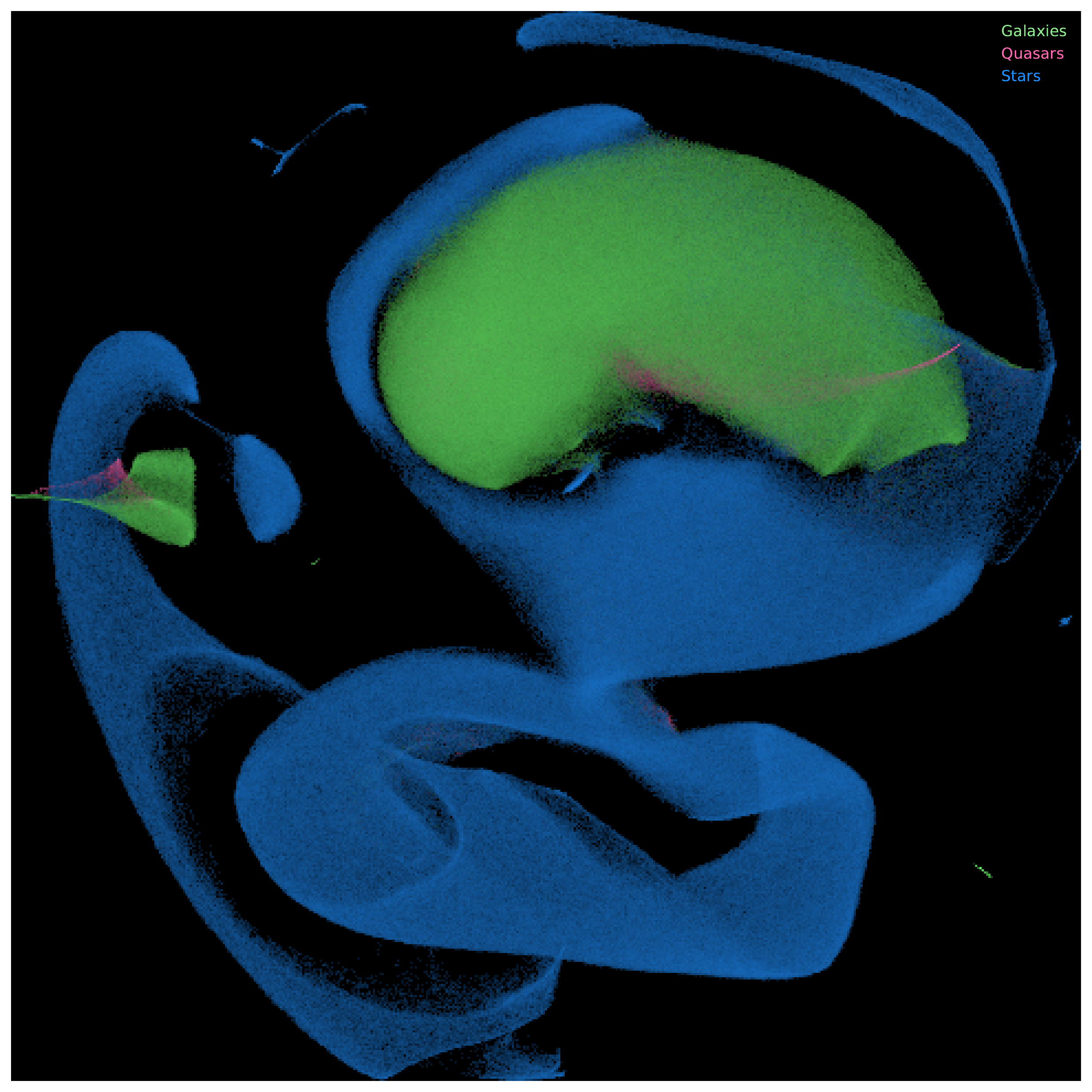}
\includegraphics[width=0.33\hsize]{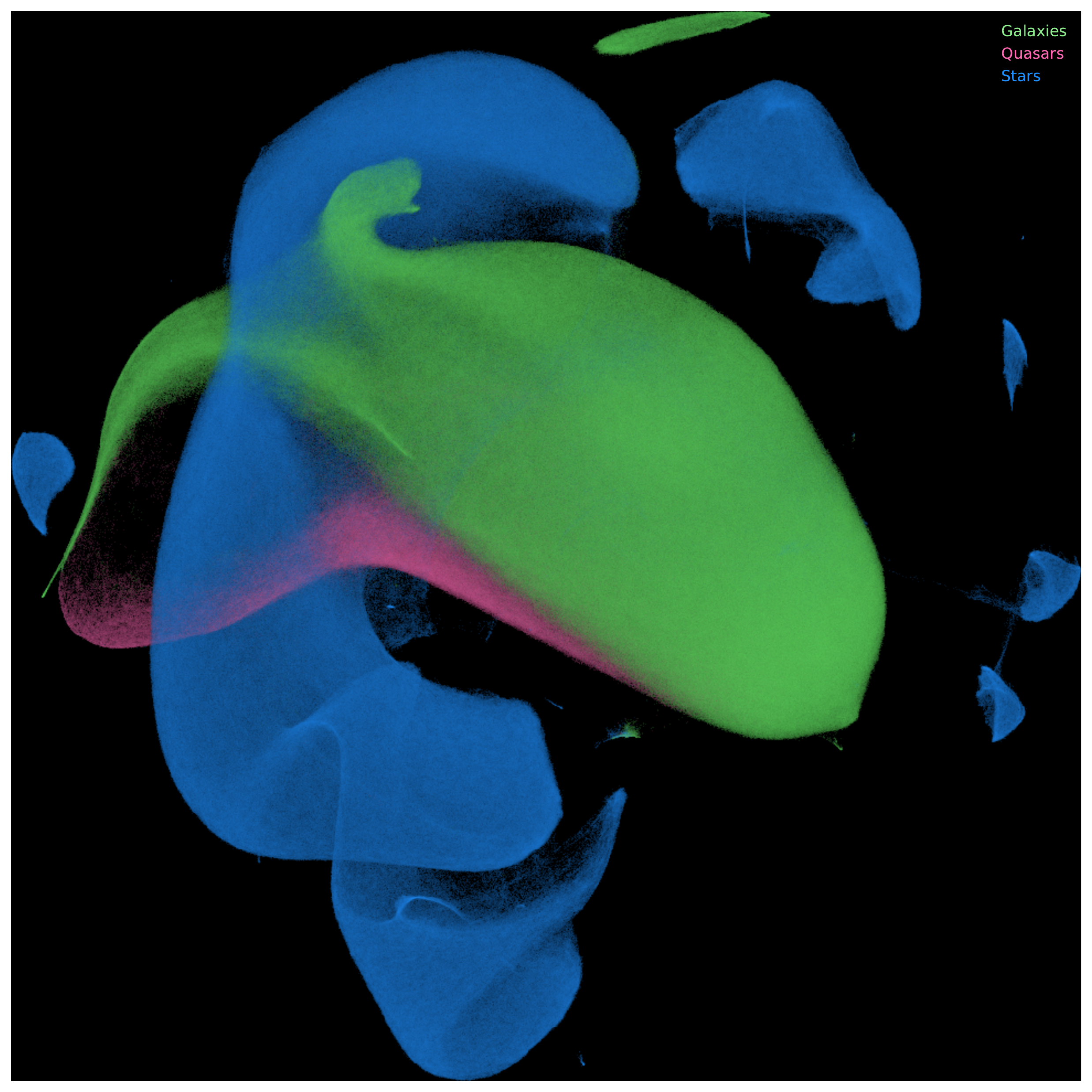}
\includegraphics[width=0.33\hsize]{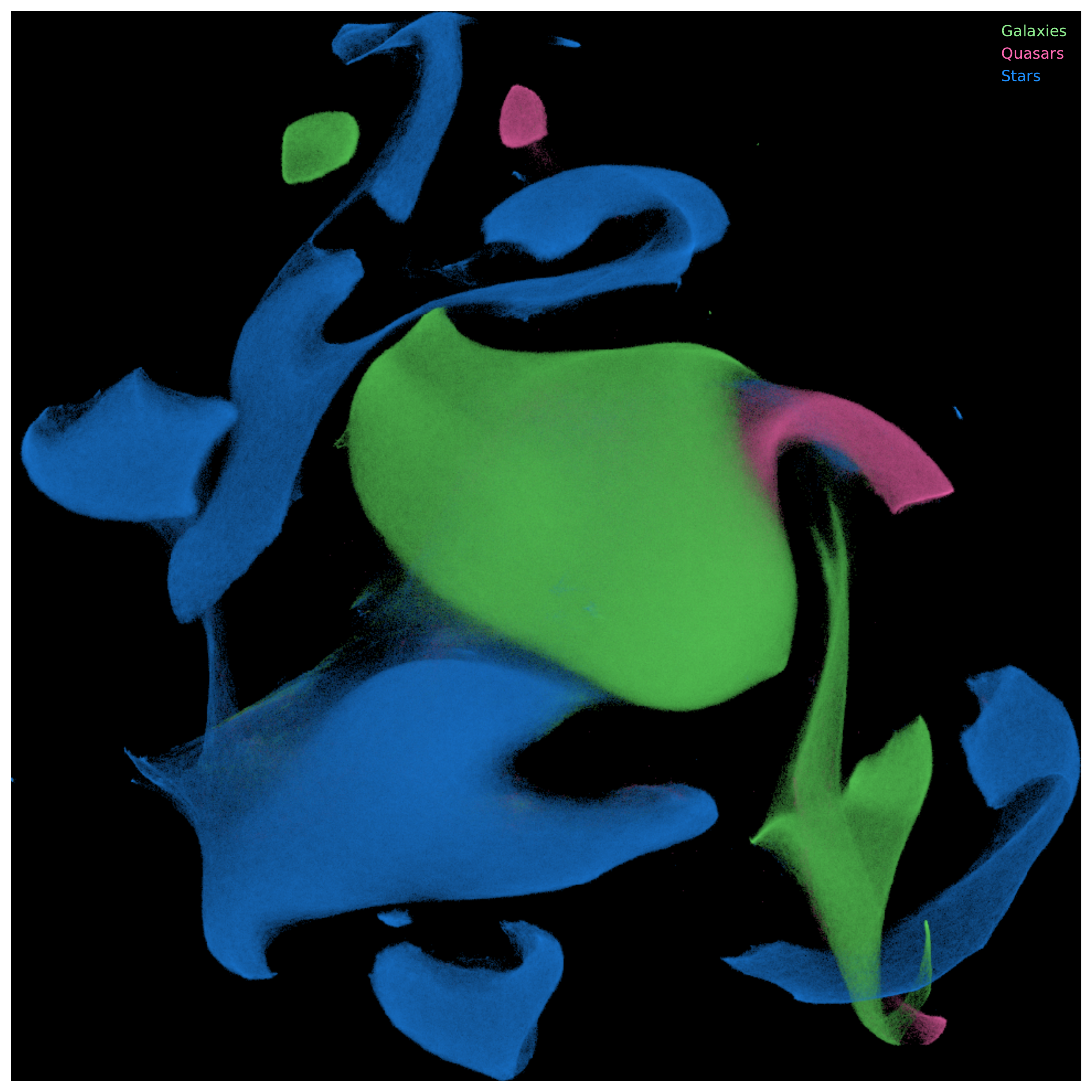}

\caption{UMAP applied to various subsets of data, reducing the original ten features to the two which are plotted. Sources are binned per pixel, with colours combined proportional to how many of each class are in that pixel bin. The brightness corresponds to the total source count in that pixel on a logarithmic scale. Source labels for galaxies (green), quasars (pink), and stars (blue) were derived from our random forest model. \textit{Left}: 11 million photometrically observed sources without spectra run in an unsupervised scheme. \textit{middle}: All spectroscopically observed sources in our dataset, plus 11 million photometrically observed sources, run in an unsupervised scheme. \textit{Right}: Same as the middle plot but with the class labels of spectroscopically observed sources passed to the UMAP algorithm in a semi-supervised scheme. Including the spectroscopically observed sources in unsupervised (middle plot), and semi-supervised schemes (right plot) UMAP helps UMAP separate out the classes without them overlapping (particularly the minority class of quasars). Furthermore, our labels assigned from the random forest model show that even when classes overlap in this 2-D space, the structures picked out by UMAP are consistent with our class labels.}
\label{figure:umap-specphoto}
\end{figure*}

\begin{figure*}
\includegraphics[width=0.5\hsize]{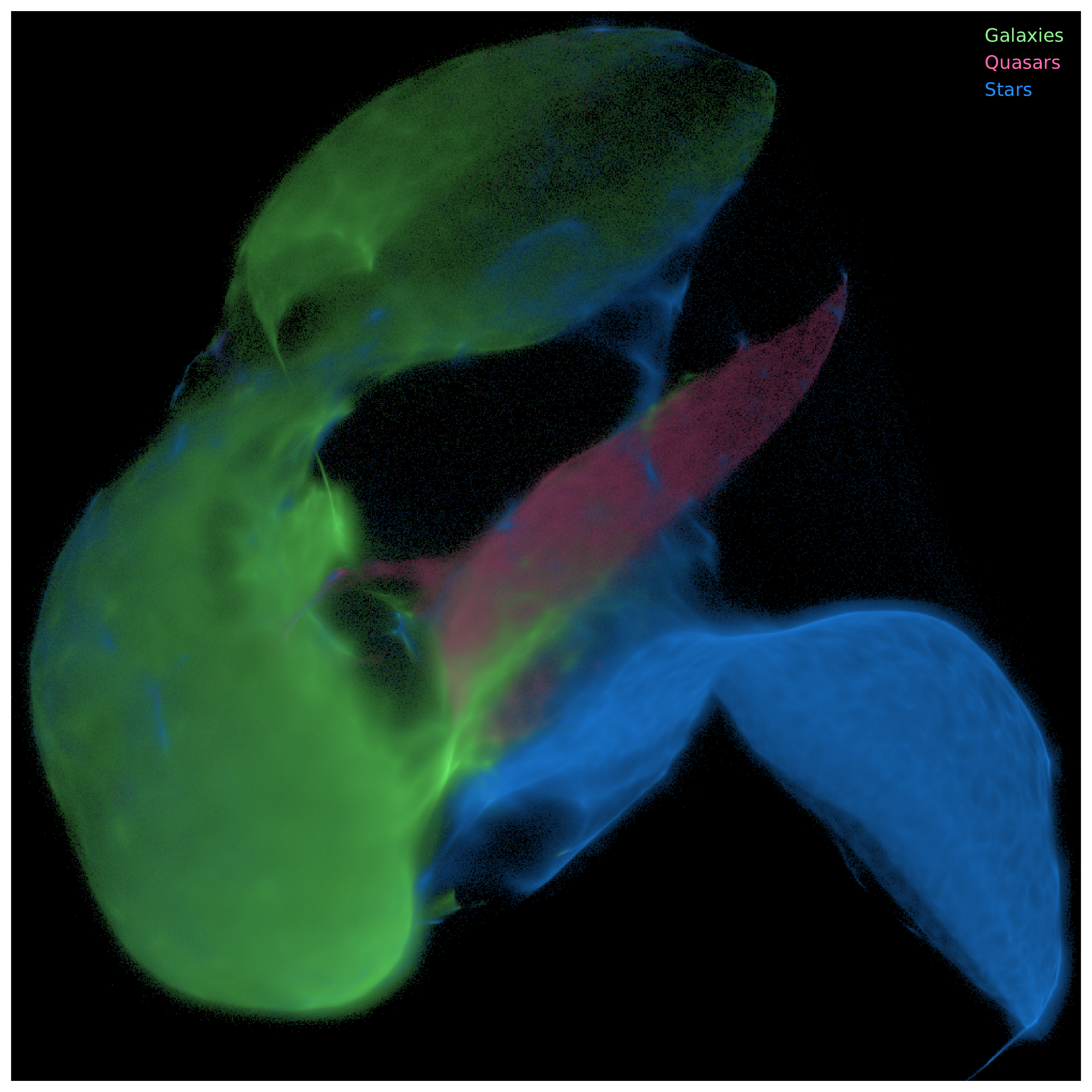}
\includegraphics[width=0.5\hsize]{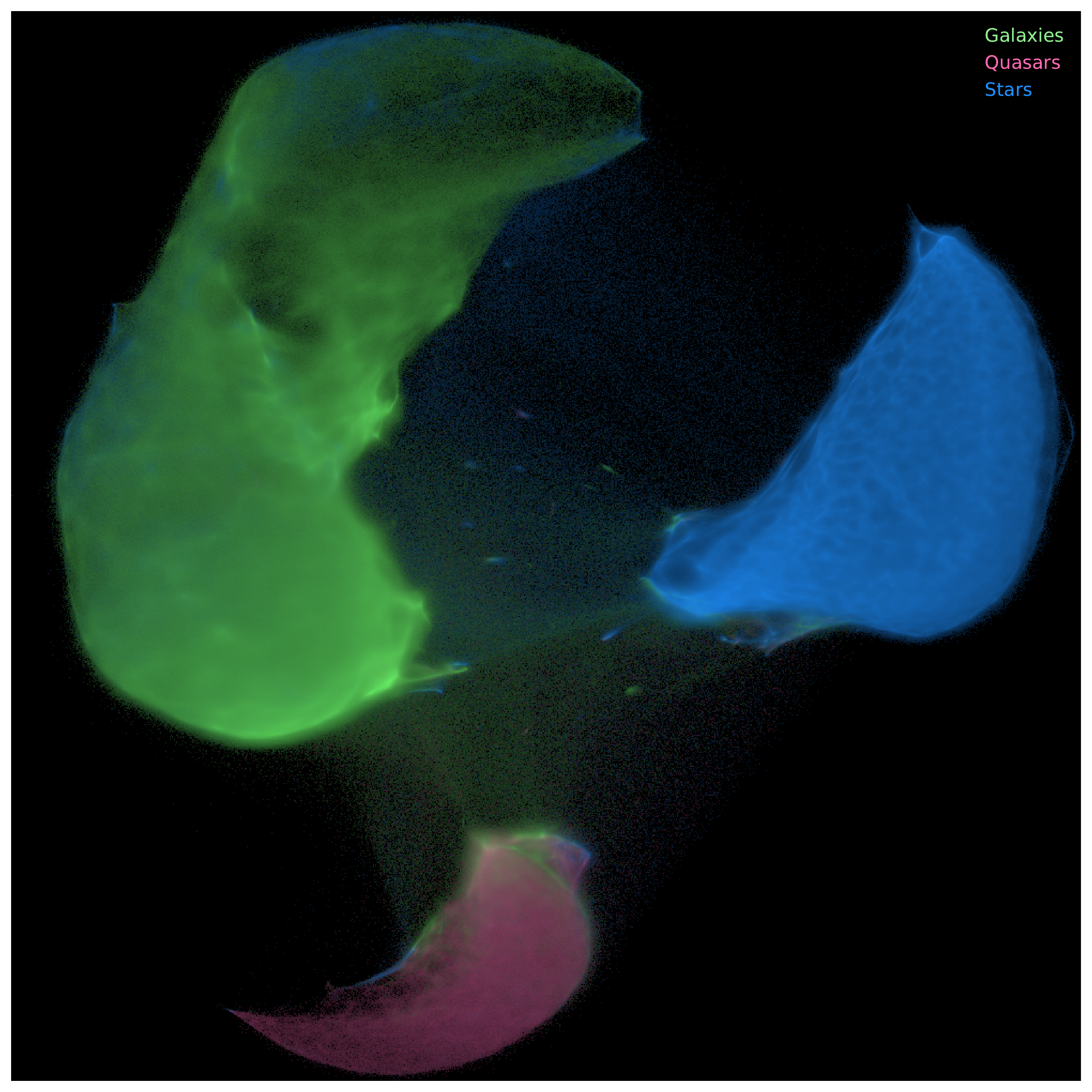}

\includegraphics[width=0.247\hsize]{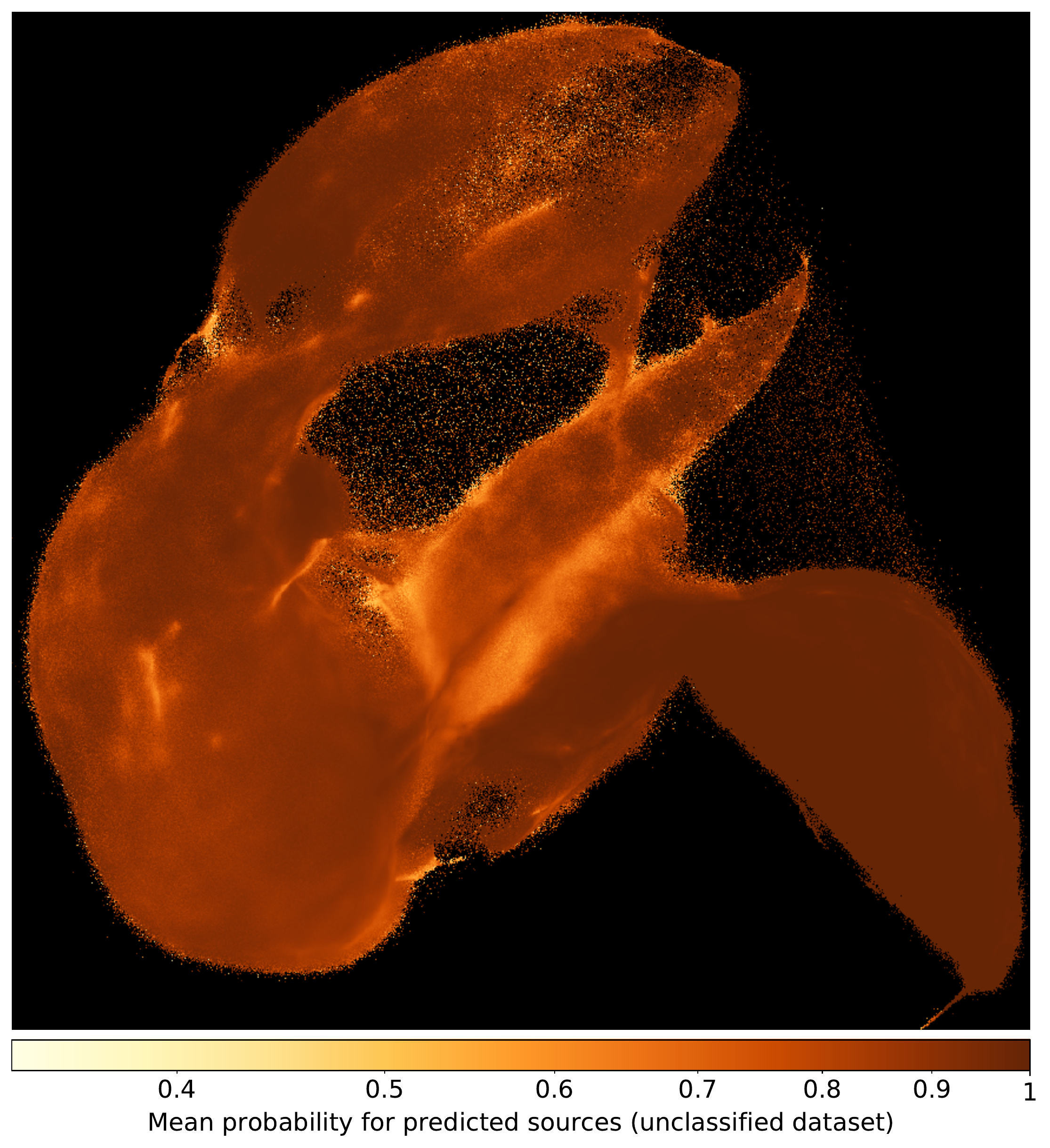}
\includegraphics[width=0.245\hsize]{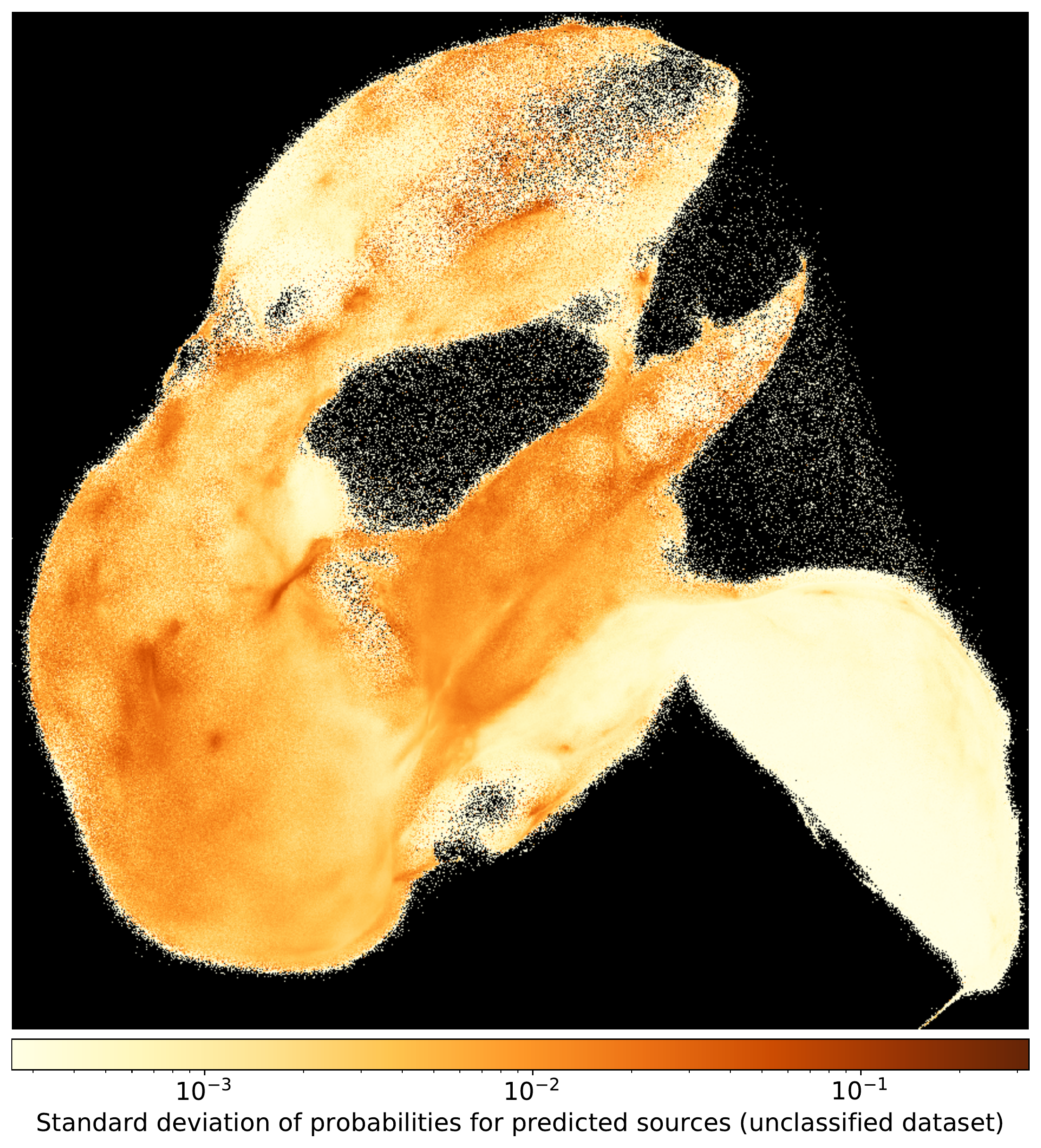}
\includegraphics[width=0.247\hsize]{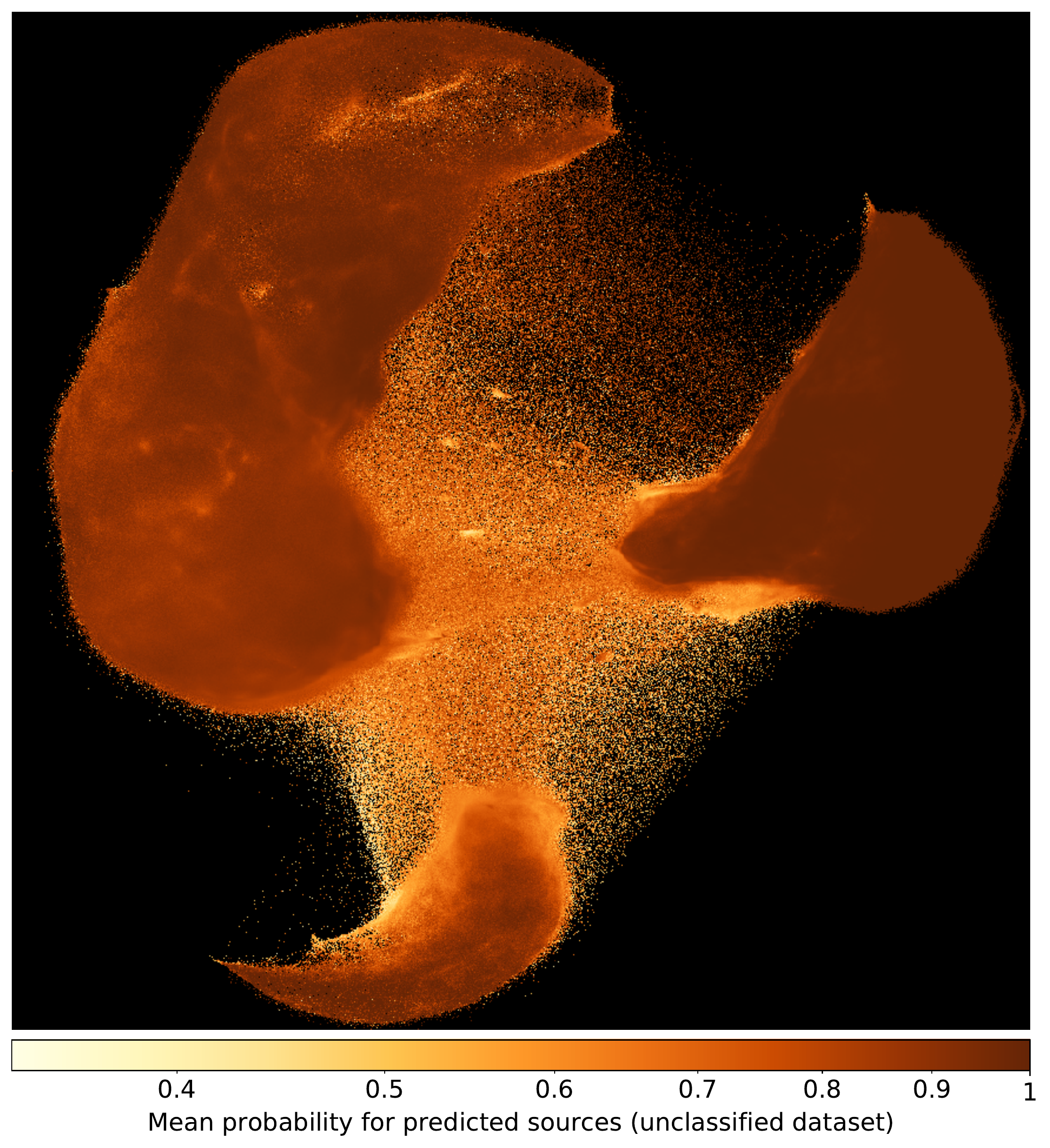}
\includegraphics[width=0.245\hsize]{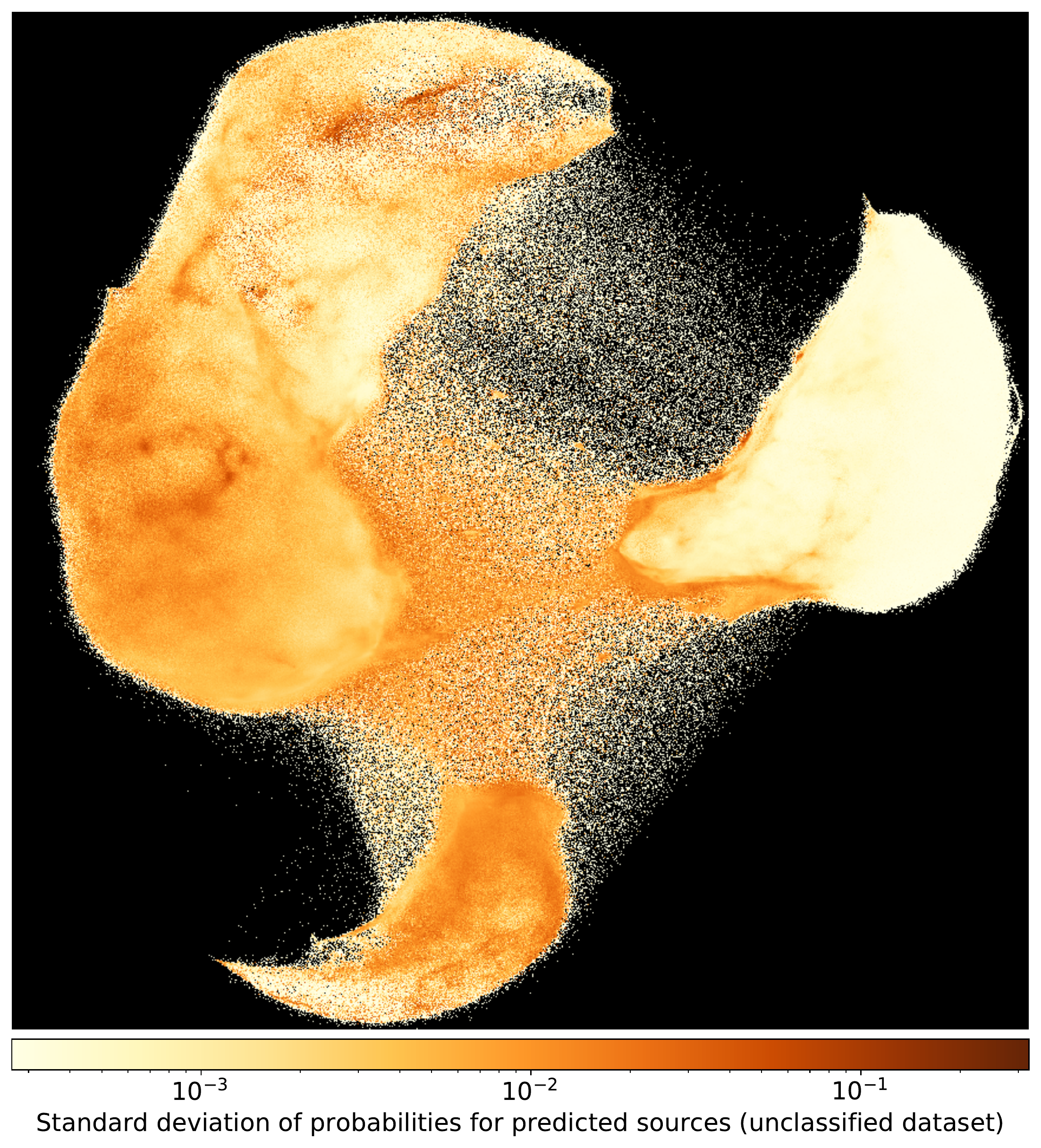}

\caption{UMAP applied to all 111 million photometrically observed sources in a metric-learning scheme, reducing the number of dimensions (features) from ten to two. UMAP models were obtained in unsupervised (\textit{top left}) and supervised (\textit{top right}) schemes using the spectroscopic training dataset (as in Figure \ref{figure:umap-spec}), and then all 111 million photometrically observed sources were embedded into this 2-D space. Colours were added afterwards for galaxies (green), quasars (pink), and stars (blue), derived from the random forest model trained on the same spectroscopically observed sources. Source are binned per pixel, and colours are combined proportional to how many of each class are in that pixel bin. The brightness corresponds to the total source count in that pixel on a logarithmic scale. UMAP effectively separates the classes in the same way as with the spectroscopic dataset (Figure~\ref{figure:umap-spec}) and is in a strong agreement with the random forest. The bottom row shows the mean and standard deviation of the random forest classification probabilities.}
\label{figure:umap-embed-photo}
\end{figure*}

\begin{figure*}
\includegraphics[width=0.33\hsize]{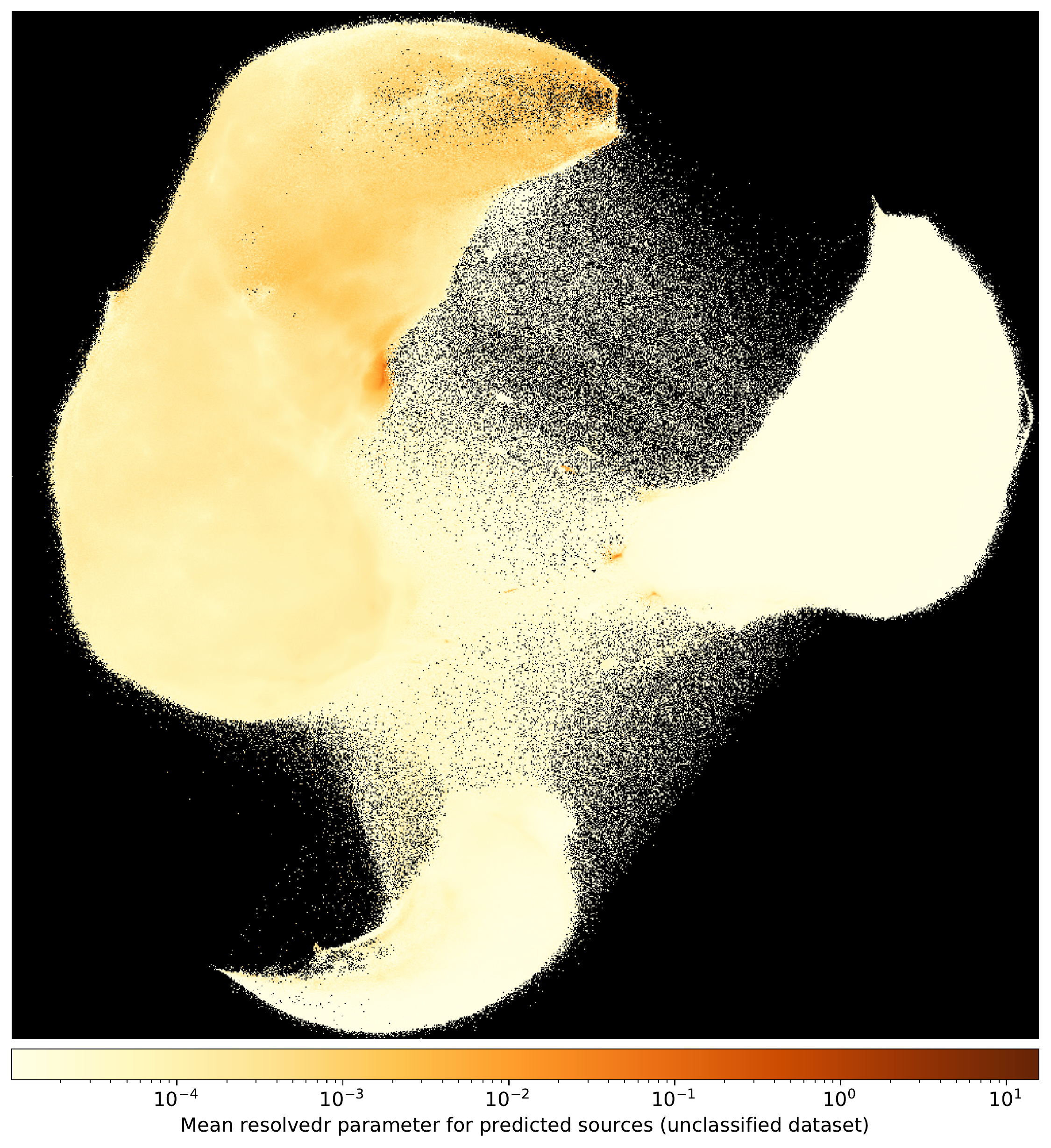}
\includegraphics[width=0.33\hsize]{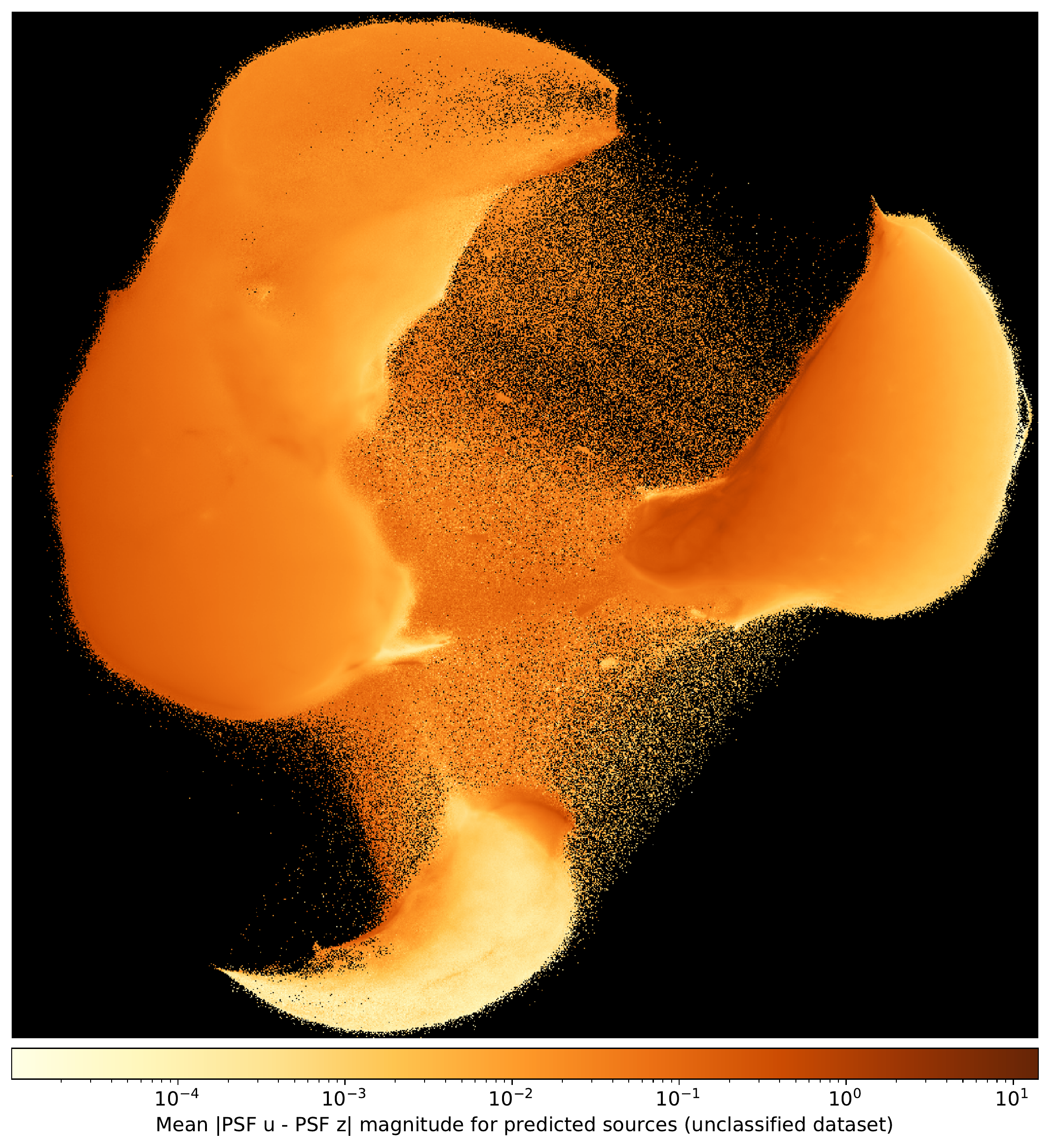}
\includegraphics[width=0.33\hsize]{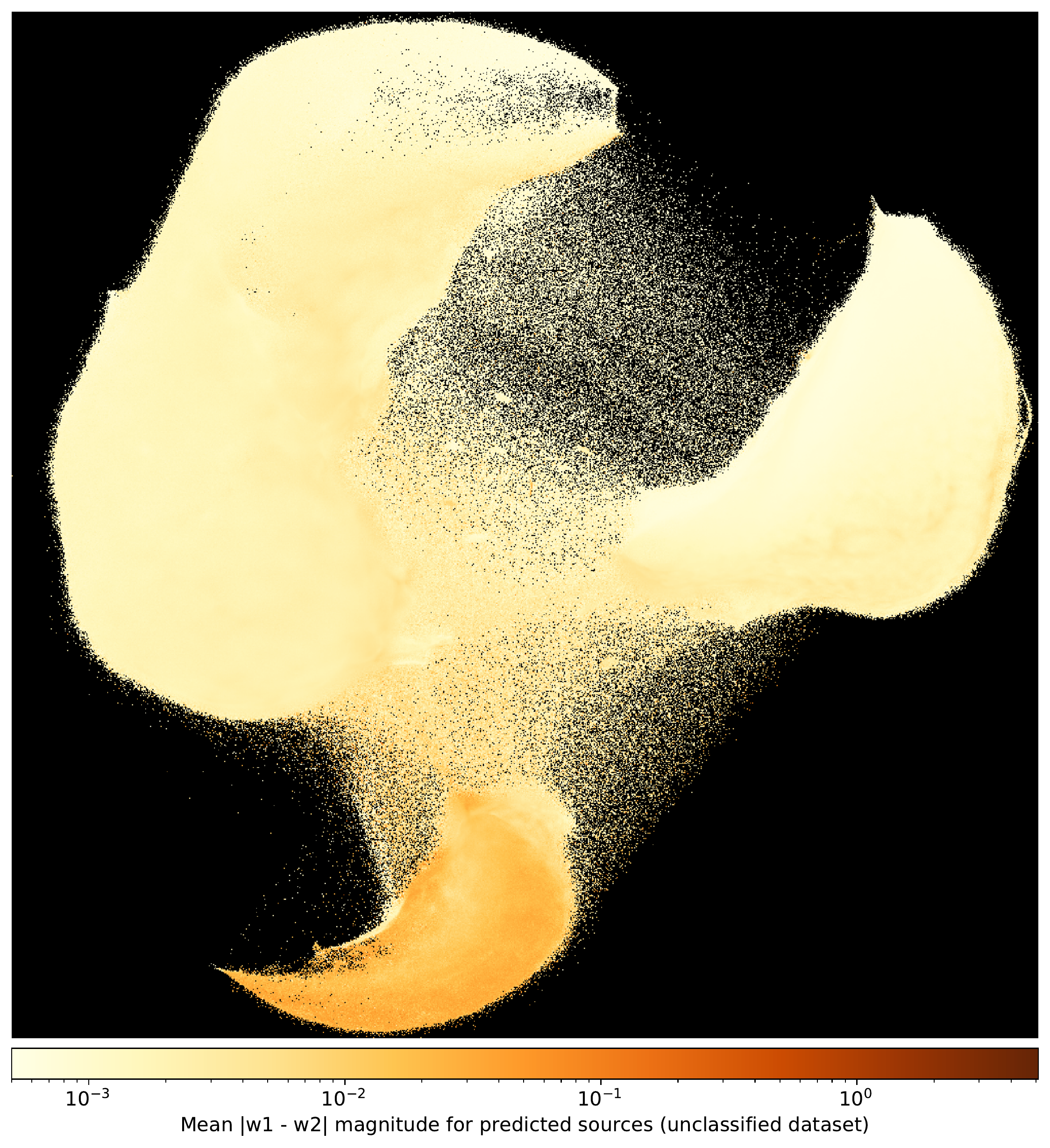}
\includegraphics[width=0.33\hsize]{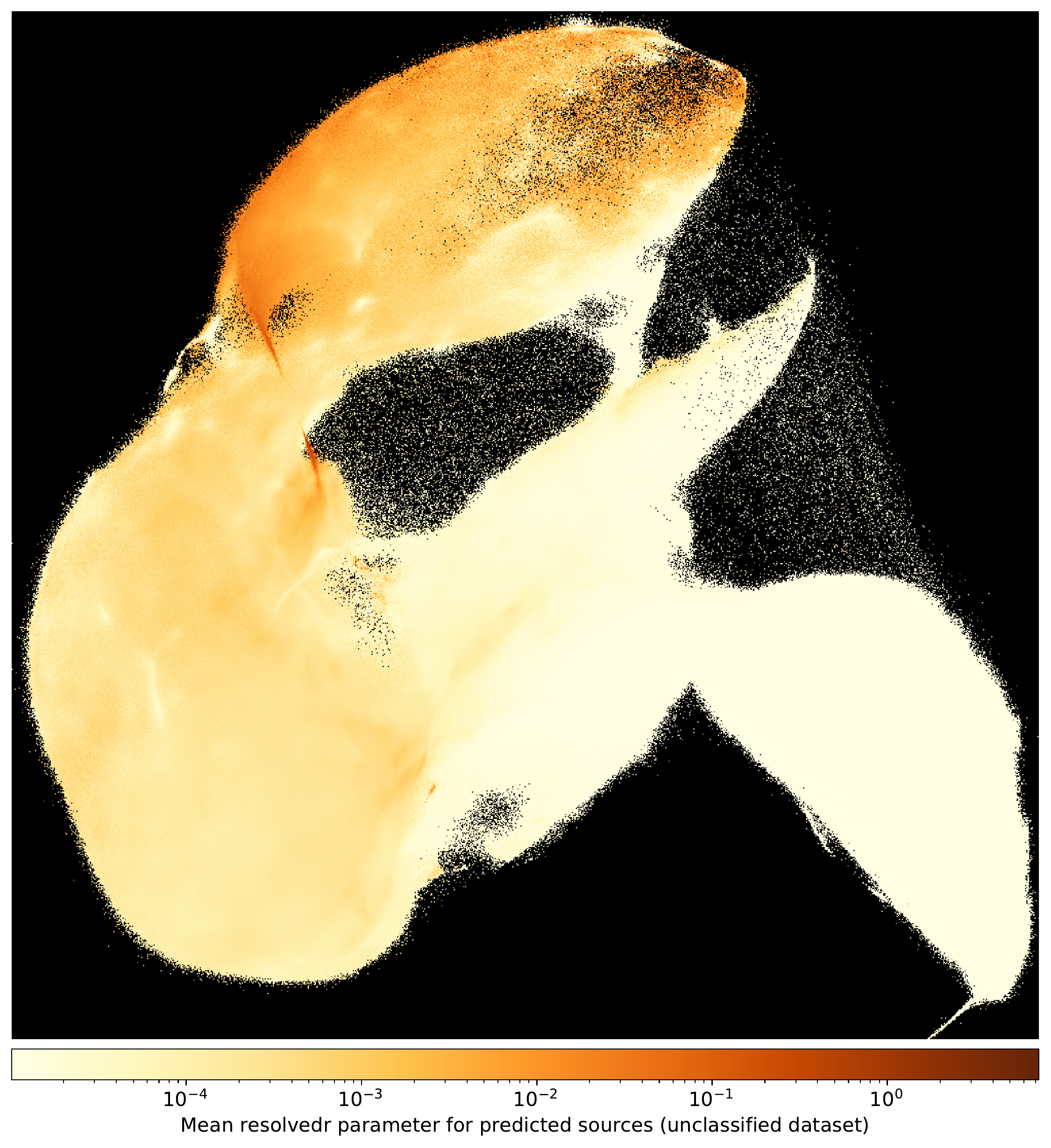}
\includegraphics[width=0.33\hsize]{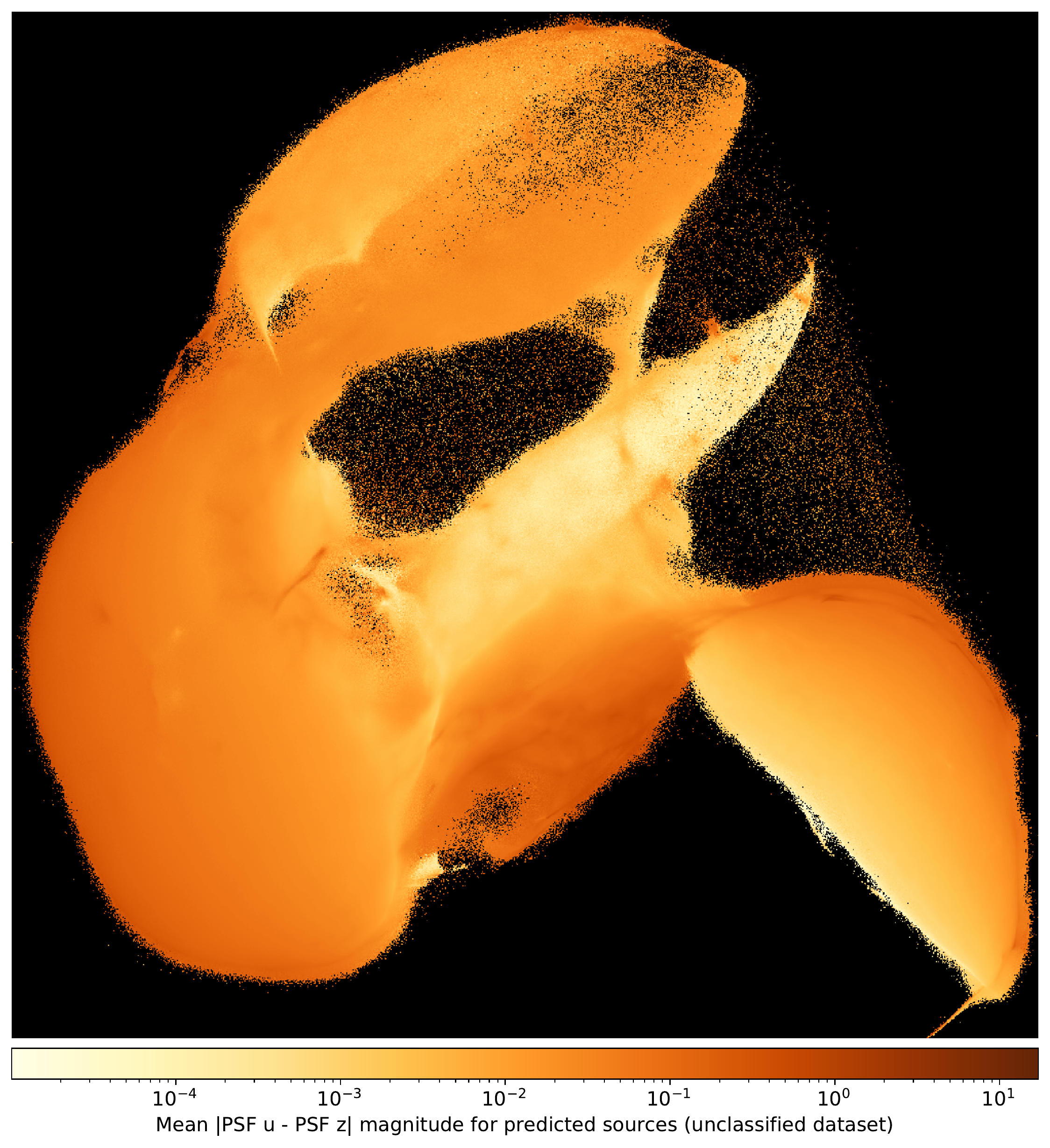}
\includegraphics[width=0.33\hsize]{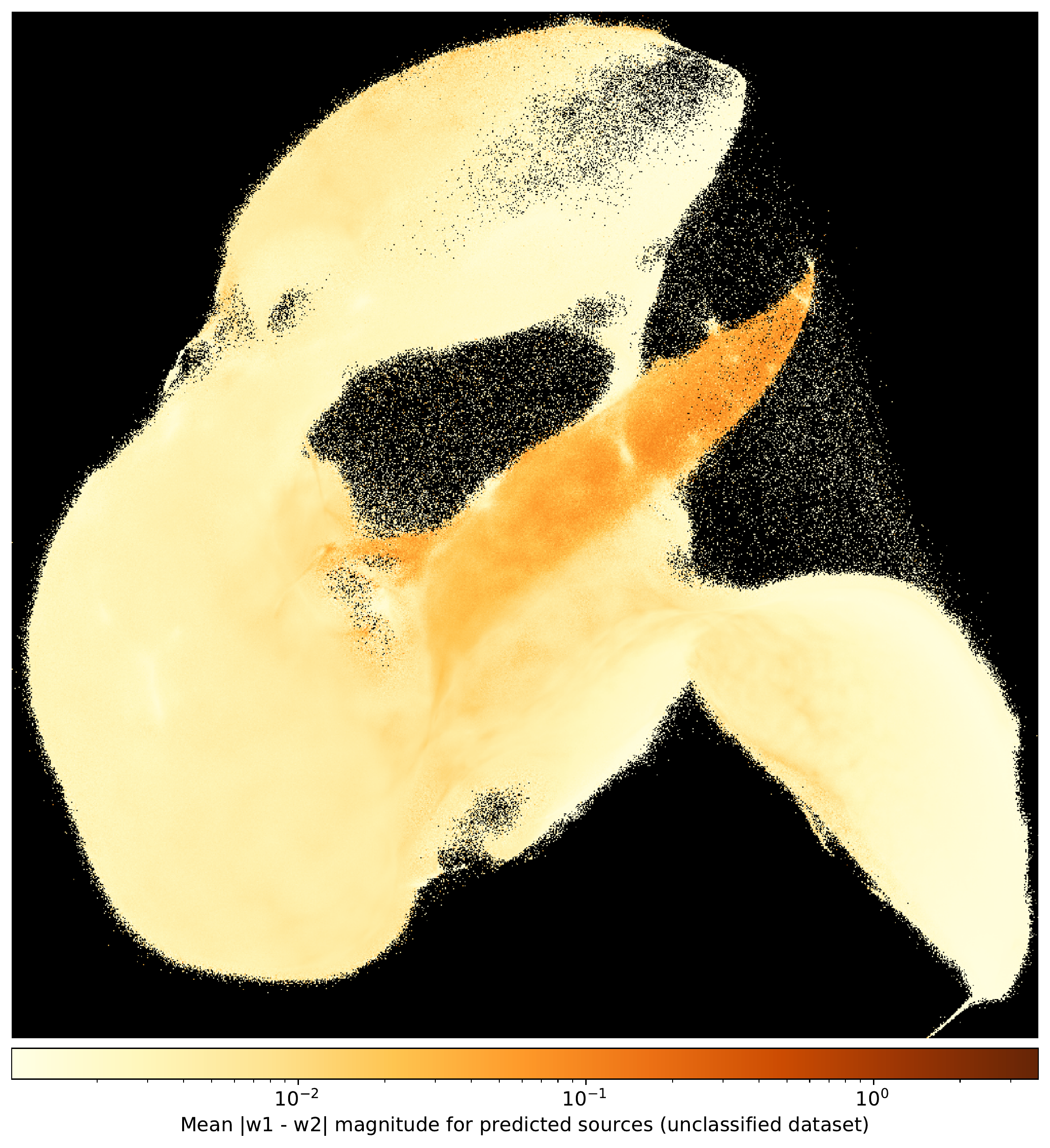}

\caption{Photometric dataset of 111 million sources processed with UMAP to reduce from ten features to two in supervised (\textit{top row}) and unsupervised (\textit{bottom row}) schemes, using models trained on the spectroscopic test dataset. The resulting two dimensions are plotted, with sources binned per pixel and coloured by the $resolvedr$ parameter (\textit{left column}), an SDSS colour $|\mathrm{PSF}\, u - \mathrm{PSF}\, z|$ (\textit{middle column}) and a WISE colour $|w1 - w2|$ (\textit{right plots}). Structures are as in Figure~\ref{figure:umap-spec-colours}, but the plot is more densely populated.}
\label{figure:umap-photo-colours}
\end{figure*}

We applied UMAP to 10\% of the 111\,395\,468 million photometrically observed sources in an unsupervised scheme, shown in the left plot of Figure~\ref{figure:umap-specphoto}, where sources are coloured by the classification labels assigned by the random forest model. UMAP effectively separates the galaxies and stars as labelled by our random forest model, but does not pick out quasars as a separate cluster due to the significantly lower number of sources. However, quasars are one continuous structure mainly mixed in with the cluster of galaxies. Stars are clustered into various groups with complex structure. Galaxies are clustered as a single group which broadly resembles the global structures displayed in the spectroscopic sources, seen in Figure~\ref{figure:umap-spec}. The middle plot in Figure~\ref{figure:umap-specphoto} includes the 3.1\,million spectroscopically observed sources without their labels, combined with 10\% of the unlabelled photometric sources. This unsupervised approach guided by the inclusion of unlabelled spectroscopically observed sources improves the clustering by increasing the number in the minority class. It particularly helps to separate out quasars which now have much less overlap with other classes.
The right hand plot in Figure~\ref{figure:umap-specphoto} includes the same sources as the middle plot, but in this case the labels for the spectroscopically observed sources were explicitly used by the UMAP clustering algorithm. This semi-supervised approach helps UMAP achieve a clearer clustering as defined by the galaxy, quasar, star classification scheme. The galaxy-star separation is very distinct in this case, although there is still some overlap with quasars. Even when classes overlap in the embedded 2-D space shown in Figure~\ref{figure:umap-specphoto}, the structures visualised in overlapping regions are coherent and consistent with the labels from the random forest. This can be interpreted as non-overlapping clusters of sources being distinguished in the higher 10-D space by the random forest, but overlapping in this 2-D space due to the embedding derived by UMAP.

We also implement supervised dimension reduction by deriving UMAP models from the spectroscopic training dataset in supervised and unsupervised schemes, and using these fitted models to transform all unlabelled photometric sources into the resulting 2-D space. This form of metric-learning allows us to efficiently apply the UMAP transformation derived from 1.55\,million spectroscopically observed sources (as seen in Figure~\ref{figure:umap-spec}) to all 111 million photometrically observed sources. This is computationally advantageous compared to running UMAP on 111 million photometric sources in an un- or semi-supervised scheme. The results are shown in Figure~\ref{figure:umap-embed-photo}, using the unsupervised model in the top left plot, and the supervised model in the top right plot, where sources are colour coded post-projection with the labels assigned by the random forest algorithm. As expected in this metric learning scheme, the global and local clustering patterns appear the same as those from the spectroscopically observed sources shown in Figure~\ref{figure:umap-spec}. There is a clear separation of the three classes as designated by the random forest algorithm, similar to that seen for the spectroscopically observed sources. Given the significant increase in source density in this figure, regions that were previously unpopulated in Figure~\ref{figure:umap-spec} are now filled. There is significant overlap in the unsupervised case where each of the three classes intersect, and many more stars labelled by the random forest appearing in the quasar and galaxies clusters. In the supervised scheme there is significantly less overlap of the three classes, with UMAP and the random forest in a much stronger agreement. Nevertheless in the supervised scheme there are still regions where stars labelled by the random forest are co-located with the main cluster of galaxies and on the edge of the cluster of quasars. We do not see quasars labelled by the random forest noticeably in the main cluster of galaxies or stars, although there are only 2.1~million quasars compared to the 50~million galaxies and 59~million stars. Galaxies labelled by the random forest overlap the edge of the cluster of quasars in the supervised scheme where their colour is much redder (see the top middle plot of Figure~\ref{figure:umap-spec-colours}), indicating a disagreement between UMAP and the random forest for a small number of these unresolved red galaxies or quasars. Although these objects have much lower random forest classification probabilities. Galaxies labelled by the random forest do not overlap the cluster of stars with any significance.

The bottom row of Figure~\ref{figure:umap-embed-photo} shows the mean and standard deviation of the random forest classification probabilities in the supervised and unsupervised metric-learning schemes for the 111 million photometric sources. Whilst the mean classification probabilities are lower than in the test dataset (as seen in Figures~\ref{figure:umap-spec} and \ref{figure:new-sources-probhist}), the mean remains high throughout this 2-D space. It decreases in regions where classes overlap in the unsupervised case, however there are small clusters of stars between the quasar and galaxy clusters which still have high classification probabilities. This suggests that this unsupervised run of the metric-learning UMAP scheme is not the optimal separation in this 2-D space. In the supervised case the probabilities mainly only decrease in the region between the three clusters of sources, though there are noticeable regions within the cluster of galaxies where the probabilities are lower. The standard deviation of the classification probabilities varies significantly throughout the galaxy and quasar clusters, but remains low for most of the cluster of stars in both supervised and unsupervised cases. Figure~\ref{figure:umap-photo-colours} shows the same UMAP projections coloured by features, as in Figure~\ref{figure:umap-spec-colours}, but with two orders of magnitude more sources much more of the 2-D space is filled in.

The metric-learning schemes (Figure~\ref{figure:umap-embed-photo}) and direct application of UMAP to the photometric data in un-supervised or semi-supervised schemes (Figure~\ref{figure:umap-specphoto}) achieve similar results, but appear to have biases towards different classes. When applying UMAP directly to the photometric data in a semi-supervised scheme including photometric data (the right plot in Figure~\ref{figure:umap-specphoto}), UMAP effectively separate stars from galaxies. However, in the unsupervised metric-learning scheme (Figure~\ref{figure:umap-embed-photo}) there is significant contamination from stars in the galaxy and quasar clusters. Quasars are most effectively separated in the supervised metric-learning schemes. Whilst the un-supervised and semi-supervised implementations only use 10\% of the photometric sources and the metric-learning scheme uses all photometric sources, the same result is achieved when limiting the metric-learning method to the same subset of sources. 

Overall, the labels assigned to photometrically observed sources by the random forest model are consistent with the clusters derived from UMAP.
The metric-learning scheme gives the most consistent and interpretable results with the best separation of the classes. Whilst much slower, applying UMAP directly to a subset of the photometric dataset can be effective, and its effectiveness at separating clusters increases when including spectroscopically observed sources in an unsupervised scheme, and further increases in a semi-supervised scheme.
The metric-learning scheme has the advantage that it can be efficiently scaled to embed any number of previously unseen sources using a parallel or distributed computing system, making it a viable algorithm to use with the next generation of surveys from telescopes such as LOFAR, SKA and LSST.
While we could use UMAP as a pre-processing tool in anticipation of a supervised clustering scheme in two dimensions, for example a nearest neighbour approach, this will always be less accurate than a supervised learning scheme in the higher 10-D feature space. Hence, our use of UMAP is primarily as a diagnostic tool for this dataset, in particular for visualising the effectiveness of the classification labels assigned to photometrically observed sources by the random forest.

\subsection{Catalogue description}
\label{sec:catalogue}

The catalogue resulting from this work contains 111\,395\,468 labelled sources. These sources all have associated WISE matches but do not have associated spectra in SDSS. We provide the catalogue under a digital object identifier: \url{https://www.doi.org/10.5281/zenodo.3459293}. The whole catalogue is in a single Pandas \citep{pandas} data frame, saved as a Pickle file (.pkl). We also provide it as Pandas data frames per class, which are sorted by classification probability in descending order. The column names are described in Table~\ref{tab:catdesc}. An extract from the full catalogue is shown in Appendix~\ref{sec:catext} for illustration. There are 50\,417\,547 galaxies, 2\,137\,839 quasars, and 58\,840\,082 stars. Of these, 6\,683\,526 galaxies (13\%), 330\,666 quasars (15\%) and 41\,279\,349 stars (70\%) have classification probabilities greater than 0.99. At the same time, 35\,075\,918 galaxies (70\%), 722\,159 quasars (34\%), and 54\,673\,689 stars (93\%) with classification probabilities greater than 0.9.

\begin{table}[h!]
    \centering
    \caption{Description of catalogue columns for our 111 million classified objects. \label{tab:catdesc}}
    \resizebox{0.5\textwidth}{!}{
    \begin{tabular}{|r|l|}
    \hline
    \textbf{Column Name}  & \textbf{Description}  \\\hline
    \textit{objid} & SDSS object ID from the \textit{PhotoPrimary} table \\
    \textit{ra}    & Right Ascension\\
    \textit{dec}   & Declination \\
    \textit{psf\_u} & SDSS PSF magnitude in the \textit{u}-band \\
    \textit{psf\_g} & SDSS PSF magnitude in the \textit{g}-band \\
    \textit{psf\_r} & SDSS PSF magnitude in the \textit{r}-band \\
    \textit{psf\_i} & SDSS PSF magnitude in the \textit{i}-band \\
    \textit{psf\_z} & SDSS PSF magnitude in the \textit{z}-band \\
    \textit{w1} & WISE band 1 magnitude \\
    \textit{w2} & WISE band 2 magnitude \\
    \textit{w3} & WISE band 3 magnitude \\
    \textit{w4} & WISE band 4 magnitude \\
    \textit{resolvedr} & A measure of how extended a source is: $| psf_r - cmod_r |$ \\
    \textit{class\_pred}   & Predicted class \\
    \textit{class\_prob\_galaxy} & Probability of prediction as a galaxy \\
    \textit{class\_prob\_quasar} & Probability of prediction as a quasar \\
    \textit{class\_prob\_star} & Probability of prediction as a star \\ \hline
    \end{tabular}}
\end{table}

\section{Discussion} \label{section:discussion}

\subsection{Feature selection}
We have used a set of ten features that are readily available in order to predict class labels for a significant fraction of SDSS sources. As random forests compare random sets of features in each decision tree estimator, they effectively interpret colours in different bands naturally throughout the forest. This method helps remove bias and over-fitting, which could arise when only using a few colours to identify objects. Furthermore, utilising all SDSS and WISE wavebands, and the $resolved_r$ parameter together in this way provides a greater confidence in the classifications over using selected colours alone. We observe that removing absolute magnitude dependence from the model, i.e. taking the difference between the photometry from each band and the SDSS \textit{r}-band photometry, resulting in a set of nine total features, did not alter the results. 

The feature rankings returned from the random forest show that the $resolved_r$ parameter separates the classes very well. This is also shown by the histograms in the right panel of Figure \ref{figure:features}. The next best features are the SDSS magnitudes, followed by the WISE magnitudes, although the W4 band provides very little additional information in this classification scheme. Whilst higher signal to noise colours can be calculated using \textit{modelMag}, we found that this made no difference to the performance of our classifier. The higher signal to noise colours calculated from \textit{modelMag} are therefore likely not significant compared to the ensemble approach of biased estimators from the random forest.

In our classification scheme, where we do not know if a source is galactic (a star) or extra-galactic (a galaxy or quasar), we cannot apply galactic extinction corrections to sources with unknown class labels, such as our photometric catalogue. Whilst this was a small correction for our dataset and made no difference to the results, we envisage that accounting for galactic extinction will be an important consideration for surveys finding quasars near the galactic plane where extinction would be significant \citep{dustquasars1998}. Furthermore, the same problem will occur for quasars at high redshifts due to extinction from the inter-galactic medium.
Such dust-obscured quasars will appear much redder in the SDSS bands (or not detected altogether) although they will be rare in our dataset of optically selected quasars. Furthermore, the inclusion of the WISE bands helps significantly when selecting quasars (shown by Table~\ref{table:performance}), and such longer wavelength observations will be important for classifying dust-obscured quasars.

SDSS provide a binary classification of star or galaxy depending on their $resolved_r$ parameter being greater or less than 0.145 \citep{SDSSDR152019}. Due to its simplicity, in many cases this classification is incorrect (for example the middle right panel of Figure \ref{figure:features}) and will mainly incorrectly identify unresolved galaxies as stars. Overall, our method of leveraging information from the photometry alongside the $resolved_r$ parameter provides a more accurate classification label, along with a quantified estimation of its probability.

\subsection{Resolution differences}

The lower resolution of the WISE data compared to the SDSS data allows for multiple SDSS sources to be matched to the same WISE counterpart. When two or more SDSS sources are matched with the same WISE counterpart, they are never more than 8~arcseconds apart, with a mean separation of 2.36~arcseconds in the spectroscopic dataset and 2.40~arcseconds in the photometric dataset. In our spectroscopic dataset this problem is minor, with 13\,904 sources having a WISE counterpart that is matched to more than one source. This occurs for a very low fraction of the training data and is not significant enough to bias the model. Table~\ref{table:performance} shows that including the WISE data as a feature improves the classification of quasars and stars (with an increase in the F1 score of 0.016) more so than galaxies (with an increase in F1 score of 0.004).
In our photometric dataset there are 14\,768\,549 sources that have a WISE counterpart matched to more than one SDSS source, with 1.4 million of these having WISE counterpart matched to more than two sources. In such cases the WISE emission could be from any one of the SDSS sources, or a combination of all of them. 
The mean $resolved_r$ parameter for these sources is 0.18, indicating they are slightly resolved sources, potentially more likely to be interacting if at a mean separation of 2.40~arcseconds.
We make no attempt to interpret if such sources are associated or not in our models.

In the datasets we have used, the resolution difference between SDSS and WISE is relatively small. Our model is dominated by information from the higher resolution SDSS data (see Figure~\ref{figure:feature-ranking}), and in cases where there is incorrect WISE data that suggests an alternate class label, the overall classification probability will be lower. When using data from various surveys with much larger resolution differences and implementing more complex classification schemes, if the information from a lower resolution survey is the only way to make a confident source classification then this would lead to a substantial number of false positives. Additional features that quantify the likelihood of a correct match would help in such scenarios. Finally we note that we do not account for the chance alignment of a WISE source on top of a single SDSS source where the two are not of the same origin. Such a scenario becomes more problematic as the resolution difference between surveys increases, and the source density increases.

\subsection{Class imbalance}

In our new labelled photometric catalogue there are 23 times more galaxies, 6 times more quasars and 115 times more stars than in the spectroscopic dataset we used for training and testing. It is notable that the ratio between the three classes is different in the spectroscopic catalogue used for training compared to the resulting photometric catalogue. This bias is expected as a result of spectroscopic surveys prioritising particular sources at different magnitude depths. For example there is greater scientific demand for galaxy spectra over those of stars, and a rare occurrence of bright quasars to target. It is therefore expected that the ratio of the three classes will be different in our new photometric catalogue compared to the ratio in the spectroscopic sources training data. Figure~\ref{figure:train-vs-f1score} demonstrates that the model still performs well when trained on several orders of magnitude fewer sources, and that sub-sampling the galaxy class hinders the model. However, under-sampling any classes in such a way that restricts the distribution in magnitude space does affect the results, as is described in Section~\ref{section:training-mag-limit}. We note that training the random forest model on only 1\% or 10\% of the spectroscopically observed sources still results in the same ratio of source classifications in the resulting catalogue, though the classification probabilities are slightly lower.

\subsection{Spectroscopic follow-up of quasars}

Our new catalogue contains 2.1\,million quasars, which is four times the 526\,356 spectroscopically and visually confirmed quasars previously identified from SDSS \citep{SDSS-dr14-quasars-2018}. Spectroscopic follow-up observations of these new quasars could be done most efficiently by prioritising those with high classification probabilities. In particular there are 330\,666 sources that have classification probabilities greater than 0.99. As an increasing number of quasars are spectroscopically confirmed in future observations, these could be progressively included into the training set to improve the model's performance. As discussed in Section~\ref{section:10d}, increasing the density of sources in the 10-D feature space by spectroscopically confirming more quasars will increase the random forest classification probabilities for other quasars in similar areas of the feature space. This would be particularly effective at fainter magnitudes where the training set has lower source counts. We note one example seen in the literature where a spectroscopic survey of hot white dwarf stars confirmed a quasar classification in our new catalogue \citep{newbluequasar2015}, with an \textit{r}-band magnitude of $17.32 \pm 0.01$.

\subsection{Future surveys using machine learning}
The next generation of telescopes performing large surveys are expected to take observations over many years before completion, using increasingly more efficient and sensitive instruments. Given the cadence of data acquisition and processing, waiting until each survey is complete before building a classification pipeline will not maximise the scientific output from the instrument to the community. Instead, an iterative scheme of re-training and updating an existing model as more data become available will be very important for the telescopes which aim to survey large portions of the sky over many years. Furthermore, the efficiency of such an approach would be improved were models developed specifically with the aim of increasing a training set gradually over the course of the survey. In such schemes, it is very important to have metrics that can assess the reliability of classifications as a function of multiple variables that change over the lifetime of the survey. In this paper, we demonstrate some of these aspects, providing classification metrics as a function of feature space, as well as variables not included as features, such as magnitude error and redshift.

\section{Conclusions} \label{section:conclusions}

We have trained a random forest machine learning model on 1.55\,million spectroscopically confirmed sources from SDSS in order to classify sources as galaxies, quasars, and stars. As features we have used photometry from both SDSS and WISE bands, plus a measure source extension. We have used cross-validation to tune the hyper-parameters in the model, and ensure that the class imbalance in the training data does not affect the performance of the model. Using a test dataset of a further 1.55\,million spectroscopically confirmed sources we determine that the random forest achieves F$_1$ scores of 0.990, 0.953, and 0.977 for galaxies, quasars, and stars, respectively. Precision, recall, and F$_1$ score are also derived as a function of individual features as well as other parameters, in order to illustrate how the performance of the classifier varies across the three classes. We use the classification probabilities from the random forest as a measure of the likelihood of an individual classification, and we show that this value is in agreement with the F$_1$ score derived from a nearest neighbour search around each source in one and ten dimensions.

We applied the random forest model to 111\,395\,468 previously unlabelled photometrically observed sources from SDSS. Our model returns 50\,417\,547 galaxies, 2\,137\,839 quasars, and 58\,840\,082 stars. Each source has an associated classification probability per class. 6\,683\,526 galaxies (13\%), 330\,666 quasars (15\%), and 41\,279\,349 stars (70\%) have classification probabilities greater than 0.99. At the same time, 35\,075\,918 galaxies (70\%), 722\,159 quasars (34\%), and 54\,673\,689 stars (93\%) have classification probabilities greater than 0.9.

As a way to validate our model we use a non-linear dimension reduction technique (UMAP) in supervised, unsupervised, and semi-supervised schemes to reduce the number of features from ten to two, in order that the result can be plotted and the class labels visualised. In all three of these schemes UMAP clearly separates the galaxy, quasar, and star classes in two dimensions for the spectroscopic sources. 

We applied UMAP to 10\% of our new photometric catalogue of labelled sources (11 million sources) in three different schemes: (i) we used UMAP in an unsupervised scheme, where it is in a strong agreement with the star and galaxy class labels assigned by the random forest, but is poor at separating quasars; (ii) we used UMAP in an unsupervised scheme, but also including 3.1 million spectroscopically observed sources (without labels) in the dataset. This gives a superior performance, separating quasars from stars and galaxies more effectively; (iii) we use UMAP in a semi-supervised scheme, including 3.1 million spectroscopically labelled sources in the dataset. This gives a further improvement in the result, clearly separating the classes. When the three classes overlap in any of these three scenarios, the labels from the random forest show coherent structures per class, demonstrating that in the original 10-D space there is a clearer separation of the classes than in the 2-D space. However, these three schemes are very computationally expensive.

We further utilised the ability of UMAP to perform metric-learning by training it on 1.55\,million spectroscopically labelled sources in supervised and unsupervised schemes to embed the dataset into two dimensions. We used the resulting models to embed the test dataset of 1.55\,million spectroscopically labelled sources, and all 111 million unlabelled photometric sources into this 2-D space.
The supervised metric-learning scheme gave the optimal performance, clearly separating out all galaxies, quasars, and stars in a 2-D space. Furthermore, this clustering was in strong agreement with the random forest class labels.
A significant advantage of the metric-learning schemes is that it can be efficiently scaled in a parallel or distributed computing system to embed any number of unlabelled sources using a model built from a significantly smaller, partially labelled training dataset. This makes it a viable algorithm to run on the next generation of surveys (e.g. LSST, SKA, LOFAR) for both classification and data visualisation.

\begin{acknowledgements}
The authors gratefully acknowledge support from the UK Research \& Innovation Science \& Technology Facilities Council (UKRI-STFC). We thank Justin Bray for numerous discussions helping to guide this work. We thank the anonymous referee for providing feedback that has improved this paper. We thank the language editors at the journal for providing corrections. We made use of the Python packages Numpy \citep{numpy2006}, Scikit-learn \citep{scikit-learn}, Matplotlib \citep{mpl2007}, Pandas \citep{pandas} and Datashader: \url{https://www.datashader.org}.
\end{acknowledgements}

\begin{appendix}

\section{Supplementary plots}

\begin{figure*}
\includegraphics[width=\hsize]{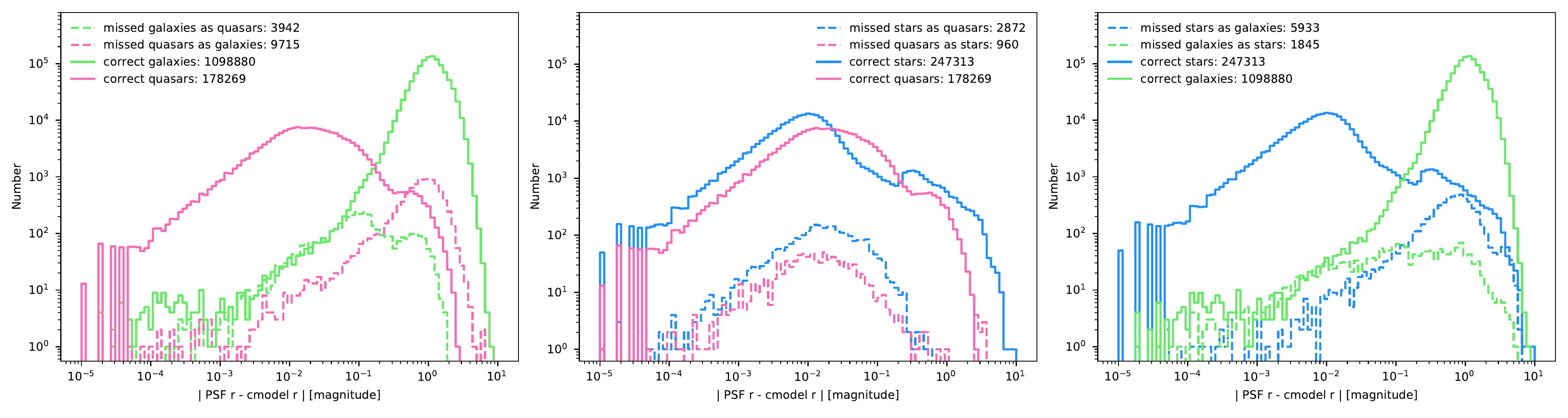}
\includegraphics[width=\hsize]{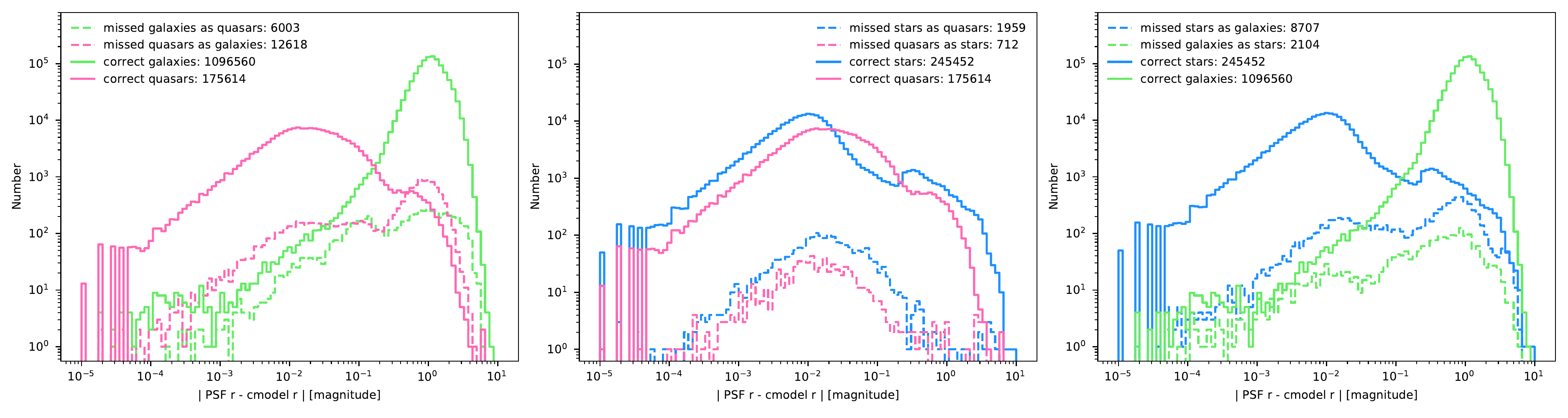}
\caption{Correct and missed spectroscopically observed sources per class when including the $resolved_r$ parameter as a feature (\textit{top}) and when excluding it as a feature (\textit{bottom}). Both use SDSS PSF magnitudes and WISE magnitudes as features. The F$_1$ score per class increases from 0.988 to 0.991 (galaxy), 0.942 to 0.952 (quasar), 0.974 to 0.978 (star).}
\label{figure:resolved-features}
\end{figure*}

\begin{landscape}
\thispagestyle{lscape}
\pagestyle{lscape}
\begin{figure}
\includegraphics[width=\hsize]{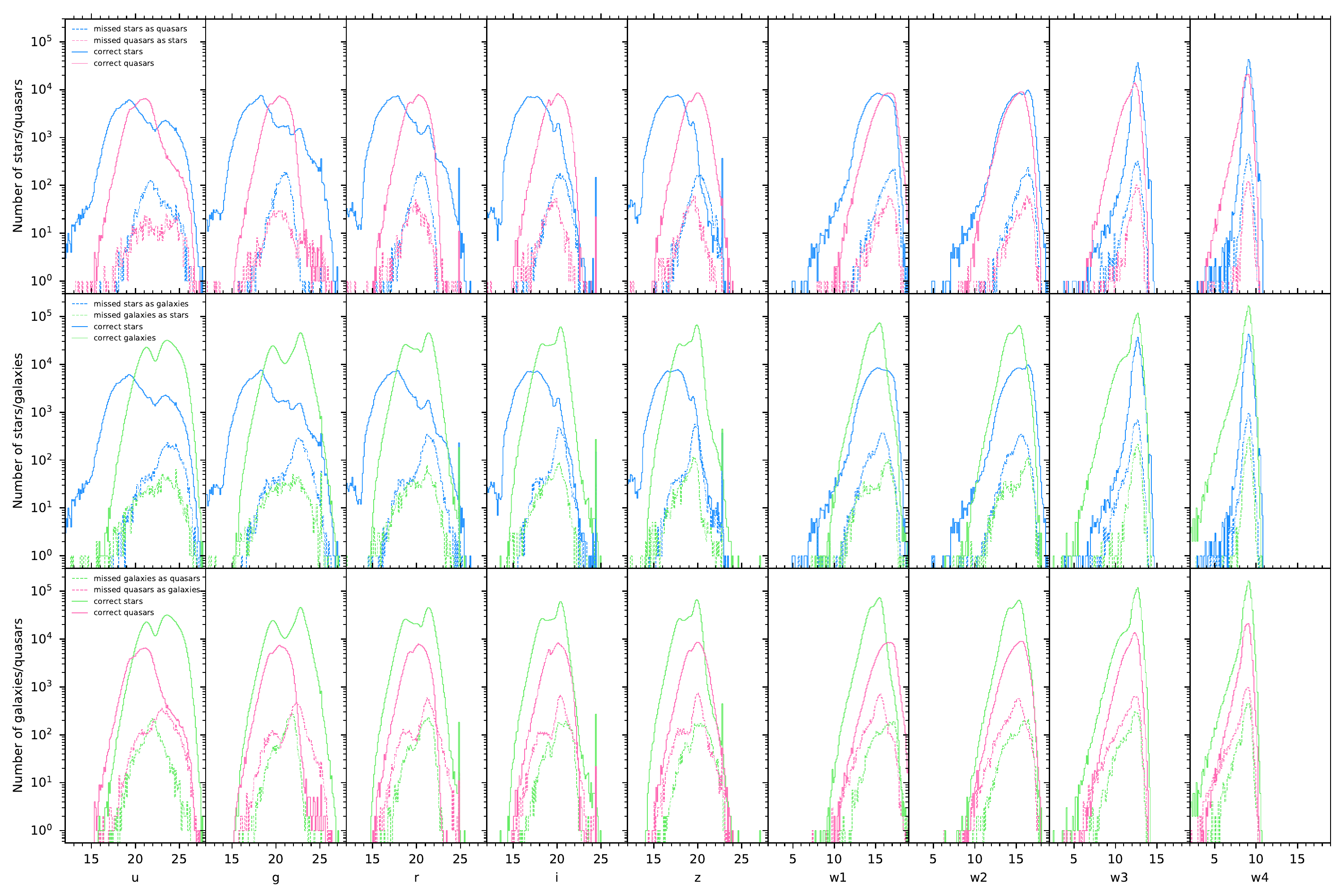}
\caption{Histogram of each magnitude feature, per class, for correct and misclassified sources from the random forest model applied to the test dataset of 1.55 million spectroscopically confirmed sources. Galaxies are shown in green, quasars in pink and stars in blue.} 
\label{figure:histmatrix-mag}
\end{figure}
\end{landscape}

\begin{landscape}
\thispagestyle{lscape}
\pagestyle{lscape}
\begin{figure}
\includegraphics[width=\hsize]{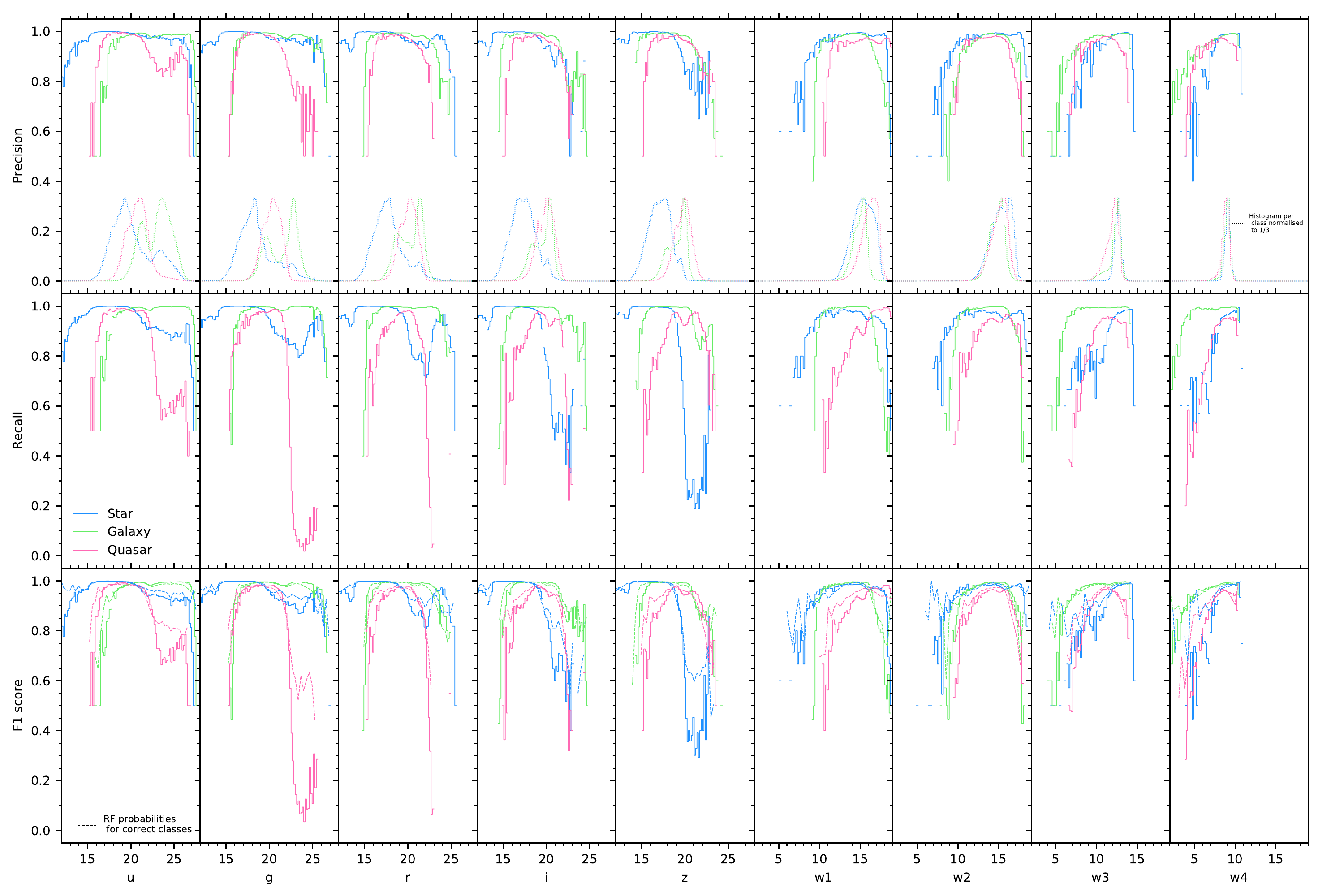}
\caption{Precision, recall, and F$_1$ score plotted as a function of each magnitude feature per class, as derived from the random forest model applied to the test dataset of 1.55 million spectroscopically confirmed sources. In the top row, a histogram of each magnitude feature is also shown as a dotted line per class (normalised to one third) to show the source density. In the bottom row the classification probabilities from the random forest model for the assigned classes are shown by dashed lines. Galaxies are shown in green, quasars in pink, and stars in blue.}
\label{figure:histmatrix-mag-metrics}
\end{figure}
\end{landscape}

\begin{figure*}
\centering
\includegraphics[width=0.16\hsize]{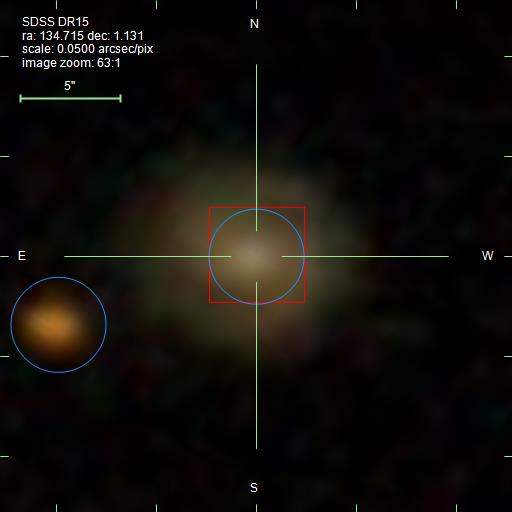}
\includegraphics[width=0.16\hsize]{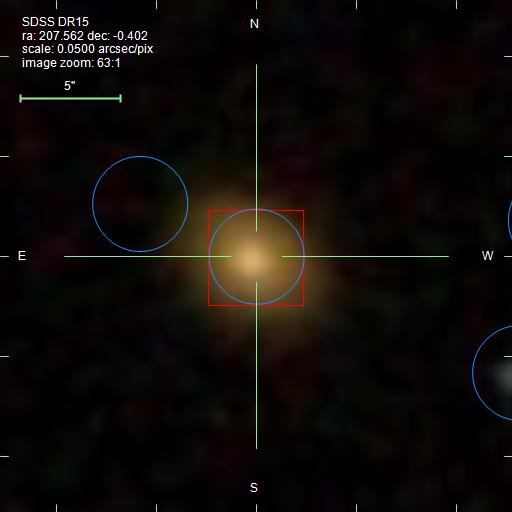}
\includegraphics[width=0.16\hsize]{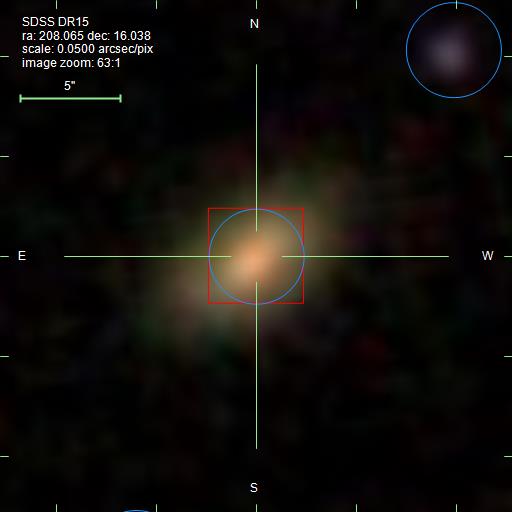}
\includegraphics[width=0.16\hsize]{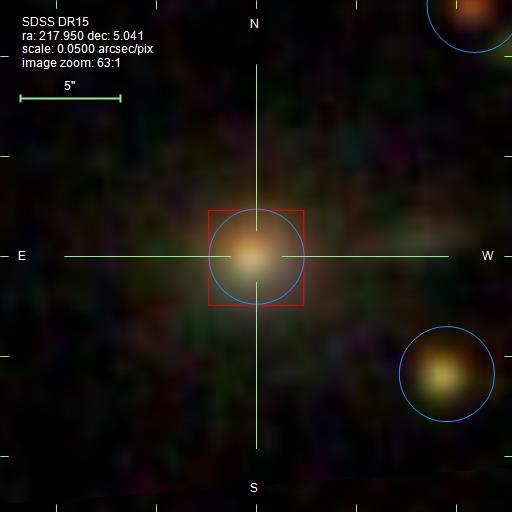}
\includegraphics[width=0.16\hsize]{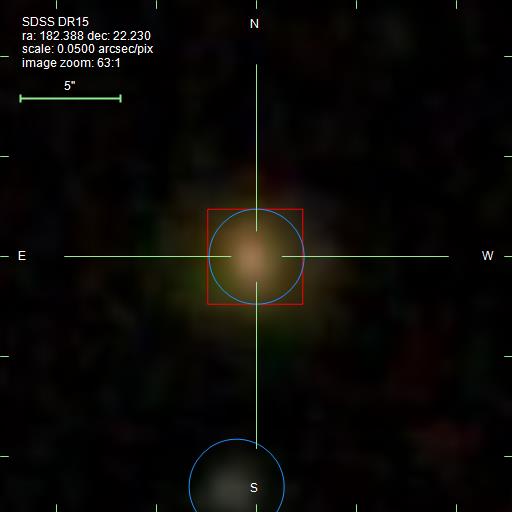}
\includegraphics[width=0.16\hsize]{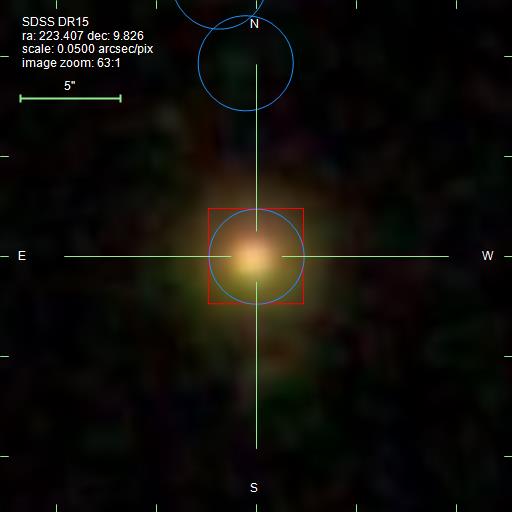}
\includegraphics[width=0.16\hsize]{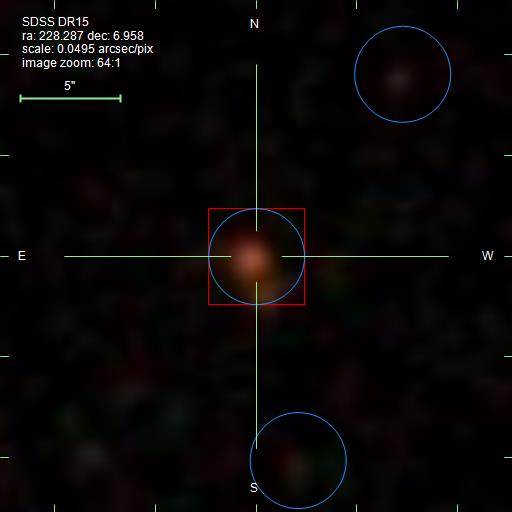}
\includegraphics[width=0.16\hsize]{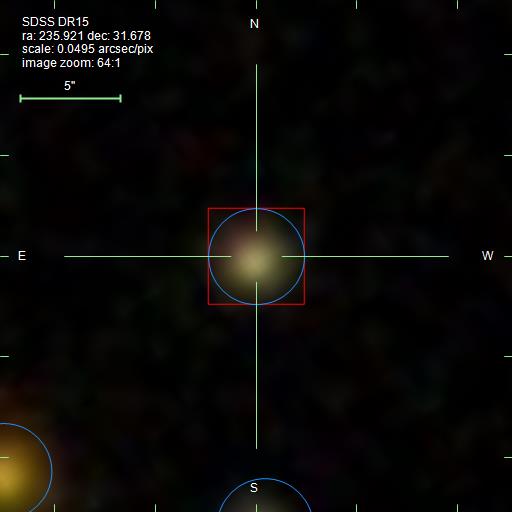}
\includegraphics[width=0.16\hsize]{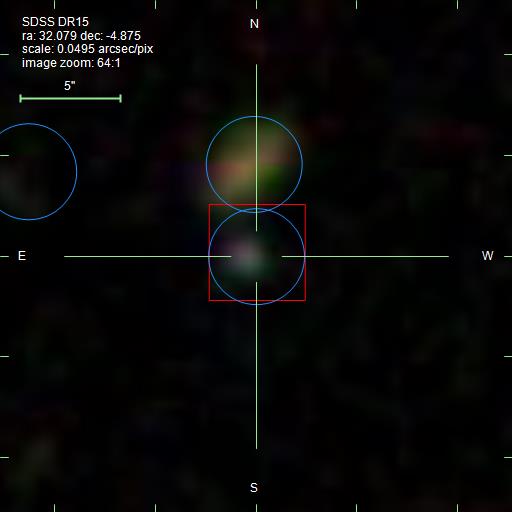}
\includegraphics[width=0.16\hsize]{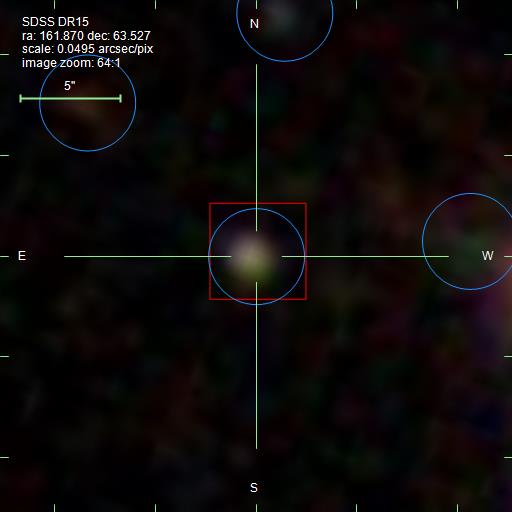}
\includegraphics[width=0.16\hsize]{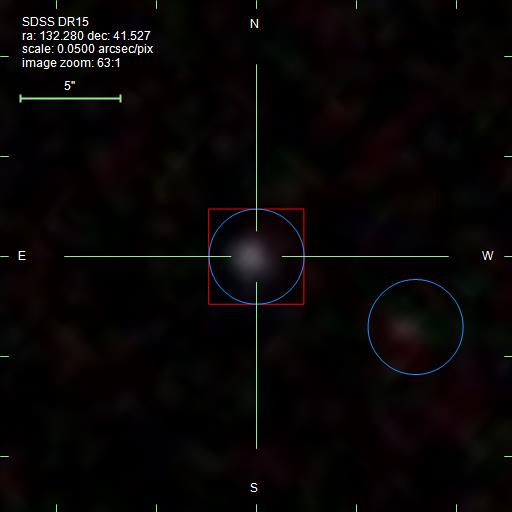}
\includegraphics[width=0.16\hsize]{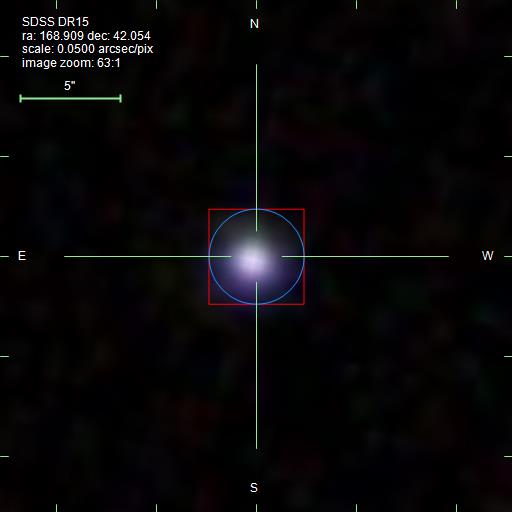}
\caption{Examples of galaxies missed as quasars. The spectroscopically observed target is at the centre in a red box, and photometrically observed sources are circled in blue. From top left to bottom right in order of how resolved they are, their SpecObjIDs, $resolved_r$ parameters and redshifts are:
527093262050158592, 1.627 (z=0.162); 337809312191637504, 1.298 (z=0.186); 3087234932888594432, 1.137 (z=0.094); 658795790411524096, 1.129 (z=0.199); 2976929452103067648, 1.064 (z=0.303); 1928687220470867968, 0.919 (z=0.162); 
5492327033231876096, 0.742 (z=0.882); 1780082440165943296, 0.296 (z=0.280);
4946128968019320832, 0.090 (z=0.640); 7990778382286561280, 0.083 (z=0.460); 9339371187510157312, 0.030 (z=1.078) and 1621402570222233600, 0.022 (z=0.051)}
\label{figure:GasQexamples}
\end{figure*}


\begin{figure*}
\centering
\includegraphics[width=0.16\hsize]{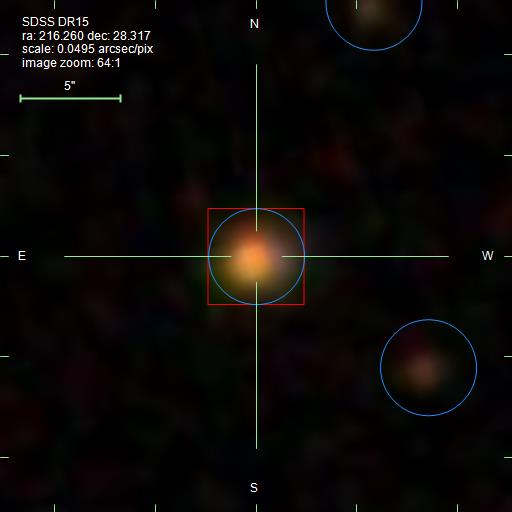}
\includegraphics[width=0.16\hsize]{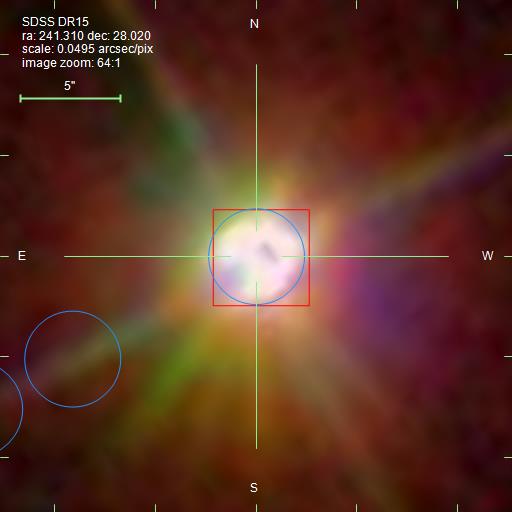}
\includegraphics[width=0.16\hsize]{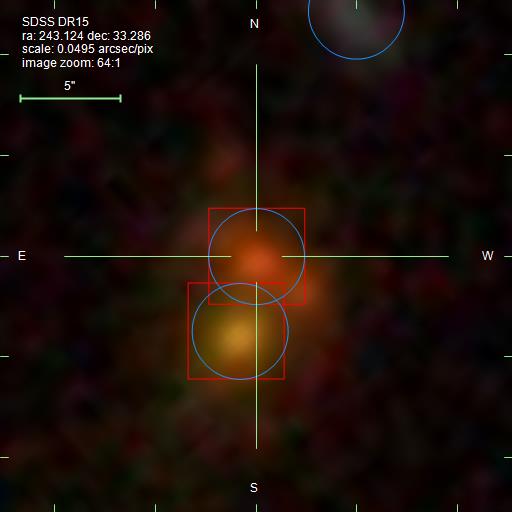}
\includegraphics[width=0.16\hsize]{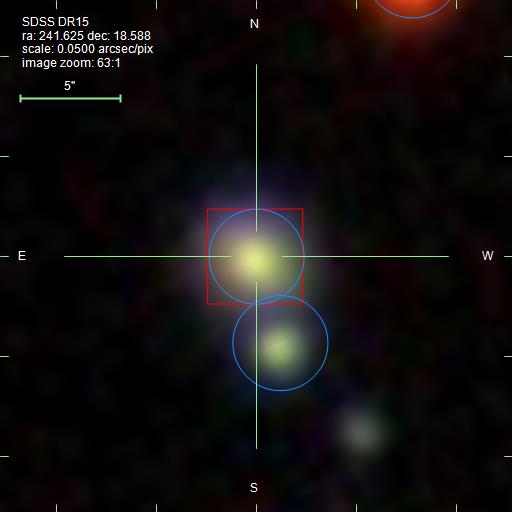}
\includegraphics[width=0.16\hsize]{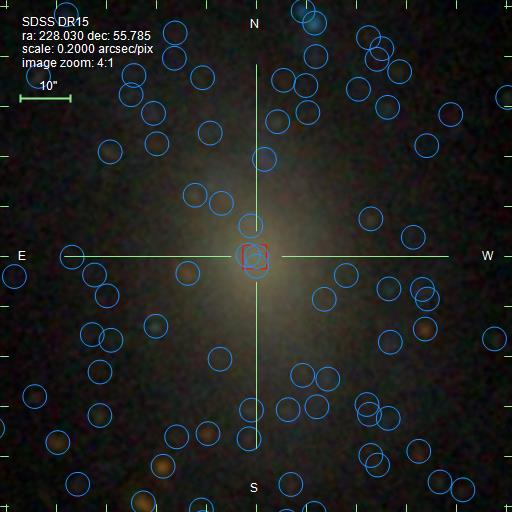}
\includegraphics[width=0.16\hsize]{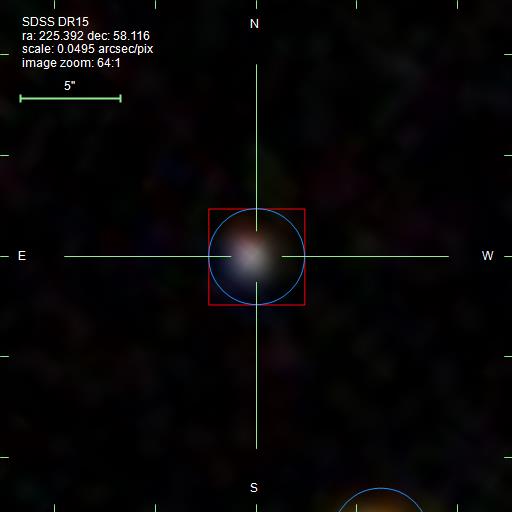}
\caption{Examples of galaxies missed as stars. The spectroscopically observed target is at the centre in a red box, and photometrically observed sources are circled in blue. From top left to bottom right in order of how resolved they are, their SpecObjIDs, $resolved_r$ parameters and redshifts are: 892985450690537472, 4.685 (z=0.003); 3383360912827195392, 2.357 (z=0.0), 5584568912597458944, 1.858 (z=0.724); 3340604199258843136, 0.386 (z=0.038); 2402769422277175296, 0.120 (z=0.089) and 686843782766815232, 0.004 (z=0.656). Note that the second source is actually a real star, incorrectly labelled by the SDSS classification pipeline.}
\label{figure:GasSexamples}
\end{figure*}

\begin{figure*}
\centering
\includegraphics[width=0.16\hsize]{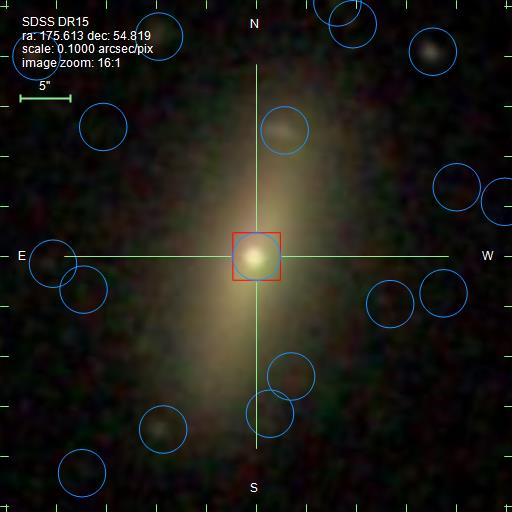}
\includegraphics[width=0.16\hsize]{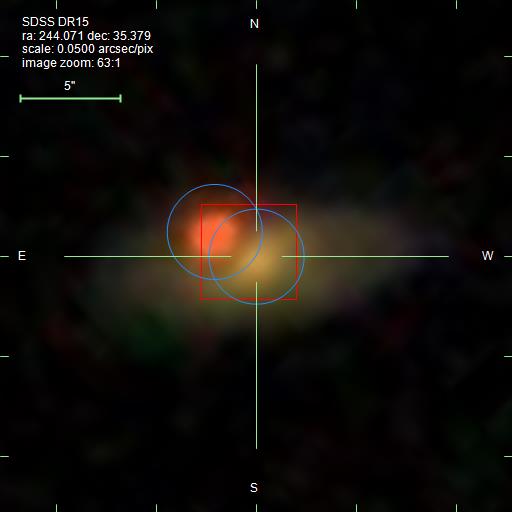}
\includegraphics[width=0.16\hsize]{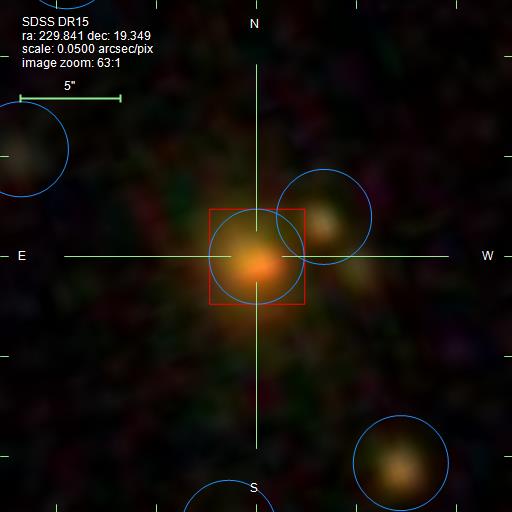}
\includegraphics[width=0.16\hsize]{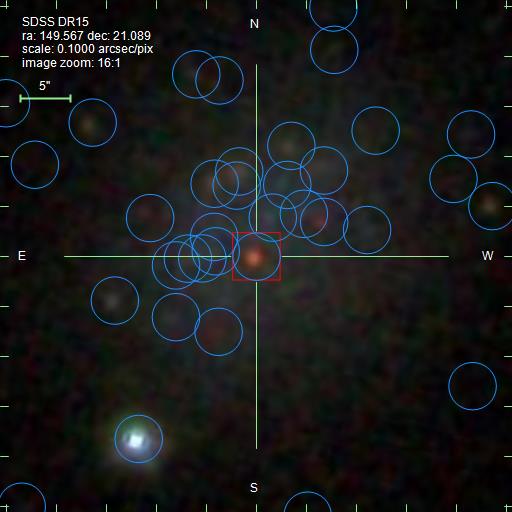}
\includegraphics[width=0.16\hsize]{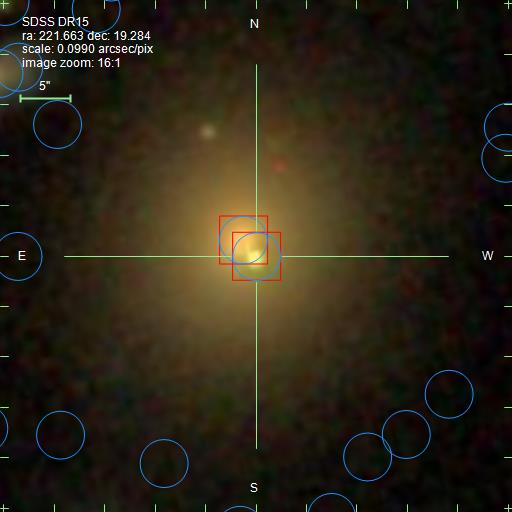}
\includegraphics[width=0.16\hsize]{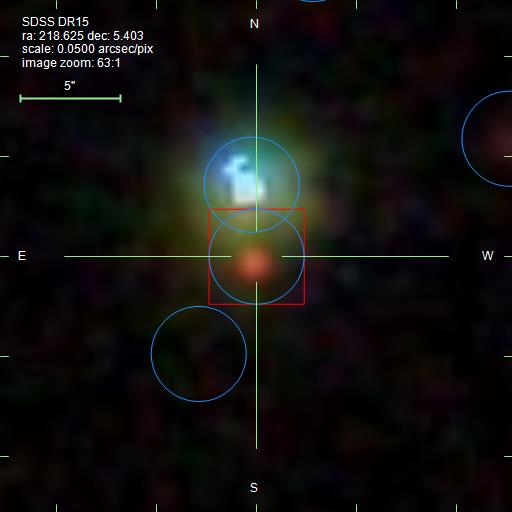}
\includegraphics[width=0.16\hsize]{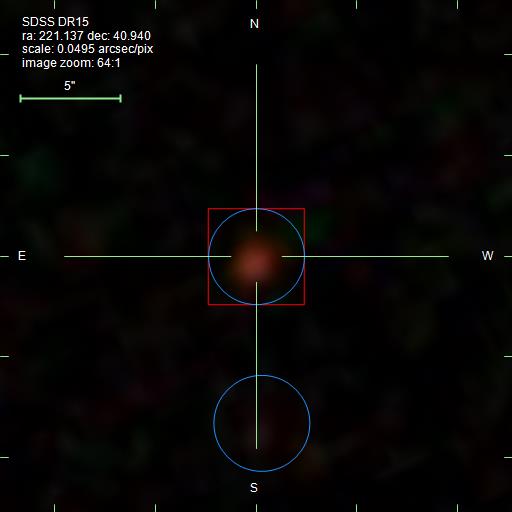}
\includegraphics[width=0.16\hsize]{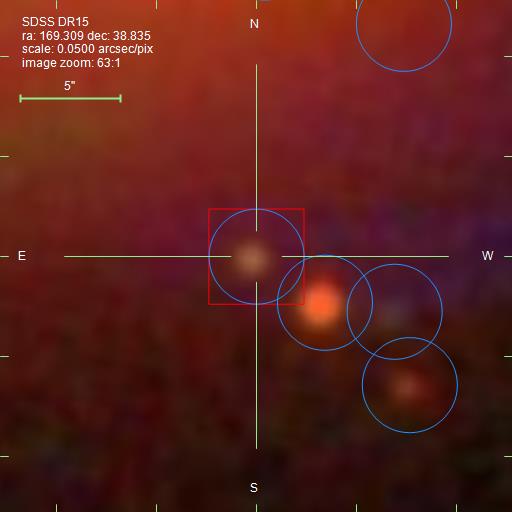}
\includegraphics[width=0.16\hsize]{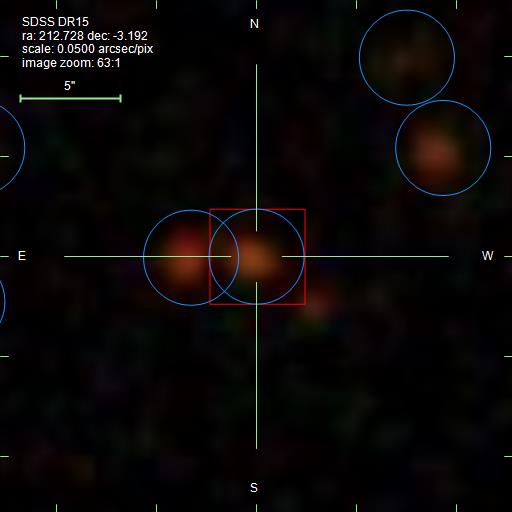}
\includegraphics[width=0.16\hsize]{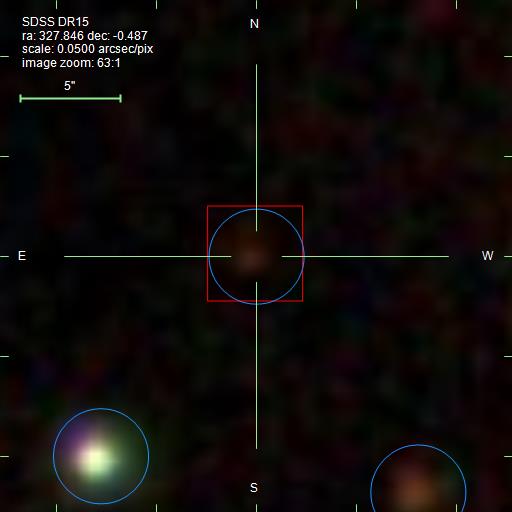}
\includegraphics[width=0.16\hsize]{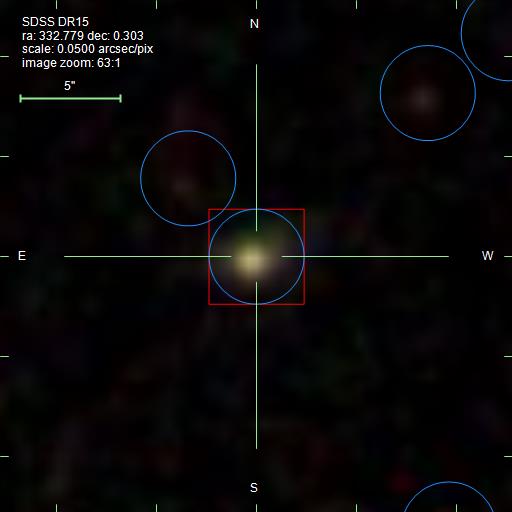}
\includegraphics[width=0.16\hsize]{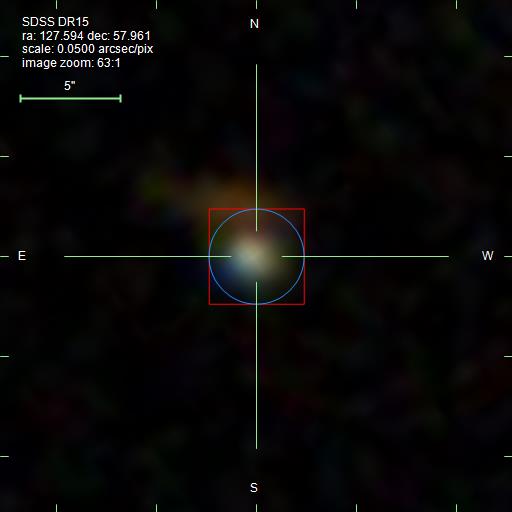}
\caption{Examples of stars missed as galaxies. The spectroscopically observed target is at the centre in a red box, and photometrically observed sources are circled in blue. The top row are resolved sources with SpecObjIDs and $resolved_r$ parameters of: 1142953103180457984, 2.323;
1190170526966900736, 1.609; 2430897686078056448, 1.133; 2660651076583188480, 5.279; 3141329535987902464, 1.498 and
5382020452850835456, 1.897. Top bottom row are unresolved sources with SpecObjIDs and $resolved_r$ parameters of: 9568057781371510784, 0.268; 9929352360244518912, 0.063; 4543105995102986240, 0.294; 4725488312604663808, 0.270; 4728959746027003904, 0.286 and 9197691961909157888, 0.097.}
\label{figure:SasGexamples}
\end{figure*}

\begin{figure*}
\centering
\includegraphics[width=0.16\hsize]{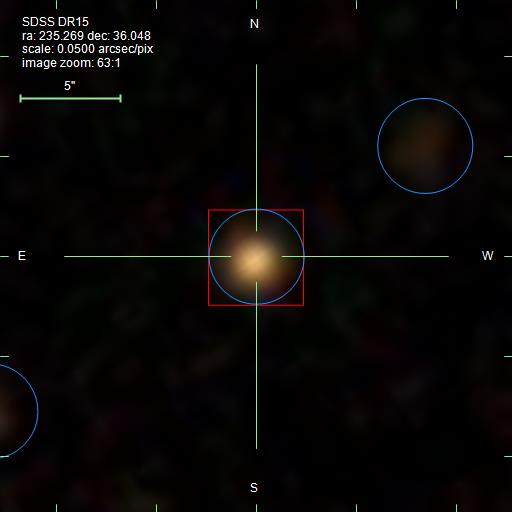}
\includegraphics[width=0.16\hsize]{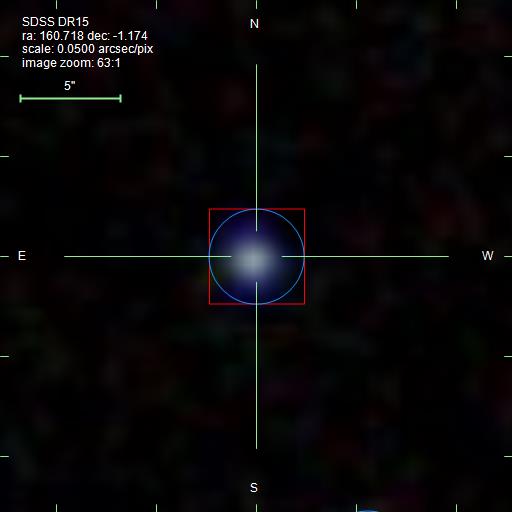}
\includegraphics[width=0.16\hsize]{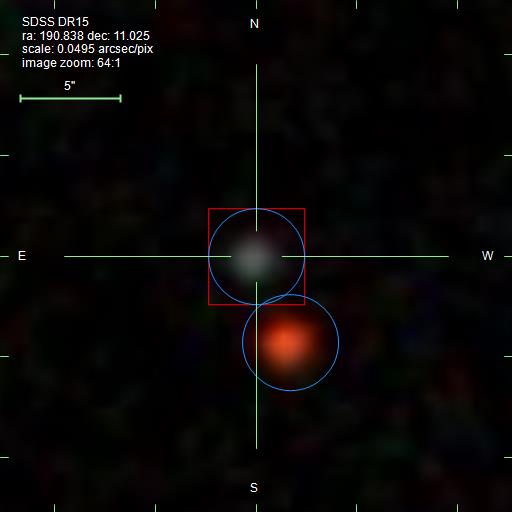}
\includegraphics[width=0.16\hsize]{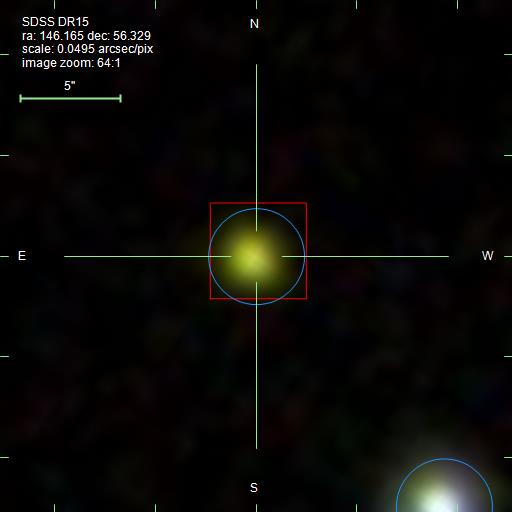}
\includegraphics[width=0.16\hsize]{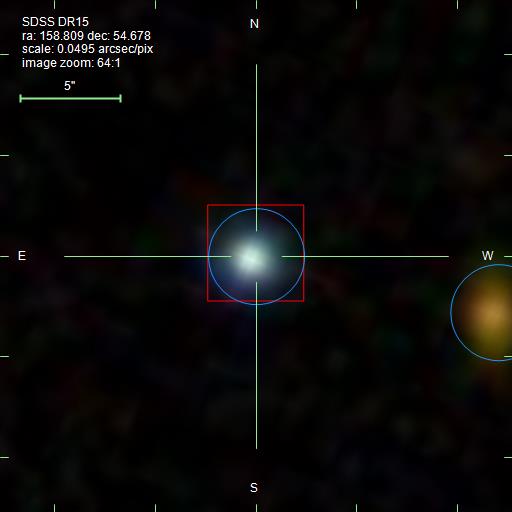}
\includegraphics[width=0.16\hsize]{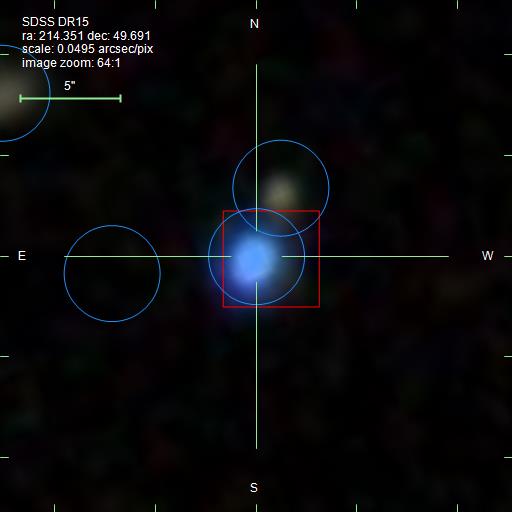}
\caption{Examples of quasars missed as stars. The spectroscopically observed target is at the centre in a red box, and photometrically observed sources are circled in blue. Their SpecObjIDs and redshifts from left to right are: 1578634038246139904 (z=4.450), 4316783906617532416 (z=2.257), 6092406673766326272 (z=3.087), 6434584599746027520 (z=3.998), 7539033580141256704 (z=2.995), 7595591358561099776 (z=1.818)}
\label{figure:QasSexamples}
\end{figure*}

\begin{figure*}
\includegraphics[width=\hsize]{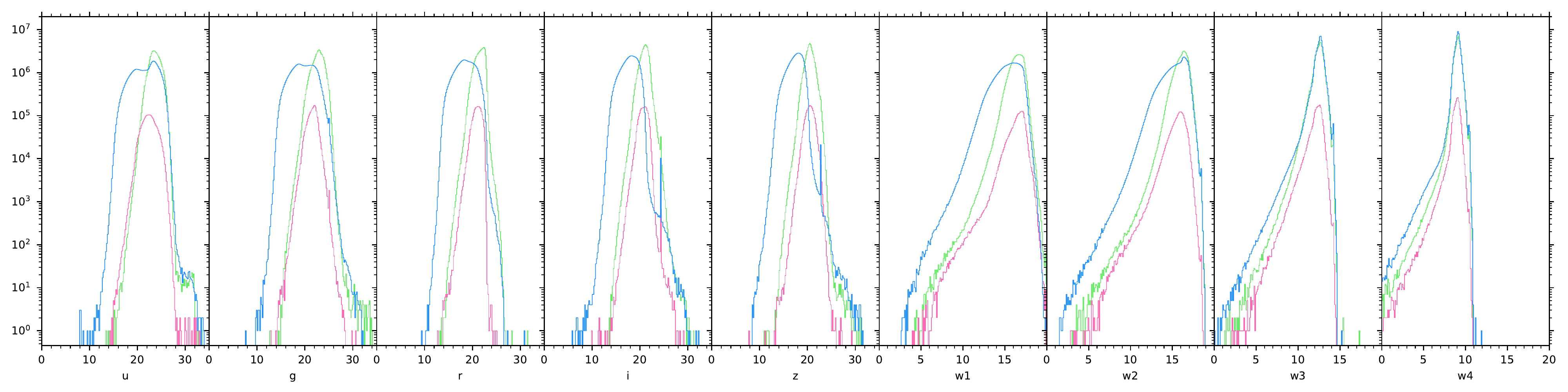}
\caption{Histogram of each magnitude band for all 111 million newly classified sources. Galaxies are plotted in green, quasars in pink, and stars in blue.}
\label{figure:new-sources-hist}
\end{figure*}


\begin{landscape}
\section{Catalogue Extract}
\label{sec:catext}

\begin{table}[!h]
\caption{Sample of 20 galaxies, 20 quasars, and 20 stars from our new catalogue of classified sources. The full catalogue can be downloaded using our digital object identifier: \url{https://www.doi.org/10.5281/zenodo.3459293}.}
\resizebox{1.3\textwidth}{!}{
\begin{tabular}{lllllllllllllllll}
{\bf objid}               & {\bf ra}        & {\bf dec}       & {\bf psf\_u} & {\bf psf\_g} & {\bf psf\_r} & {\bf psf\_i} & {\bf psf\_z} & {\bf w1}     & {\bf w2}     & {\bf w3}     & {\bf w4}    & {\bf resolvedr} & {\bf class\_pred} & {\bf class\_prob\_galaxy} & {\bf class\_prob\_quasar} & {\bf class\_prob\_star} \\
1237663782600311620 & 30.97881  & -1.1961   & 24.05239 & 22.94274 & 21.72948 & 20.89259 & 20.18513 & 17.004 & 16.64  & 13.011 & 9.35  & 0.17654   & GALAXY      & 0.85                & 0.035               & 0.115             \\
1237662224094987748 & 253.04789 & 16.46796  & 23.38386 & 23.02948 & 21.74218 & 21.1817  & 20.92808 & 17.121 & 16.255 & 12.591 & 9.03  & 1.0353    & GALAXY      & 0.995               & 0.005               & 0.0               \\
1237667212670927199 & 169.1246  & 29.96615  & 23.87665 & 23.06762 & 21.51385 & 20.906   & 20.21203 & 17.263 & 15.922 & 12.425 & 8.955 & 0.3124    & GALAXY      & 0.845               & 0.055               & 0.1               \\
1237657066330456647 & 19.49697  & -2.5862   & 24.27969 & 23.02265 & 21.38321 & 20.85167 & 20.26984 & 15.531 & 15.433 & 12.627 & 9.001 & 1.10773   & GALAXY      & 0.995               & 0.0                 & 0.005             \\
1237679439352693622 & 37.94041  & -6.12846  & 24.72696 & 23.59927 & 22.73186 & 21.93435 & 21.46056 & 17.329 & 16.524 & 12.602 & 9.573 & 0.9788    & GALAXY      & 0.98                & 0.02                & 0.0               \\
1237663916799493064 & 114.34315 & 45.58042  & 22.7852  & 21.79741 & 21.06585 & 20.85956 & 20.17542 & 17.015 & 16.844 & 12.216 & 8.763 & 1.0378    & GALAXY      & 0.99                & 0.0                 & 0.01              \\
1237668658984258502 & 263.24811 & 3.58866   & 25.02485 & 21.90447 & 20.5326  & 20.00715 & 19.70523 & 17.191 & 17.006 & 12.799 & 9.292 & 0.57036   & GALAXY      & 0.63                & 0.015               & 0.355             \\
1237676440385880795 & 5.63369   & 30.80671  & 22.69767 & 23.41565 & 22.46992 & 22.44775 & 21.79413 & 17.484 & 16.698 & 11.819 & 8.791 & 0.03125   & GALAXY      & 0.605               & 0.355               & 0.04              \\
1237659324960998467 & 255.46213 & 29.25622  & 22.68677 & 22.64796 & 21.78658 & 21.11527 & 20.73625 & 16.631 & 16.39  & 12.614 & 9.12  & 0.8028    & GALAXY      & 1.0                 & 0.0                 & 0.0               \\
1237661069789889052 & 160.9454  & 13.80896  & 23.55087 & 23.38648 & 22.82657 & 22.02799 & 21.7818  & 16.131 & 15.653 & 11.989 & 8.35  & 0.39083   & GALAXY      & 0.96                & 0.035               & 0.005             \\
1237654398616732586 & 172.03828 & 63.61214  & 23.22376 & 23.29302 & 22.29113 & 21.53194 & 20.72159 & 17.037 & 16.951 & 13.063 & 9.291 & 0.82094   & GALAXY      & 0.99                & 0.01                & 0.0               \\
1237661950270506022 & 222.08511 & 10.39907  & 24.16395 & 24.06727 & 22.9681  & 22.29608 & 21.48033 & 17.512 & 17.092 & 13.105 & 9.481 & 0.36943   & GALAXY      & 0.885               & 0.1                 & 0.015             \\
1237648721250354405 & 224.83936 & -0.17729  & 25.11269 & 24.19453 & 22.86048 & 21.46819 & 20.51785 & 16.366 & 16.746 & 12.97  & 9.464 & 0.25505   & GALAXY      & 0.895               & 0.07                & 0.035             \\
1237665566072701160 & 220.53691 & 20.68538  & 22.41847 & 20.729   & 19.7746  & 19.33793 & 19.02539 & 16.111 & 15.915 & 13.398 & 9.211 & 0.53491   & GALAXY      & 0.98                & 0.0                 & 0.02              \\
1237651250455708581 & 265.56557 & 18.2352   & 23.1753  & 21.76556 & 21.44754 & 21.03805 & 20.988   & 15.66  & 16.0   & 12.879 & 8.822 & 0.71257   & GALAXY      & 0.82                & 0.105               & 0.075             \\
1237659324414165143 & 236.03141 & 46.14449  & 21.12239 & 19.61248 & 18.82504 & 18.55415 & 18.21965 & 15.21  & 15.187 & 13.755 & 9.428 & 0.96891   & GALAXY      & 1.0                 & 0.0                 & 0.0               \\
1237665582188594993 & 261.75976 & 32.83714  & 24.12317 & 23.18384 & 21.37892 & 20.65753 & 20.16027 & 16.567 & 16.439 & 13.322 & 9.683 & 0.38477   & GALAXY      & 0.95                & 0.02                & 0.03              \\
1237655744013992825 & 213.85594 & 6.70062   & 24.50207 & 24.28458 & 22.33565 & 21.34448 & 20.89184 & 17.256 & 16.933 & 13.146 & 9.16  & 0.12349   & GALAXY      & 0.765               & 0.14                & 0.095             \\
1237678893897614163 & 42.55299  & 29.19873  & 25.57469 & 25.11423 & 22.41595 & 22.91776 & 22.07617 & 16.586 & 16.551 & 12.04  & 9.108 & 0.49736   & GALAXY      & 0.79                & 0.195               & 0.015             \\
1237664667372749288 & 187.63352 & 39.05828  & 22.92461 & 21.93413 & 20.80021 & 20.4253  & 20.10496 & 16.11  & 15.998 & 12.939 & 8.668 & 0.8793    & GALAXY      & 0.995               & 0.005               & 0.0               \\
1237656241167991166 & 350.40227 & 13.89792  & 21.11566 & 21.03429 & 20.60135 & 20.56733 & 20.76451 & 16.613 & 15.446 & 12.312 & 8.366 & 0.02375   & QSO         & 0.0                 & 1.0                 & 0.0               \\
1237655374111572446 & 250.81746 & 38.11814  & 20.96288 & 21.02744 & 20.72233 & 20.82046 & 20.75615 & 16.322 & 15.557 & 12.999 & 8.86  & 0.02256   & QSO         & 0.0                 & 1.0                 & 0.0               \\
1237678597013897372 & 353.531   & 2.3162    & 19.81265 & 19.62335 & 19.39883 & 19.08751 & 18.86235 & 15.768 & 14.974 & 11.58  & 8.916 & 0.01079   & QSO         & 0.0                 & 1.0                 & 0.0               \\
1237660412643115212 & 122.40555 & 5.06184   & 23.44366 & 21.35939 & 20.07348 & 19.68628 & 19.32755 & 16.383 & 15.825 & 12.13  & 8.472 & 0.00001     & QSO         & 0.16                & 0.59                & 0.25              \\
1237662238005461695 & 189.9602  & 10.30672  & 24.41866 & 23.85092 & 22.40509 & 22.09653 & 21.42698 & 17.228 & 16.825 & 12.62  & 8.923 & 0.03813   & QSO         & 0.475               & 0.51                & 0.015             \\
1237657775010021675 & 145.29387 & 45.03001  & 22.37506 & 22.23012 & 21.56038 & 21.09927 & 21.11072 & 17.167 & 16.552 & 12.82  & 9.206 & 0.11814   & QSO         & 0.495               & 0.505               & 0.0               \\
1237679478546759920 & 6.76944   & 23.14234  & 20.17334 & 19.28211 & 19.11902 & 19.04537 & 18.94314 & 16.015 & 15.103 & 11.6   & 9.086 & 0.00113   & QSO         & 0.0                 & 1.0                 & 0.0               \\
1237678619550680055 & 326.50424 & 2.42972   & 25.33274 & 21.72988 & 20.53654 & 20.07195 & 19.75681 & 17.133 & 16.399 & 12.269 & 8.559 & 0.00623   & QSO         & 0.105               & 0.57                & 0.325             \\
1237668289087865466 & 165.22677 & 18.72498  & 23.25236 & 21.80137 & 20.95243 & 20.63299 & 20.51833 & 17.142 & 16.933 & 12.312 & 8.864 & 0.01473   & QSO         & 0.045               & 0.785               & 0.17              \\
1237660765908108179 & 122.53009 & 26.0469   & 24.24844 & 22.17328 & 21.72695 & 21.38462 & 21.4842  & 16.032 & 15.546 & 12.326 & 9.049 & 0.06148   & QSO         & 0.305               & 0.67                & 0.025             \\
1237665532798370519 & 243.72118 & 15.97169  & 22.06984 & 22.22394 & 22.54231 & 21.42486 & 20.81908 & 15.873 & 14.047 & 10.139 & 7.575 & 0.0522    & QSO         & 0.48                & 0.515               & 0.005             \\
1237652935645987674 & 322.61939 & -6.88957  & 21.0438  & 21.16118 & 20.80203 & 20.76583 & 20.87947 & 16.46  & 15.738 & 12.642 & 9.211 & 0.0064    & QSO         & 0.005               & 0.995               & 0.0               \\
1237672024628400965 & 319.33005 & 51.397    & 23.33645 & 21.81444 & 20.02815 & 19.22479 & 18.62908 & 14.463 & 14.365 & 10.459 & 8.041 & 0.01449   & QSO         & 0.13                & 0.565               & 0.305             \\
1237673700739908103 & 33.44822  & -11.88112 & 22.59563 & 21.9656  & 21.69    & 21.3438  & 21.32575 & 16.388 & 15.743 & 12.396 & 9.436 & 0.31421   & QSO         & 0.225               & 0.77                & 0.005             \\
1237665429175861425 & 226.01481 & 23.96105  & 20.2236  & 19.99413 & 19.76245 & 19.80832 & 19.6857  & 15.384 & 14.331 & 11.631 & 9.197 & 0.00625   & QSO         & 0.0                 & 1.0                 & 0.0               \\
1237680310158557420 & 8.3174    & 31.78577  & 21.49847 & 21.41801 & 21.61896 & 21.76059 & 21.80273 & 17.693 & 16.943 & 13.0   & 9.387 & 0.01924   & QSO         & 0.065               & 0.82                & 0.115             \\
1237664673255325936 & 202.65973 & 36.21947  & 23.30307 & 22.47356 & 21.94134 & 21.61565 & 22.68182 & 17.556 & 16.262 & 13.139 & 8.905 & 0.04033   & QSO         & 0.19                & 0.78                & 0.03              \\
1237678777928975115 & 331.88532 & 4.53833   & 22.63443 & 20.66668 & 19.81801 & 19.46239 & 19.2608  & 16.423 & 16.038 & 12.237 & 9.023 & 0.0107    & QSO         & 0.125               & 0.455               & 0.42              \\
1237673843542197677 & 100.78405 & 10.8244   & 22.6022  & 20.73836 & 19.64029 & 19.11246 & 18.7769  & 15.627 & 15.117 & 11.035 & 9.054 & 0.00417   & QSO         & 0.085               & 0.645               & 0.27              \\
1237668681000682493 & 287.16176 & 39.46109  & 21.21755 & 19.28199 & 18.42947 & 18.13732 & 18.01754 & 15.151 & 13.937 & 10.853 & 8.643 & 0.00471   & QSO         & 0.055               & 0.85                & 0.095             \\
1237666464250266474 & 55.54909  & 48.24496  & 24.52368 & 21.66051 & 19.87054 & 18.89041 & 18.24001 & 15.916 & 16.261 & 12.78  & 9.1   & 0.01944   & STAR        & 0.0                 & 0.0                 & 1.0               \\
1237673709329514753 & 133.53233 & -1.4081   & 21.65659 & 19.08867 & 17.77378 & 17.21752 & 16.92153 & 14.822 & 14.855 & 12.169 & 8.72  & 0.00148   & STAR        & 0.0                 & 0.0                 & 1.0               \\
1237646647834575470 & 72.46913  & 0.49403   & 24.72099 & 21.65013 & 20.06832 & 19.07889 & 18.60951 & 16.731 & 16.803 & 12.333 & 9.195 & 0.00753   & STAR        & 0.0                 & 0.0                 & 1.0               \\
1237668571473707146 & 259.54026 & 8.32209   & 21.90678 & 20.01839 & 19.03246 & 18.63957 & 18.34537 & 16.928 & 16.714 & 12.672 & 8.478 & 0.00514   & STAR        & 0.0                 & 0.0                 & 1.0               \\
1237661059576693833 & 55.69782  & 38.37885  & 20.63609 & 18.74323 & 17.91564 & 17.63848 & 17.4431  & 16.093 & 16.429 & 12.003 & 9.135 & 0.01573   & STAR        & 0.0                 & 0.0                 & 1.0               \\
1237659931072987776 & 12.20406  & 42.41898  & 16.92939 & 15.77489 & 15.40054 & 15.1995  & 15.1552  & 13.973 & 14.001 & 12.781 & 9.435 & 0.06922   & STAR        & 0.0                 & 0.0                 & 1.0               \\
1237673756037284051 & 99.86375  & 26.85257  & 17.91034 & 16.36815 & 15.76871 & 15.56511 & 15.43967 & 13.943 & 14.0   & 12.231 & 8.72  & 0.00315   & STAR        & 0.0                 & 0.0                 & 1.0               \\
1237658298448019952 & 130.44034 & 3.64197   & 22.22529 & 19.88207 & 18.64938 & 18.20052 & 17.98668 & 15.195 & 15.053 & 12.669 & 9.143 & 0.02333   & STAR        & 0.015               & 0.0                 & 0.985             \\
1237651191884743535 & 112.09385 & 33.34835  & 25.01766 & 21.33149 & 19.88368 & 19.04117 & 18.57499 & 16.44  & 16.859 & 12.591 & 9.162 & 0.03535   & STAR        & 0.0                 & 0.0                 & 1.0               \\
1237664669507584470 & 151.23338 & 35.65406  & 22.34285 & 22.49298 & 20.88751 & 19.52134 & 18.73    & 16.143 & 15.82  & 12.334 & 9.133 & 0.25618   & STAR        & 0.345               & 0.01                & 0.645             \\
1237656241157702978 & 326.50066 & 11.24917  & 23.11194 & 22.41932 & 20.89891 & 19.27414 & 18.38841 & 15.919 & 15.997 & 12.785 & 8.992 & 0.09163   & STAR        & 0.0                 & 0.0                 & 1.0               \\
1237665567689933347 & 236.2179  & 17.4867   & 23.00173 & 21.77308 & 20.343   & 19.56532 & 19.14472 & 17.295 & 16.72  & 13.205 & 9.505 & 0.00894   & STAR        & 0.005               & 0.0                 & 0.995             \\
1237658300064203395 & 143.12281 & 6.1885    & 24.32115 & 22.6965  & 20.97186 & 19.3615  & 18.48424 & 16.169 & 15.842 & 12.502 & 8.979 & 0.00234   & STAR        & 0.0                 & 0.0                 & 1.0               \\
1237673739927290076 & 89.06887  & 22.25195  & 18.52332 & 16.89002 & 16.07351 & 15.71227 & 15.50633 & 13.85  & 13.953 & 12.306 & 8.354 & 0.00048   & STAR        & 0.0                 & 0.0                 & 1.0               \\
1237671941433463064 & 330.48767 & 46.69989  & 19.97013 & 18.11446 & 17.39923 & 17.01703 & 16.9614  & 15.249 & 15.942 & 12.718 & 9.433 & 0.01104   & STAR        & 0.0                 & 0.0                 & 1.0               \\
1237668271378595910 & 241.74229 & 10.66511  & 16.82092 & 15.54588 & 15.07946 & 14.92202 & 14.87713 & 13.564 & 13.658 & 12.582 & 8.571 & 0.00397   & STAR        & 0.0                 & 0.0                 & 1.0               \\
1237657630043996309 & 121.34553 & 30.71609  & 18.27571 & 17.31759 & 16.99278 & 16.87686 & 16.83438 & 15.646 & 15.934 & 12.69  & 8.605 & 0.01256   & STAR        & 0.0                 & 0.0                 & 1.0               \\
1237667246472495228 & 25.0478   & -18.36897 & 17.10527 & 16.06291 & 15.71597 & 15.60779 & 15.58288 & 14.442 & 14.621 & 12.243 & 9.17  & 0.00426   & STAR        & 0.0                 & 0.0                 & 1.0               \\
1237671067926528629 & 25.974    & 57.09697  & 17.08796 & 15.06386 & 14.31758 & 14.12652 & 13.86313 & 12.146 & 12.232 & 12.212 & 9.116 & 0.02518   & STAR        & 0.01                & 0.0                 & 0.99              \\
1237666227515491616 & 307.94848 & 5.11788   & 21.7546  & 19.44613 & 18.17798 & 17.63606 & 17.36406 & 15.302 & 15.152 & 12.694 & 9.122 & 0.02043   & STAR        & 0.015               & 0.0                 & 0.985         
\end{tabular}}
\end{table}
\end{landscape}

\end{appendix}


\bibliographystyle{aa}
\bibliography{ml_astro}

\begin{thebibliography}{66}
\expandafter\ifx\csname natexlab\endcsname\relax\def\natexlab#1{#1}\fi

\bibitem[{{Aguado} {et~al.}(2019){Aguado}, {Ahumada}, {Almeida}, {Anderson},
  {Andrews}, {Anguiano}, {Aquino Ort{\'\i}z}, {Arag{\'o}n-Salamanca},
  {Argudo-Fern{\'a}ndez}, {Aubert}, {Avila-Reese}, {Badenes}, {Barboza
  Rembold}, {Barger}, {Barrera-Ballesteros}, {Bates}, {Bautista}, {Beaton},
  {Beers}, {Belfiore}, {Bernardi}, {Bershady}, {Beutler}, {Bird}, {Bizyaev},
  {Blanc}, {Blanton}, {Blomqvist}, {Bolton}, {Boquien}, {Borissova}, {Bovy},
  {Brand t}, {Brinkmann}, {Brownstein}, {Bundy}, {Burgasser}, {Byler}, {Cano
  Diaz}, {Cappellari}, {Carrera}, {Cervantes Sodi}, {Chen}, {Cherinka}, {Choi},
  {Chung}, {Coffey}, {Comerford}, {Comparat}, {Covey}, {da Silva Ilha}, {da
  Costa}, {Dai}, {Damke}, {Darling}, {Davies}, {Dawson}, {de Sainte Agathe},
  {Deconto Machado}, {Del Moro}, {De Lee}, {Diamond-Stanic}, {Dom{\'\i}nguez
  S{\'a}nchez}, {Donor}, {Drory}, {du Mas des Bourboux}, {Duckworth}, {Dwelly},
  {Ebelke}, {Emsellem}, {Escoffier}, {Fern{\'a}ndez-Trincado}, {Feuillet},
  {Fischer}, {Fleming}, {Fraser-McKelvie}, {Freischlad}, {Frinchaboy}, {Fu},
  {Galbany}, {Garcia-Dias}, {Garc{\'\i}a-Hern{\'a}ndez}, {Garma Oehmichen},
  {Geimba Maia}, {Gil-Mar{\'\i}n}, {Grabowski}, {Gu}, {Guo}, {Ha},
  {Harrington}, {Hasselquist}, {Hayes}, {Hearty}, {Hernandez Toledo}, {Hicks},
  {Hogg}, {Holley-Bockelmann}, {Holtzman}, {Hsieh}, {Hunt}, {Hwang},
  {Ibarra-Medel}, {Jimenez Angel}, {Johnson}, {Jones}, {J{\"o}nsson},
  {Kinemuchi}, {Kollmeier}, {Krawczyk}, {Kreckel}, {Kruk}, {Lacerna}, {Lan},
  {Lane}, {Law}, {Lee}, {Li}, {Lian}, {Lin}, {Lin}, {Lintott}, {Long},
  {Longa-Pe{\~n}a}, {Mackereth}, {de la Macorra}, {Majewski}, {Malanushenko},
  {Manchado}, {Maraston}, {Mariappan}, {Marinelli}, {Marques-Chaves},
  {Masseron}, {Masters}, {McDermid}, {Medina Pe{\~n}a}, {Meneses-Goytia},
  {Merloni}, {Merrifield}, {Meszaros}, {Minniti}, {Minsley}, {Muna}, {Myers},
  {Nair}, {Correa do Nascimento}, {Newman}, {Nitschelm}, {Olmstead}, {Oravetz},
  {Oravetz}, {Ortega Minakata}, {Pace}, {Padilla}, {Palicio}, {Pan}, {Pan},
  {Parikh}, {Parker}, {Peirani}, {Penny}, {Percival}, {Perez-Fournon},
  {Peterken}, {Pinsonneault}, {Prakash}, {Raddick}, {Raichoor}, {Riffel},
  {Riffel}, {Rix}, {Robin}, {Roman-Lopes}, {Rose}, {Ross}, {Rossi}, {Rowlands},
  {Rubin}, {S{\'a}nchez}, {S{\'a}nchez-Gallego}, {Sayres}, {Schaefer},
  {Schiavon}, {Schimoia}, {Schlafly}, {Schlegel}, {Schneider}, {Schultheis},
  {Seo}, {Shamsi}, {Shao}, {Shen}, {Shetty}, {Simonian}, {Smethurst}, {Sobeck},
  {Souter}, {Spindler}, {Stark}, {Stassun}, {Steinmetz}, {Storchi-Bergmann},
  {Stringfellow}, {Su{\'a}rez}, {Sun}, {Taghizadeh-Popp}, {Talbot}, {Tayar},
  {Thakar}, {Thomas}, {Tissera}, {Tojeiro}, {Troup}, {Unda-Sanzana},
  {Valenzuela}, {Vargas-Maga{\~n}a}, {V{\'a}zquez-Mata}, {Wake}, {Weaver},
  {Weijmans}, {Westfall}, {Wild}, {Wilson}, {Woods}, {Yan}, {Yang}, {Zamora},
  {Zasowski}, {Zhang}, {Zheng}, {Zheng}, {Zhu}, {Zinn}, \&
  {Zou}}]{SDSSDR152019}
{Aguado}, D.~S., {Ahumada}, R., {Almeida}, A., {et~al.} 2019, \apjs, 240, 23

\bibitem[{{Angthopo} {et~al.}(2019){Angthopo}, {Ferreras}, \&
  {Silk}}]{greenvalley2019}
{Angthopo}, J., {Ferreras}, I., \& {Silk}, J. 2019, \mnras, 488, L99

\bibitem[{{Antonucci}(2012)}]{seyfert2012}
{Antonucci}, R. 2012, Astronomical and Astrophysical Transactions, 27, 557

\bibitem[{{Bai} {et~al.}(2019){Bai}, {Liu}, {Wang}, \& {Yang}}]{rf-qsg-2019}
{Bai}, Y., {Liu}, J., {Wang}, S., \& {Yang}, F. 2019, \aj, 157, 9

\bibitem[{{Baldry} {et~al.}(2010){Baldry}, {Robotham}, {Hill}, {Driver},
  {Liske}, {Norberg}, {Bamford}, {Hopkins}, {Loveday}, {Peacock}, {Cameron},
  {Croom}, {Cross}, {Doyle}, {Dye}, {Frenk}, {Jones}, {van Kampen}, {Kelvin},
  {Nichol}, {Parkinson}, {Popescu}, {Prescott}, {Sharp}, {Sutherland},
  {Thomas}, \& {Tuffs}}]{Baldry2010}
{Baldry}, I.~K., {Robotham}, A.~S.~G., {Hill}, D.~T., {et~al.} 2010, \mnras,
  404, 86

\bibitem[{{Beck} {et~al.}(2016){Beck}, {Dobos}, {Budav{\'a}ri}, {Szalay}, \&
  {Csabai}}]{beck2016photoz}
{Beck}, R., {Dobos}, L., {Budav{\'a}ri}, T., {Szalay}, A.~S., \& {Csabai}, I.
  2016, \mnras, 460, 1371

\bibitem[{{Begelman} {et~al.}(1984){Begelman}, {Blandford}, \&
  {Rees}}]{AGN-radio-1984}
{Begelman}, M.~C., {Blandford}, R.~D., \& {Rees}, M.~J. 1984, Reviews of Modern
  Physics, 56, 255

\bibitem[{{Bolton} {et~al.}(2012){Bolton}, {Schlegel}, {Aubourg}, {Bailey},
  {Bhardwaj}, {Brownstein}, {Burles}, {Chen}, {Dawson}, {Eisenstein}, {Gunn},
  {Knapp}, {Loomis}, {Lupton}, {Maraston}, {Muna}, {Myers}, {Olmstead},
  {Padmanabhan}, {P{\^a}ris}, {Percival}, {Petitjean}, {Rockosi}, {Ross},
  {Schneider}, {Shu}, {Strauss}, {Thomas}, {Tremonti}, {Wake}, {Weaver}, \&
  {Wood-Vasey}}]{SDSS_class_pipeline2012}
{Bolton}, A.~S., {Schlegel}, D.~J., {Aubourg}, {\'E}., {et~al.} 2012, \aj, 144,
  144

\bibitem[{{Burbidge} {et~al.}(1963){Burbidge}, {Burbidge}, \&
  {Sandage}}]{AGN-1963}
{Burbidge}, G.~R., {Burbidge}, E.~M., \& {Sandage}, A.~R. 1963, Reviews of
  Modern Physics, 35, 947

\bibitem[{{Carrasco} {et~al.}(2015){Carrasco}, {Barrientos}, {Pichara},
  {Anguita}, {Murphy}, {Gilbank}, {Gladders}, {Yee}, {Hsieh}, \&
  {L{\'o}pez}}]{rf-quasars-2015}
{Carrasco}, D., {Barrientos}, L.~F., {Pichara}, K., {et~al.} 2015, \aap, 584,
  A44

\bibitem[{{Fix} {et~al.}(2015){Fix}, {Smith}, {Tucker}, {Wester}, \&
  {Annis}}]{newbluequasar2015}
{Fix}, M.~B., {Smith}, J.~A., {Tucker}, D.~L., {Wester}, W., \& {Annis}, J.
  2015, Astronomische Nachrichten, 336, 614

\bibitem[{{Francis} {et~al.}(1991){Francis}, {Hewett}, {Foltz}, {Chaffee},
  {Weymann}, \& {Morris}}]{quasarspec-1991}
{Francis}, P.~J., {Hewett}, P.~C., {Foltz}, C.~B., {et~al.} 1991, \apj, 373,
  465

\bibitem[{{Greenstein} \& {Matthews}(1963)}]{3C48-1963}
{Greenstein}, J.~L. \& {Matthews}, T.~A. 1963, \aj, 68, 279

\bibitem[{{Greenstein} \& {Schmidt}(1964)}]{3C48-3C273-1964}
{Greenstein}, J.~L. \& {Schmidt}, M. 1964, \apj, 140, 1

\bibitem[{{G{\"u}rkan} {et~al.}(2019){G{\"u}rkan}, {Hardcastle}, {Best},
  {Morabito}, {Prandoni}, {Jarvis}, {Duncan}, {Calistro Rivera}, {Callingham},
  {Cochrane}, {Croston}, {Heald}, {Mingo}, {Mooney}, {Sabater},
  {R{\"o}ttgering}, {Shimwell}, {Smith}, {Tasse}, \&
  {Williams}}]{lofar-quasars-2019}
{G{\"u}rkan}, G., {Hardcastle}, M.~J., {Best}, P.~N., {et~al.} 2019, \aap, 622,
  A11

\bibitem[{{Herschel}(1789)}]{herschel1789}
{Herschel}, W. 1789, Philosophical Transactions of the Royal Society of London
  Series I, 79, 212

\bibitem[{{Hubble}(1929)}]{hubble1929}
{Hubble}, E.~P. 1929, \apj, 69, 103

\bibitem[{Hunter(2007)}]{mpl2007}
Hunter, J.~D. 2007, Computing in Science \& Engineering, 9, 90

\bibitem[{{Hutsem{\'e}kers} {et~al.}(2005){Hutsem{\'e}kers}, {Cabanac}, {Lamy},
  \& {Sluse}}]{quasar-pol-alignment-2005}
{Hutsem{\'e}kers}, D., {Cabanac}, R., {Lamy}, H., \& {Sluse}, D. 2005, \aap,
  441, 915

\bibitem[{{Ivezi{\'c}} {et~al.}(2019){Ivezi{\'c}}, {Kahn}, {Tyson}, {Abel},
  {Acosta}, {Allsman}, {Alonso}, {AlSayyad}, {Anderson}, {Andrew}, {Angel},
  {Angeli}, {Ansari}, {Antilogus}, {Araujo}, {Armstrong}, {Arndt}, {Astier},
  {Aubourg}, {Auza}, {Axelrod}, {Bard}, {Barr}, {Barrau}, {Bartlett}, {Bauer},
  {Bauman}, {Baumont}, {Bechtol}, {Bechtol}, {Becker}, {Becla}, {Beldica},
  {Bellavia}, {Bianco}, {Biswas}, {Blanc}, {Blazek}, {Bland ford}, {Bloom},
  {Bogart}, {Bond}, {Booth}, {Borgland}, {Borne}, {Bosch}, {Boutigny},
  {Brackett}, {Bradshaw}, {Brand t}, {Brown}, {Bullock}, {Burchat}, {Burke},
  {Cagnoli}, {Calabrese}, {Callahan}, {Callen}, {Carlin}, {Carlson}, {Chand
  rasekharan}, {Charles-Emerson}, {Chesley}, {Cheu}, {Chiang}, {Chiang},
  {Chirino}, {Chow}, {Ciardi}, {Claver}, {Cohen-Tanugi}, {Cockrum}, {Coles},
  {Connolly}, {Cook}, {Cooray}, {Covey}, {Cribbs}, {Cui}, {Cutri}, {Daly},
  {Daniel}, {Daruich}, {Daubard}, {Daues}, {Dawson}, {Delgado}, {Dellapenna},
  {de Peyster}, {de Val-Borro}, {Digel}, {Doherty}, {Dubois},
  {Dubois-Felsmann}, {Durech}, {Economou}, {Eifler}, {Eracleous}, {Emmons},
  {Fausti Neto}, {Ferguson}, {Figueroa}, {Fisher-Levine}, {Focke}, {Foss},
  {Frank}, {Freemon}, {Gangler}, {Gawiser}, {Geary}, {Gee}, {Geha}, {Gessner},
  {Gibson}, {Gilmore}, {Glanzman}, {Glick}, {Goldina}, {Goldstein}, {Goodenow},
  {Graham}, {Gressler}, {Gris}, {Guy}, {Guyonnet}, {Haller}, {Harris},
  {Hascall}, {Haupt}, {Hernand ez}, {Herrmann}, {Hileman}, {Hoblitt},
  {Hodgson}, {Hogan}, {Howard}, {Huang}, {Huffer}, {Ingraham}, {Innes},
  {Jacoby}, {Jain}, {Jammes}, {Jee}, {Jenness}, {Jernigan}, {Jevremovi{\'c}},
  {Johns}, {Johnson}, {Johnson}, {Jones}, {Juramy-Gilles}, {Juri{\'c}},
  {Kalirai}, {Kallivayalil}, {Kalmbach}, {Kantor}, {Karst}, {Kasliwal},
  {Kelly}, {Kessler}, {Kinnison}, {Kirkby}, {Knox}, {Kotov}, {Krabbendam},
  {Krughoff}, {Kub{\'a}nek}, {Kuczewski}, {Kulkarni}, {Ku}, {Kurita}, {Lage},
  {Lambert}, {Lange}, {Langton}, {Le Guillou}, {Levine}, {Liang}, {Lim},
  {Lintott}, {Long}, {Lopez}, {Lotz}, {Lupton}, {Lust}, {MacArthur}, {Mahabal},
  {Mand elbaum}, {Markiewicz}, {Marsh}, {Marshall}, {Marshall}, {May},
  {McKercher}, {McQueen}, {Meyers}, {Migliore}, {Miller}, {Mills}, {Miraval},
  {Moeyens}, {Moolekamp}, {Monet}, {Moniez}, {Monkewitz}, {Montgomery},
  {Morrison}, {Mueller}, {Muller}, {Mu{\~n}oz Arancibia}, {Neill}, {Newbry},
  {Nief}, {Nomerotski}, {Nordby}, {O'Connor}, {Oliver}, {Olivier}, {Olsen},
  {O'Mullane}, {Ortiz}, {Osier}, {Owen}, {Pain}, {Palecek}, {Parejko},
  {Parsons}, {Pease}, {Peterson}, {Peterson}, {Petravick}, {Libby Petrick},
  {Petry}, {Pierfederici}, {Pietrowicz}, {Pike}, {Pinto}, {Plante}, {Plate},
  {Plutchak}, {Price}, {Prouza}, {Radeka}, {Rajagopal}, {Rasmussen},
  {Regnault}, {Reil}, {Reiss}, {Reuter}, {Ridgway}, {Riot}, {Ritz}, {Robinson},
  {Roby}, {Roodman}, {Rosing}, {Roucelle}, {Rumore}, {Russo}, {Saha},
  {Sassolas}, {Schalk}, {Schellart}, {Schindler}, {Schmidt}, {Schneider},
  {Schneider}, {Schoening}, {Schumacher}, {Schwamb}, {Sebag}, {Selvy},
  {Sembroski}, {Seppala}, {Serio}, {Serrano}, {Shaw}, {Shipsey}, {Sick},
  {Silvestri}, {Slater}, {Smith}, {Smith}, {Sobhani}, {Soldahl},
  {Storrie-Lombardi}, {Stover}, {Strauss}, {Street}, {Stubbs}, {Sullivan},
  {Sweeney}, {Swinbank}, {Szalay}, {Takacs}, {Tether}, {Thaler}, {Thayer},
  {Thomas}, {Thornton}, {Thukral}, {Tice}, {Trilling}, {Turri}, {Van Berg},
  {Vanden Berk}, {Vetter}, {Virieux}, {Vucina}, {Wahl}, {Walkowicz}, {Walsh},
  {Walter}, {Wang}, {Wang}, {Warner}, {Wiecha}, {Willman}, {Winters},
  {Wittman}, {Wolff}, {Wood-Vasey}, {Wu}, {Xin}, {Yoachim}, \&
  {Zhan}}]{LSST2019}
{Ivezi{\'c}}, {\v{Z}}., {Kahn}, S.~M., {Tyson}, J.~A., {et~al.} 2019, \apj,
  873, 111

\bibitem[{{Jarvis} {et~al.}(2015){Jarvis}, {Bacon}, {Blake}, {Brown},
  {Lindsay}, {Raccanelli}, {Santos}, \& {Schwarz}}]{SKA2015}
{Jarvis}, M., {Bacon}, D., {Blake}, C., {et~al.} 2015, Advancing Astrophysics
  with the Square Kilometre Array (AASKA14), 18

\bibitem[{{Jones}(2014)}]{MLreview-2014}
{Jones}, N. 2014, \nat, 505, 146

\bibitem[{{Kang} {et~al.}(2019){Kang}, {Fan}, {Mao}, {Wu}, {Feng}, \&
  {Yin}}]{rf-BCUs-2019}
{Kang}, S.-J., {Fan}, J.-H., {Mao}, W., {et~al.} 2019, \apj, 872, 189

\bibitem[{{Kauffmann} \& {Haehnelt}(2000)}]{galaxyquasar-evolution2000}
{Kauffmann}, G. \& {Haehnelt}, M. 2000, \mnras, 311, 576

\bibitem[{{Leistedt} \& {Peiris}(2014)}]{quasar-powerspectrum-2014}
{Leistedt}, B. \& {Peiris}, H.~V. 2014, \mnras, 444, 2

\bibitem[{{Lintott} {et~al.}(2011){Lintott}, {Schawinski}, {Bamford}, {Slosar},
  {Land}, {Thomas}, {Edmondson}, {Masters}, {Nichol}, {Raddick}, {Szalay},
  {Andreescu}, {Murray}, \& {Vandenberg}}]{galaxyzoo2011}
{Lintott}, C., {Schawinski}, K., {Bamford}, S., {et~al.} 2011, \mnras, 410, 166

\bibitem[{{Lintott} {et~al.}(2008){Lintott}, {Schawinski}, {Slosar}, {Land},
  {Bamford}, {Thomas}, {Raddick}, {Nichol}, {Szalay}, {Andreescu}, {Murray}, \&
  {Vandenberg}}]{galaxyzoo2008}
{Lintott}, C.~J., {Schawinski}, K., {Slosar}, A., {et~al.} 2008, \mnras, 389,
  1179

\bibitem[{{Louppe}(2014)}]{RF2014}
{Louppe}, G. 2014, arXiv e-prints, arXiv:1407.7502

\bibitem[{{LSST Dark Energy Science Collaboration}(2012)}]{LSST2012}
{LSST Dark Energy Science Collaboration}. 2012, arXiv e-prints, arXiv:1211.0310

\bibitem[{{LSST Science Collaboration} {et~al.}(2009){LSST Science
  Collaboration}, {Abell}, {Allison}, {Anderson}, {Andrew}, {Angel}, {Armus},
  {Arnett}, {Asztalos}, {Axelrod}, {Bailey}, {Ballantyne}, {Bankert},
  {Barkhouse}, {Barr}, {Barrientos}, {Barth}, {Bartlett}, {Becker}, {Becla},
  {Beers}, {Bernstein}, {Biswas}, {Blanton}, {Bloom}, {Bochanski}, {Boeshaar},
  {Borne}, {Bradac}, {Brandt}, {Bridge}, {Brown}, {Brunner}, {Bullock},
  {Burgasser}, {Burge}, {Burke}, {Cargile}, {Chand rasekharan}, {Chartas},
  {Chesley}, {Chu}, {Cinabro}, {Claire}, {Claver}, {Clowe}, {Connolly}, {Cook},
  {Cooke}, {Cooray}, {Covey}, {Culliton}, {de Jong}, {de Vries}, {Debattista},
  {Delgado}, {Dell'Antonio}, {Dhital}, {Di Stefano}, {Dickinson}, {Dilday},
  {Djorgovski}, {Dobler}, {Donalek}, {Dubois-Felsmann}, {Durech},
  {Eliasdottir}, {Eracleous}, {Eyer}, {Falco}, {Fan}, {Fassnacht}, {Ferguson},
  {Fernandez}, {Fields}, {Finkbeiner}, {Figueroa}, {Fox}, {Francke}, {Frank},
  {Frieman}, {Fromenteau}, {Furqan}, {Galaz}, {Gal-Yam}, {Garnavich},
  {Gawiser}, {Geary}, {Gee}, {Gibson}, {Gilmore}, {Grace}, {Green}, {Gressler},
  {Grillmair}, {Habib}, {Haggerty}, {Hamuy}, {Harris}, {Hawley}, {Heavens},
  {Hebb}, {Henry}, {Hileman}, {Hilton}, {Hoadley}, {Holberg}, {Holman},
  {Howell}, {Infante}, {Ivezic}, {Jacoby}, {Jain}, {R}, {Jedicke}, {Jee},
  {Garrett Jernigan}, {Jha}, {Johnston}, {Jones}, {Juric}, {Kaasalainen},
  {Styliani}, {Kafka}, {Kahn}, {Kaib}, {Kalirai}, {Kantor}, {Kasliwal},
  {Keeton}, {Kessler}, {Knezevic}, {Kowalski}, {Krabbendam}, {Krughoff},
  {Kulkarni}, {Kuhlman}, {Lacy}, {Lepine}, {Liang}, {Lien}, {Lira}, {Long},
  {Lorenz}, {Lotz}, {Lupton}, {Lutz}, {Macri}, {Mahabal}, {Mandelbaum},
  {Marshall}, {May}, {McGehee}, {Meadows}, {Meert}, {Milani}, {Miller},
  {Miller}, {Mills}, {Minniti}, {Monet}, {Mukadam}, {Nakar}, {Neill}, {Newman},
  {Nikolaev}, {Nordby}, {O'Connor}, {Oguri}, {Oliver}, {Olivier}, {Olsen},
  {Olsen}, {Olszewski}, {Oluseyi}, {Padilla}, {Parker}, {Pepper}, {Peterson},
  {Petry}, {Pinto}, {Pizagno}, {Popescu}, {Prsa}, {Radcka}, {Raddick},
  {Rasmussen}, {Rau}, {Rho}, {Rhoads}, {Richards}, {Ridgway}, {Robertson},
  {Roskar}, {Saha}, {Sarajedini}, {Scannapieco}, {Schalk}, {Schindler},
  {Schmidt}, {Schmidt}, {Schneider}, {Schumacher}, {Scranton}, {Sebag},
  {Seppala}, {Shemmer}, {Simon}, {Sivertz}, {Smith}, {Allyn Smith}, {Smith},
  {Spitz}, {Stanford}, {Stassun}, {Strader}, {Strauss}, {Stubbs}, {Sweeney},
  {Szalay}, {Szkody}, {Takada}, {Thorman}, {Trilling}, {Trimble}, {Tyson}, {Van
  Berg}, {Vand en Berk}, {VanderPlas}, {Verde}, {Vrsnak}, {Walkowicz}, {Wand
  elt}, {Wang}, {Wang}, {Warner}, {Wechsler}, {West}, {Wiecha}, {Williams},
  {Willman}, {Wittman}, {Wolff}, {Wood-Vasey}, {Wozniak}, {Young}, {Zentner},
  \& {Zhan}}]{LSST2009}
{LSST Science Collaboration}, {Abell}, P.~A., {Allison}, J., {et~al.} 2009,
  arXiv e-prints, arXiv:0912.0201

\bibitem[{{Masci}(1998)}]{dustquasars1998}
{Masci}, F.~J. 1998, PhD thesis, -

\bibitem[{{Matthews} \& {Sandage}(1963)}]{3C48-1963B}
{Matthews}, T.~A. \& {Sandage}, A.~R. 1963, \apj, 138, 30

\bibitem[{{McDonald} \& {Eisenstein}(2007)}]{quasar-lyaforest-2007}
{McDonald}, P. \& {Eisenstein}, D.~J. 2007, \prd, 76, 063009

\bibitem[{{McInnes} {et~al.}(2018){McInnes}, {Healy}, \& {Melville}}]{UMAP}
{McInnes}, L., {Healy}, J., \& {Melville}, J. 2018, arXiv e-prints,
  arXiv:1802.03426

\bibitem[{McKinney {et~al.}(2010)}]{pandas}
McKinney, W. {et~al.} 2010, in Proceedings of the 9th Python in Science
  Conference, Vol. 445, Austin, TX, 51--56

\bibitem[{{Morice-Atkinson} {et~al.}(2018){Morice-Atkinson}, {Hoyle}, \&
  {Bacon}}]{xan2018}
{Morice-Atkinson}, X., {Hoyle}, B., \& {Bacon}, D. 2018, \mnras, 481, 4194

\bibitem[{Mosteller \& Tukey(1968)}]{cross-validation1968}
Mosteller, F. \& Tukey, J.~W. 1968, in Handbook of Social Psychology, Vol. 2,
  ed. G.~Lindzey \& E.~Aronson (Addison-Wesley)

\bibitem[{{Nakoneczny} {et~al.}(2019){Nakoneczny}, {Bilicki}, {Solarz},
  {Pollo}, {Maddox}, {Spiniello}, {Brescia}, \&
  {Napolitano}}]{rf-quasars-KIDS-2019}
{Nakoneczny}, S., {Bilicki}, M., {Solarz}, A., {et~al.} 2019, \aap, 624, A13

\bibitem[{{Nikutta} {et~al.}(2014){Nikutta}, {Hunt-Walker}, {Nenkova},
  {Ivezi{\'c}}, \& {Elitzur}}]{WISE-colours-2014}
{Nikutta}, R., {Hunt-Walker}, N., {Nenkova}, M., {Ivezi{\'c}}, {\v{Z}}., \&
  {Elitzur}, M. 2014, \mnras, 442, 3361

\bibitem[{Oliphant(2006)}]{numpy2006}
Oliphant, T.~E. 2006, A guide to NumPy, Vol.~1 (Trelgol Publishing USA)

\bibitem[{{Opik}(1922)}]{Opik1922}
{Opik}, E. 1922, \apj, 55, 406

\bibitem[{{P{\^a}ris} {et~al.}(2018){P{\^a}ris}, {Petitjean}, {Aubourg},
  {Myers}, {Streblyanska}, {Lyke}, {Anderson}, {Armengaud}, {Bautista},
  {Blanton}, {Blomqvist}, {Brinkmann}, {Brownstein}, {Brand t}, {Burtin},
  {Dawson}, {de la Torre}, {Georgakakis}, {Gil-Mar{\'\i}n}, {Green}, {Hall},
  {Kneib}, {LaMassa}, {Le Goff}, {MacLeod}, {Mariappan}, {McGreer}, {Merloni},
  {Noterdaeme}, {Palanque-Delabrouille}, {Percival}, {Ross}, {Rossi},
  {Schneider}, {Seo}, {Tojeiro}, {Weaver}, {Weijmans}, {Y{\`e}che}, {Zarrouk},
  \& {Zhao}}]{SDSS-dr14-quasars-2018}
{P{\^a}ris}, I., {Petitjean}, P., {Aubourg}, {\'E}., {et~al.} 2018, \aap, 613,
  A51

\bibitem[{Pedregosa {et~al.}(2011)Pedregosa, Varoquaux, Gramfort, Michel,
  Thirion, Grisel, Blondel, Prettenhofer, Weiss, Dubourg, Vanderplas, Passos,
  Cournapeau, Brucher, Perrot, \& Duchesnay}]{scikit-learn}
Pedregosa, F., Varoquaux, G., Gramfort, A., {et~al.} 2011, Journal of Machine
  Learning Research, 12, 2825

\bibitem[{{Peters} {et~al.}(2015){Peters}, {Richards}, {Myers}, {Strauss},
  {Schmidt}, {Ivezi{\'c}}, {Ross}, {MacLeod}, \& {Riegel}}]{quasarcolours2015}
{Peters}, C.~M., {Richards}, G.~T., {Myers}, A.~D., {et~al.} 2015, \apj, 811,
  95

\bibitem[{Pratt {et~al.}(1991)Pratt, Mostow, Kamm, \& Kamm}]{pratt1991}
Pratt, L.~Y., Mostow, J., Kamm, C.~A., \& Kamm, A.~A. 1991, in AAAI, Vol.~91,
  584--589

\bibitem[{{Rauch}(1998)}]{quasar-lyaforest-1998}
{Rauch}, M. 1998, \araa, 36, 267

\bibitem[{{Rees}(1984)}]{AGN-1984}
{Rees}, M.~J. 1984, \araa, 22, 471

\bibitem[{{Salim}(2014)}]{greenvalley2014}
{Salim}, S. 2014, Serbian Astronomical Journal, 189, 1

\bibitem[{{Sanders} {et~al.}(1988){Sanders}, {Soifer}, {Elias}, {Madore},
  {Matthews}, {Neugebauer}, \& {Scoville}}]{quasar-origins1988}
{Sanders}, D.~B., {Soifer}, B.~T., {Elias}, J.~H., {et~al.} 1988, \apj, 325, 74

\bibitem[{{Schindler} {et~al.}(2019){Schindler}, {Fan}, {Huang}, {Yue}, {Yang},
  {Hall}, {Wenzl}, {Hughes}, {Litke}, \& {Rees}}]{rf-quasars-2019}
{Schindler}, J.-T., {Fan}, X., {Huang}, Y.-H., {et~al.} 2019, \apjs, 243, 5

\bibitem[{{Schlegel} {et~al.}(1998){Schlegel}, {Finkbeiner}, \&
  {Davis}}]{extinction1998}
{Schlegel}, D.~J., {Finkbeiner}, D.~P., \& {Davis}, M. 1998, \apj, 500, 525

\bibitem[{{Schmidt}(1963)}]{3C273-1963}
{Schmidt}, M. 1963, \nat, 197, 1040

\bibitem[{{Schmidt} \& {Green}(1983)}]{quasar-evolution1983}
{Schmidt}, M. \& {Green}, R.~F. 1983, \apj, 269, 352

\bibitem[{{Scranton} {et~al.}(2005){Scranton}, {M{\'e}nard}, {Richards},
  {Nichol}, {Myers}, {Jain}, {Gray}, {Bartelmann}, {Brunner}, {Connolly},
  {Gunn}, {Sheth}, {Bahcall}, {Brinkman}, {Loveday}, {Schneider}, {Thakar}, \&
  {York}}]{quasar-cosmicmagnetism-2005}
{Scranton}, R., {M{\'e}nard}, B., {Richards}, G.~T., {et~al.} 2005, \apj, 633,
  589

\bibitem[{{Shimwell} {et~al.}(2019){Shimwell}, {Tasse}, {Hardcastle}, {Mechev},
  {Williams}, {Best}, {R{\"o}ttgering}, {Callingham}, {Dijkema}, {de Gasperin},
  {Hoang}, {Hugo}, {Mirmont}, {Oonk}, {Prandoni}, {Rafferty}, {Sabater},
  {Smirnov}, {van Weeren}, {White}, {Atemkeng}, {Bester}, {Bonnassieux},
  {Br{\"u}ggen}, {Brunetti}, {Chy{\.z}y}, {Cochrane}, {Conway}, {Croston},
  {Danezi}, {Duncan}, {Haverkorn}, {Heald}, {Iacobelli}, {Intema}, {Jackson},
  {Jamrozy}, {Jarvis}, {Lakhoo}, {Mevius}, {Miley}, {Morabito}, {Morganti},
  {Nisbet}, {Orr{\'u}}, {Perkins}, {Pizzo}, {Schrijvers}, {Smith}, {Vermeulen},
  {Wise}, {Alegre}, {Bacon}, {van Bemmel}, {Beswick}, {Bonafede}, {Botteon},
  {Bourke}, {Brienza}, {Calistro Rivera}, {Cassano}, {Clarke}, {Conselice},
  {Dettmar}, {Drabent}, {Dumba}, {Emig}, {En{\ss}lin}, {Ferrari}, {Garrett},
  {G{\'e}nova-Santos}, {Goyal}, {G{\"u}rkan}, {Hale}, {Harwood}, {Heesen},
  {Hoeft}, {Horellou}, {Jackson}, {Kokotanekov}, {Kondapally},
  {Kunert-Bajraszewska}, {Mahatma}, {Mahony}, {Mandal}, {McKean}, {Merloni},
  {Mingo}, {Miskolczi}, {Mooney}, {Nikiel-Wroczy{\'n}ski}, {O'Sullivan},
  {Quinn}, {Reich}, {Roskowi{\'n}ski}, {Rowlinson}, {Savini}, {Saxena},
  {Schwarz}, {Shulevski}, {Sridhar}, {Stacey}, {Urquhart}, {van der Wiel},
  {Varenius}, {Webster}, \& {Wilber}}]{LOTSS2019}
{Shimwell}, T.~W., {Tasse}, C., {Hardcastle}, M.~J., {et~al.} 2019, \aap, 622,
  A1

\bibitem[{{Smith} \& {Hoffleit}(1961)}]{3C48-1961}
{Smith}, H.~J. \& {Hoffleit}, D. 1961, \aj, 70, 295

\bibitem[{{Stoughton} {et~al.}(2002){Stoughton}, {Lupton}, {Bernardi},
  {Blanton}, {Burles}, {Castand er}, {Connolly}, {Eisenstein}, {Frieman},
  {Hennessy}, {Hindsley}, {Ivezi{\'c}}, {Kent}, {Kunszt}, {Lee}, {Meiksin},
  {Munn}, {Newberg}, {Nichol}, {Nicinski}, {Pier}, {Richards}, {Richmond},
  {Schlegel}, {Smith}, {Strauss}, {SubbaRao}, {Szalay}, {Thakar}, {Tucker},
  {Vand en Berk}, {Yanny}, {Adelman}, {Anderson}, {Anderson}, {Annis},
  {Bahcall}, {Bakken}, {Bartelmann}, {Bastian}, {Bauer}, {Berman},
  {B{\"o}hringer}, {Boroski}, {Bracker}, {Briegel}, {Briggs}, {Brinkmann},
  {Brunner}, {Carey}, {Carr}, {Chen}, {Christian}, {Colestock}, {Crocker},
  {Csabai}, {Czarapata}, {Dalcanton}, {Davidsen}, {Davis}, {Dehnen},
  {Dodelson}, {Doi}, {Dombeck}, {Donahue}, {Ellman}, {Elms}, {Evans}, {Eyer},
  {Fan}, {Federwitz}, {Friedman}, {Fukugita}, {Gal}, {Gillespie}, {Glazebrook},
  {Gray}, {Grebel}, {Greenawalt}, {Greene}, {Gunn}, {de Haas}, {Haiman},
  {Haldeman}, {Hall}, {Hamabe}, {Hansen}, {Harris}, {Harris}, {Harvanek},
  {Hawley}, {Hayes}, {Heckman}, {Helmi}, {Henden}, {Hogan}, {Hogg}, {Holmgren},
  {Holtzman}, {Huang}, {Hull}, {Ichikawa}, {Ichikawa}, {Johnston}, {Kauffmann},
  {Kim}, {Kimball}, {Kinney}, {Klaene}, {Kleinman}, {Klypin}, {Knapp},
  {Korienek}, {Krolik}, {Kron}, {Krzesi{\'n}ski}, {Lamb}, {Leger},
  {Limmongkol}, {Lindenmeyer}, {Long}, {Loomis}, {Loveday}, {MacKinnon},
  {Mannery}, {Mantsch}, {Margon}, {McGehee}, {McKay}, {McLean}, {Menou},
  {Merelli}, {Mo}, {Monet}, {Nakamura}, {Narayanan}, {Nash}, {Neilsen},
  {Newman}, {Nitta}, {Odenkirchen}, {Okada}, {Okamura}, {Ostriker}, {Owen},
  {Pauls}, {Peoples}, {Peterson}, {Petravick}, {Pope}, {Pordes}, {Postman},
  {Prosapio}, {Quinn}, {Rechenmacher}, {Rivetta}, {Rix}, {Rockosi}, {Rosner},
  {Ruthmansdorfer}, {Sandford}, {Schneider}, {Scranton}, {Sekiguchi}, {Sergey},
  {Sheth}, {Shimasaku}, {Smee}, {Snedden}, {Stebbins}, {Stubbs}, {Szapudi},
  {Szkody}, {Szokoly}, {Tabachnik}, {Tsvetanov}, {Uomoto}, {Vogeley}, {Voges},
  {Waddell}, {Walterbos}, {Wang}, {Watanabe}, {Weinberg}, {White}, {White},
  {Wilhite}, {Wolfe}, {Yasuda}, {York}, {Zehavi}, \&
  {Zheng}}]{SDSS_catdescription2002}
{Stoughton}, C., {Lupton}, R.~H., {Bernardi}, M., {et~al.} 2002, \aj, 123, 485

\bibitem[{{Tang} {et~al.}(2019){Tang}, {Scaife}, \& {Leahy}}]{tang2019}
{Tang}, H., {Scaife}, A.~M.~M., \& {Leahy}, J.~P. 2019, \mnras, 488, 3358

\bibitem[{{Urry} \& {Padovani}(1995)}]{AGNreview1995}
{Urry}, C.~M. \& {Padovani}, P. 1995, \pasp, 107, 803

\bibitem[{van~der Maaten \& Hinton(2008)}]{tsne2008}
van~der Maaten, L. \& Hinton, G. 2008, Journal of Machine Learning Research, 9,
  2579

\bibitem[{{van Haarlem} {et~al.}(2013){van Haarlem}, {Wise}, {Gunst}, {Heald},
  {McKean}, {Hessels}, {de Bruyn}, {Nijboer}, {Swinbank}, {Fallows},
  {Brentjens}, {Nelles}, {Beck}, {Falcke}, {Fender}, {H{\"o}randel},
  {Koopmans}, {Mann}, {Miley}, {R{\"o}ttgering}, {Stappers}, {Wijers},
  {Zaroubi}, {van den Akker}, {Alexov}, {Anderson}, {Anderson}, {van Ardenne},
  {Arts}, {Asgekar}, {Avruch}, {Batejat}, {B{\"a}hren}, {Bell}, {Bell}, {van
  Bemmel}, {Bennema}, {Bentum}, {Bernardi}, {Best}, {B{\^\i}rzan}, {Bonafede},
  {Boonstra}, {Braun}, {Bregman}, {Breitling}, {van de Brink}, {Broderick},
  {Broekema}, {Brouw}, {Br{\"u}ggen}, {Butcher}, {van Cappellen}, {Ciardi},
  {Coenen}, {Conway}, {Coolen}, {Corstanje}, {Damstra}, {Davies}, {Deller},
  {Dettmar}, {van Diepen}, {Dijkstra}, {Donker}, {Doorduin}, {Dromer}, {Drost},
  {van Duin}, {Eisl{\"o}ffel}, {van Enst}, {Ferrari}, {Frieswijk}, {Gankema},
  {Garrett}, {de Gasperin}, {Gerbers}, {de Geus}, {Grie{\ss}meier}, {Grit},
  {Gruppen}, {Hamaker}, {Hassall}, {Hoeft}, {Holties}, {Horneffer}, {van der
  Horst}, {van Houwelingen}, {Huijgen}, {Iacobelli}, {Intema}, {Jackson},
  {Jelic}, {de Jong}, {Juette}, {Kant}, {Karastergiou}, {Koers}, {Kollen},
  {Kondratiev}, {Kooistra}, {Koopman}, {Koster}, {Kuniyoshi}, {Kramer},
  {Kuper}, {Lambropoulos}, {Law}, {van Leeuwen}, {Lemaitre}, {Loose}, {Maat},
  {Macario}, {Markoff}, {Masters}, {McFadden}, {McKay-Bukowski}, {Meijering},
  {Meulman}, {Mevius}, {Middelberg}, {Millenaar}, {Miller-Jones}, {Mohan},
  {Mol}, {Morawietz}, {Morganti}, {Mulcahy}, {Mulder}, {Munk}, {Nieuwenhuis},
  {van Nieuwpoort}, {Noordam}, {Norden}, {Noutsos}, {Offringa}, {Olofsson},
  {Omar}, {Orr{\'u}}, {Overeem}, {Paas}, {Pand ey-Pommier}, {Pandey}, {Pizzo},
  {Polatidis}, {Rafferty}, {Rawlings}, {Reich}, {de Reijer}, {Reitsma},
  {Renting}, {Riemers}, {Rol}, {Romein}, {Roosjen}, {Ruiter}, {Scaife}, {van
  der Schaaf}, {Scheers}, {Schellart}, {Schoenmakers}, {Schoonderbeek},
  {Serylak}, {Shulevski}, {Sluman}, {Smirnov}, {Sobey}, {Spreeuw}, {Steinmetz},
  {Sterks}, {Stiepel}, {Stuurwold}, {Tagger}, {Tang}, {Tasse}, {Thomas},
  {Thoudam}, {Toribio}, {van der Tol}, {Usov}, {van Veelen}, {van der Veen},
  {ter Veen}, {Verbiest}, {Vermeulen}, {Vermaas}, {Vocks}, {Vogt}, {de Vos},
  {van der Wal}, {van Weeren}, {Weggemans}, {Weltevrede}, {White}, {Wijnholds},
  {Wilhelmsson}, {Wucknitz}, {Yatawatta}, {Zarka}, {Zensus}, \& {van
  Zwieten}}]{LOFAR2013}
{van Haarlem}, M.~P., {Wise}, M.~W., {Gunst}, A.~W., {et~al.} 2013, \aap, 556,
  A2

\bibitem[{{Vanden Berk} {et~al.}(2001){Vanden Berk}, {Richards}, {Bauer},
  {Strauss}, {Schneider}, {Heckman}, {York}, {Hall}, {Fan}, {Knapp},
  {Anderson}, {Annis}, {Bahcall}, {Bernardi}, {Briggs}, {Brinkmann}, {Brunner},
  {Burles}, {Carey}, {Castander}, {Connolly}, {Crocker}, {Csabai}, {Doi},
  {Finkbeiner}, {Friedman}, {Frieman}, {Fukugita}, {Gunn}, {Hennessy},
  {Ivezi{\'c}}, {Kent}, {Kunszt}, {Lamb}, {Leger}, {Long}, {Loveday}, {Lupton},
  {Meiksin}, {Merelli}, {Munn}, {Newberg}, {Newcomb}, {Nichol}, {Owen}, {Pier},
  {Pope}, {Rockosi}, {Schlegel}, {Siegmund}, {Smee}, {Snir}, {Stoughton},
  {Stubbs}, {SubbaRao}, {Szalay}, {Szokoly}, {Tremonti}, {Uomoto}, {Waddell},
  {Yanny}, \& {Zheng}}]{quasarspec-SDSS-2001}
{Vanden Berk}, D.~E., {Richards}, G.~T., {Bauer}, A., {et~al.} 2001, \aj, 122,
  549

\bibitem[{{Weedman}(1977)}]{seyfert1977}
{Weedman}, D.~W. 1977, \araa, 15, 69

\bibitem[{Wilson(1927)}]{wilson1927}
Wilson, E.~B. 1927, Journal of the American Statistical Association, 22, 209

\bibitem[{{Wright} {et~al.}(2010){Wright}, {Eisenhardt}, {Mainzer}, {Ressler},
  {Cutri}, {Jarrett}, {Kirkpatrick}, {Padgett}, {McMillan}, {Skrutskie},
  {Stanford}, {Cohen}, {Walker}, {Mather}, {Leisawitz}, {Gautier}, {McLean},
  {Benford}, {Lonsdale}, {Blain}, {Mendez}, {Irace}, {Duval}, {Liu}, {Royer},
  {Heinrichsen}, {Howard}, {Shannon}, {Kendall}, {Walsh}, {Larsen}, {Cardon},
  {Schick}, {Schwalm}, {Abid}, {Fabinsky}, {Naes}, \& {Tsai}}]{WISE2010}
{Wright}, E.~L., {Eisenhardt}, P. R.~M., {Mainzer}, A.~K., {et~al.} 2010, \aj,
  140, 1868

\bibitem[{{Wu} {et~al.}(2016){Wu}, {Buyya}, \& {Ramamohanarao}}]{bigdata2016}
{Wu}, C., {Buyya}, R., \& {Ramamohanarao}, K. 2016, arXiv e-prints,
  arXiv:1601.03115

\end{thebibliography}

\end{document}